\documentclass[12pt]{article}
\usepackage{jheppub}
\newcommand\bC{\mathbb{C}}
\newcommand\bR{\mathbb{R}}
\newcommand\bZ{\mathbb{Z}}

\newcommand\cH{\mathcal H}
\newcommand\cG{\mathcal G}
\newcommand\cI{\mathcal I}
\newcommand\cL{\mathcal L}
\newcommand\cM{\mathcal M}
\newcommand\cN{\mathcal N}
\newcommand\cP{\mathcal P}
\newcommand\cV{\mathcal V}
\newcommand\cU{\mathcal U}

\newcommand\fsl{\mathfrak{sl}}
\newcommand\fgl{\mathfrak{gl}}
\newcommand\ff{\mathfrak{f}}
\newcommand\fg{\mathfrak{g}}
\newcommand\fh{\mathfrak{h}}
\newcommand\fso{\mathfrak{so}}
\newcommand\Tr{\mathrm{Tr}}
\newcommand\tr{\mathrm{tr}}
\newcommand\op{\mathrm{op}}
\newcommand\cl{\mathrm{cl}}
\newcommand\Vol{\mathrm{Vol}}
\newcommand\wt{\widetilde}
\title{Sphere quantization of Higgs and Coulomb branches and Analytic Symplectic Duality}
\author{Davide Gaiotto}
\affiliation{Perimeter Institute for Theoretical Physics, Waterloo, ON N2L 2Y5, Canada}

\abstract{We employ the protected sphere correlation functions of three-dimensional Super Conformal Field Theories with eight supercharges in order to define a quantization of their Higgs and Coulomb branches of vacua as real phase spaces. We also employ hemisphere correlation functions to define a quantization of certain real loci in the Higgs and Coulomb branches. Localization formulae and dualities applied to these quantizations result in a body of predictions about unitary representations of certain algebras, which may perhaps be understood as an ``analytic'' form of the symplectic duality program. In particular, the protected correlation functions in the class of theories denoted as $T[G]$ are naturally related to the theory of unitary representations of complex or real semi-simple Lie groups.}
\begin{document}
\maketitle
\section{Introduction}
The objective of this paper is define a general quantization strategy for a large class of complex symplectic manifolds: the Higgs- or Coulomb- branches of supersymmetric vacua of 3d ${\cal N}=4$ Super-Conformal Field Theories. The strategy is a three-dimensional lift of brane quantization \cite{Gukov:2008ve,Kapustin:2001ij,Bressler:2002eu,Kapustin:2005vs,Pestun:2006rj,Gualtieri:2007bq,Aldi:2005hz}, 
and employs the protected sphere correlation functions of 3d SCFTs \cite{Kapustin:2010xq,Chester:2014mea,Beem:2016cbd,Dedushenko:2016jxl,Dedushenko:2017avn,Dedushenko:2018icp}. In the rest of this paper we will refer to our approach as sphere quantization. \footnote{The name also resonates with the appearance of ``spherical'' vectors in the resulting unitary representations
for the quantized algebras of (anti)-holomorphic functions on the Higgs- or Coulomb- branches.}

Sphere quantization is fully explicit whenever the 3d SCFTs admit a Lagrangian description: sphere correlation functions 
are exactly computable via supersymmetric localization. This should be contrasted to brane quantization, 
which is defined for a larger class of manifold but is not directly computable unless it can be related to 
alternative constructions such as geometric quantization \cite{Gaiotto:2021kma}.\footnote{It may be possible to formulate a 2d analogue of the sphere setup, amenable of exact localization formulae. This would give a direct computation strategy for brane quantization.} Even when no Lagrangian description is available, 
the procedure is strongly constrained and we will often be able to give explicit answers. 

In order to formulate our proposal, we need to introduce some notation. Any 3d ${\cal N}=4$ SCFT $T$ is associated to two hyper-k\"ahler cones: the Higgs and Coulomb branches of supersymmetric vacua. We denote  the Higgs branch as a complex symplectic manifold as $\cM[T]$. The Higgs branch chiral ring is a collection of half-BPS operators with non-singular OPE, whose expectation values parameterize $\cM[T]$. We denote it as $A_\cl[T]$. We also denote the analogous Coulomb branch data as  $\cM[T^!]$ and $A_\cl[T^!]$, as a nod to the fact that the Coulomb branch of $T$ is, by definition, the Higgs branch of the 3d mirror theory $T^!$ \cite{Intriligator:1996ex}.  
 
Sphere quantization is a quantization of $\cM$ as a real phase space equipped with the imaginary part $\mathrm{Im} \, \Omega$ of the complex symplectic form $\Omega$. The (Coulomb)Higgs branch algebra $A_\cl$ has a canonical deformation quantization \cite{Yagi:2014toa,Bullimore:2015lsa,Nakajima:2015txa,Braverman:2016wma} we denote as $A$. Sphere quantization produces an Hilbert space $\cH$ equipped with an irreducible action of $A\otimes A^\op$, representing the quantization of holomorphic and anti-holomorphic functions on the phase space.

By construction, sphere quantization respects any symmetries of $\cM$ which are inherited from $T$. It only depends on $T$ and not on the specific description we give to the theory. 
In particular, it is invariant under dualities for $T$. This will have interesting mathematical consequences. 

\subsection{Quantization from positive traces}
The key observation, which goes back to the bootstrap interpretation of sphere correlation functions \cite{Chester:2014mea,Beem:2016cbd,Etingof:2019guc,Etingof:2020fls}, is that the protected sphere correlation functions take the form of a {\it positive twisted trace} on $A$: a linear map $\Tr : A \to \bC$ such that 
\begin{equation} \label{eq:potr}
	\Tr \,ab = \Tr \,\rho^2(b) \, a \qquad \qquad \Tr\, \rho(a) a >0 \, ,
\end{equation}
where $\rho$ is an anti-linear automorphism of $A$. In our setup, $\rho^2$ is a $\bZ_2$ symmetry which acts as $-1$ on operators of half-integral scaling dimension. The map $\rho$ quantizes the anti-linear map $\rho_\cl$ defined on $A_\cl$ by composing complex conjugation with a specific hyper-k\"ahler rotation mapping anti-holomorphic functions on $\cM$ back to holomorphic 
functions.

The twisted trace $\Tr$ equips $A$ itself with a positive-definite Hermitian inner product $\Tr\, \rho(a) b$. We can thus define an Hilbert space $\cH$ as the completion of $A$ with respect of this inner product. The left- and right- actions of $A$ on itself define unbounded operators with dense domain $A \subset \cH$. If we denote as $|a\rangle$ the image in $\cH$ of the element $a \in A$, we can denote the two actions as 
\begin{equation}
	a |b\rangle \equiv |a\,b\rangle \qquad \qquad \wt a |b\rangle \equiv |b\,a\rangle \, .
\end{equation}
The two action commute with each other and give an action of $A\otimes A^\op$.

We have 
\begin{equation}
	\langle b\, a |c\rangle = \Tr \,\rho(b\,a) \,c  = \langle b| \rho(a)|c\rangle \, .
\end{equation}
This implies that the domain of $\wt a^\dagger$ includes $A$ and 
\begin{equation} \label{eq:adjoint}
\wt a^\dagger = \rho(a) 
\end{equation}
when acting on $A$. The operators $\wt a$ and $\rho(a)$ commute on $A$, by definition. We thus expect $A\otimes A^\op$ to act as unbounded normal operators on $\cH$. 

The relation (\ref{eq:adjoint}) allows us to interpret $a$ and $\wt a$ as quantizations of holomorphic and anti-holomorphic functions on the phase space $\cM$. More precisely, $a$ and $\wt a$ quantize holomorphic and anti-holomorphic functions related by an appropriate hyper-K\"ahler rotation. The opposite operator ordering in $A^\op$ 
corresponds to the relative sign between $\Omega$ and $\bar \Omega$ in $\mathrm{Im} \, \Omega$. 

The  image of the identity in $A$ is a special vector $|1 \rangle \in \cH$, which intertwines the $A$ and $A^\op$ actions:
\begin{equation}
	a |1 \rangle = \wt a |1 \rangle = |a\rangle \, .
\end{equation}
We can call such a vector ``spherical'' in analogy with the notion of spherical vector in representation theory. It is cyclic and separating, as it generates the dense subspace $A \subset \cH$.

The data of the Hilbert space $\cH$, the dense subspace $A\to \cH$ and the $A\otimes A^\op$ action on $\cH$ with spherical vector $|1\rangle$ is the output of the sphere quantization of $\cM$. 

We refer to Section \ref{sec:extended} for a more detailed discussion of this setup. 

\subsection{Localization formulae and Analytic Symplectic Duality}
For Lagrangian gauge theories, the sphere correlation functions for $A[T]$ and $A[T^!]$ are both computable via localization. The resulting finite-dimensional integral expressions 
do not necessarily make evident the cyclicity and positivity properties in equation (\ref{eq:potr}). These become simple, non-trivial mathematical conjectures backed by the full machinery of 3d SCFTs. 

Our perspective is closely related to the idea that the localized path integral on $S^3$ can be factored into two hemi-sphere path integrals \cite{Drukker:2010jp,Dimofte:2011py,Dedushenko:2017avn,Dedushenko:2018tgx}. This was an important motivation for this work.  

On the Higgs branch side, the localization formulae are easily re-cast in a form which makes the trace property manifest \cite{companion}. Positivity of the trace is not immediately manifest,
but we will demonstrate it with a bit of effort. In a large class of examples, the analysis also allows us to identify $\cH$ as a space of $L^2$-normalizable half-densities on some auxiliary 
space $\cN$, with $A$($A^\op$) acting as (anti)holomorphic differential operators and a rather explicit spherical vector.\footnote{Classically, $\cM$ is identified with the ``affine closure'' of $T^* \cN$, i.e. with the spectrum $\mathrm{Spec}\,\bC[T^* \cN]$ of the algebra of holomorphic functions on $T^* \cN$.}

On the Coulomb branch side, localization formulae already take the form
\begin{equation}
	 \langle 1|a|1 \rangle
\end{equation}
for a vector $|1\rangle$ which formally interpolates the $A$ and $A_\op$ actions on an auxiliary Hilbert space. The main challenge here is to prove that $|1\rangle$ is actually in the domain of 
$a$ and $\tilde a$ and that one can legitimately employ the Hermiticity relations $\rho(a)^\dagger = \wt a$ to identify these inner products as defining a trace. The auxiliary Hilbert space can then be identified with $\cH$. We will only do so in examples, leaving a full proof to experts in the BFN formalism used to define Coulomb branches \cite{Nakajima:2015txa,Braverman:2016wma,Braverman:2017ofm,Braverman:2022zei}. 

Coulomb branch algebras of gauge theories have large commutative sub-algebras, which are manifestly diagonalized in the localization formulae. Effectively, the 
localization formulae give a spectral decomposition of $\cH$ and of the cyclic vector $|1\rangle$ for the action of this commutative sub-algebra and express the rest of the algebra as difference operators. 

We will be able to formulate natural conjectural generalizations of the localization formulae which apply to theories which only admit a partial Lagrangian description. These compute the sphere correlation functions for theories of (twisted) vectormultiplets and hypermultiplets coupled to abstract 3d SCFTs whose sphere correlation functions have been derived by other means. 

Whenever we have an explicit mirror pair $(T,T^!)$ and localization formulae for both the Higgs branch of $T$ and the Coulomb branch of $T^!$, we gain two very different realizations of $\cH$. 
The result can often be recast as a spectral decomposition of some functional space into eigenspaces of certain commuting (anti)holomorphic differential operators. 
We will recover in this manner many classical or conjectural results in representation theory. We propose the term ``Analytic Symplectic Duality'' for this collection of results.   

We refer to Section \ref{sec:ex} for a variety of examples illustrating the relation between sphere quantization and mathematical results in representation theory. 

\subsection{Higgs branch global symmetries and gauging}
The theory $T$ may admit a global symmetry group $F$ acting on Higgs branch operators and as tri-holomorphic isometries of the Higgs branch. Accordingly, $\cM$ 
will have an action of the complexified group $F_\bC$ preserving the complex symplectic form and implemented infinitesimally by moment maps $\mu_\cl$. 

The algebra $A$ will similarly carry an action of $F$ and include quantum moment maps $\mu$, giving a representation of the Lie algebra $\ff$, with $\rho(\mu) = - \mu$. 
In turn, the Hilbert space $\cH$ will also carry an action of $F$ as well as an unitary action of the complexified Lie algebra $\ff_\bC$ 
by unbounded normal operators $\mu$ and $\tilde \mu$. Notice that $\mu - \tilde \mu$ are the infinitesimal generators for the $F$ action and that $|1\rangle$ is $F$-invariant.  
This is the representation-theoretic notion of spherical vector.  

Th action of $F$  on $\cH$ can be extended to an unitary action of $F_\bC$. The appearance of unitary representations of complex Lie groups 
will lead to many points of contact with classical results in representation theory, especially when combined with ideas from S-duality \cite{Gaiotto:2008ak}.

The sphere correlation functions can be deformed to compute a twisted trace 
\begin{equation}
	\Tr \,  e^{2 \pi \beta \mu}  \,a \,,
\end{equation}
which is finite and well-defined for any Hermitean $\beta \in \ff$. \footnote{The trace is expected to have a rich analytic structure as a function of complex $\beta$, with potential poles on the weight lattice for anti-Hermitean $\beta$.} 

Barring some discrete anomalies which will be discussed in the main text, the $F$ symmetry can be gauged to produce a new 3d theory $T/F$.
The algebra $A[T/F]$ is the quantum Hamiltonian reduction of $A[T]$, i.e. the quotient of the $F$-invariant part $A[T]^F$ by the intersection with the ideal $\mu A[T]$ generated by $\mu$. 
If $T$ is a a theory of free hypermultiplets, $T/F$ is a standard gauge theory. 

If $T/F$ flows to an SCFT in a sufficiently smooth manner (not ``bad'' in the sense of \cite{Gaiotto:2008ak}), the standard localization formulae predict \cite{Dedushenko:2016jxl}:
\begin{equation}
	\Tr_{T/F} \, a \equiv \frac{1}{|W_F|} \int d\beta \, \left[\prod_\alpha  2 \sinh \pi (\alpha, \beta) \right]\,\Tr_T \, e^{2 \pi \beta \mu} \, a \, ,
\end{equation} 
where $W_F$ is the Weyl group of $F$, the product runs over roots $\alpha$ of $F$ and the integral runs over Hermitean elements in the Cartan of $\ff$. Originally, $T$ would be a theory of free hypermultiplets, but the formula makes sense for general $T$. 

The standard localization formula is not manifestly well-defined on the quotient by $\mu A[T]$. We can recast \cite{companion} the standard localization formula as an average over $F_\bC$:
\begin{equation}
	\Tr_{T/F} \, a \equiv \oint_{h \in F^+} d\Vol_h \, \Tr_T  \, h \, a \,
\end{equation} 
where $h$ is integrated over the middle-dimensional cycle $F^+$ in $F_\bC$ consisting of positive-definite Hermitean elements, represented in the trace by $\exp 2 \pi \beta \mu$ for Hermitean $\beta$. The measure $d\Vol_h$ is the invariant holomorphic top form on $F_\bC$. A simple contour deformation shows that this expression is well defined on the quotient, 
as $\Tr_T  \, h \, \mu \, a$ is a total derivative. Convergence of the integral is less obvious, but expected to hold if the theory $T/F$ flows to an SCFT with the same definition of R-symmetry as $T$, so that the sphere correlation functions exist and are computed by localization. 

Accordingly, we can give a rather explicit description of $\cH(T/F)$ in terms of $\cH(T)$. Be $\cH(T)^F$ the $F$-invariant part of $\cH(T)$. 
The space of $F_\bC$ co-invariants in $\cH(T)^F$ is defined by taking a quotient of $\cH(T)^F$ by the intersection with the image of the moment maps. 
Sphere correlation functions give the natural inner product on co-invariants: 
\begin{equation}
	\langle a|b\rangle \equiv \oint_{h \in F^+} d\Vol_h \, \langle a| h|b\rangle
\end{equation}
In the inner product, $h$ acts as $\exp 2 \pi \beta \cdot \mu$ on vectors in $\cH(T)^F$.

As a consequence, $\cH(T/F)$ is identified with the closure of the space of coinvariants in $\cH(T)^F$ under the above $L^2$ norm. \footnote{We should stress that the pairing inside the integral is {\it not} positive definite. Only the integrated expression is expected to be positive definite. We should also observe that this formula would also emerge from a standard BRST quantization of a gauged quantum-mechanics. See e.g. Appendix B of \cite{Chandrasekaran:2022cip} for a recent review and further literature. The lack of manifest positivity is a typical feature of BRST quantization.}

We refer to Section \ref{sec:gauhiggs} for a detailed discussion of Higgs branch gauging and associated functional descriptions of $\cH$. 
  
\subsection{Quantum FI parameters and Coulomb gauging.}
The theory $T$ may admit a global symmetry group $F^!$ acting on Coulomb branch operators. This always implies that $\cM$ has a family of complex structure deformations parameterized by elements $t$ of the Cartan sub-algebra $\ff^!$ of $F^!$ (modulo the action of the Weyl group), aka FI parameters. It also implies that $A$ has an analogous family of deformations $A_{t}$ parameterized by quantum FI parameters $t$. 

More precisely, one can define an extended algebra $A_{F^!}$ whose center is isomorphic to the commutative algebra of Weyl-invariant polynomials $\cP$ on $\ff^!$, such that the localization of $A_{F^!}$ at a point $t$ is $A_t$. The algebra $A_{F^!}$ has a concrete physical realization in terms of protected boundary operators if we promote $T$ to a boundary condition for a four-dimensional ${\cal N}=4$ gauge theory with gauge group $F^!$ \cite{Dedushenko:2020vgd}.

The real parts of the quantum FI parameters can be turned on in sphere correlation functions and preserve the positivity properties of the system. They give a trace $\Tr_{T,t}$ on $A_t$ as a function of $t$. In particular, the reality condition on $t$ allows the Hermiticity condition on the $A_t \times A_t^\op$ action to still make sense and we obtain a consistent sphere quantization.
This can also be interpreted as a family of traces on $A_{F^!}$ which assign specific values to the elements in the center. \footnote{If $F^!$ is non-Abelian, the  Hermiticity condition on the $A_t \times A_t^\op$ can make sense on more general loci where $\bar t$ is related to $t$ by a Weyl reflection. 
We expect the reflection positivity argument to still hold when we include the effect of the Weyl reflection, but the argument will break down outside the range of $t$ 
where the sphere correlators are defined directly rather than through analytic continuation around some poles. The characterization of the sub-regions of these alternative loci where positivity of the inner product holds is mathematically rich. We will not explore this topic in detail. }

It is possible to analytically continue the sphere partition function to complex values of $t$. The imaginary part of $t$ tends to destabilize the theory on $S^3$, so that the partition function will become singular at specific values of $t$ determined by properties of the Coulomb branch \cite{Gaiotto:2012xa}, with meromorphic analytic continuation for all values of $t$. 
The residues at certain poles can be identified with the sphere partition functions of new theories defined as endpoints of RG flows triggered by Coulomb branch vevs in $T$ \cite{Gaiotto:2012xa}.
We conjecture an analogous statement for sphere correlation functions: the residue of the sphere correlation functions of $T$ defines a pairing on $A$ with a non-trivial kernel $I$, 
such that $A/I$ is the algebra associated to the new theory and the pairing is positive-definite on $A/I$. This appears to be a non-trivial mathematical conjecture. 

It is possible to gauge a Coulomb branch global symmetry, or a subgroup thereof, using mirror vector-multiplets. 
The new Higgs branch algebra, which we can denote as $A[T \rtimes F^!]$, has an $A_{F^!}$ sub-algebra.
Standard Coulomb branch localization formulae \cite{Dedushenko:2017avn,Dedushenko:2018icp} are particularly simple for 
the trace of elements in $A_{F^!}$:
\begin{equation} \label{eq:coust}
	\Tr_{T \rtimes F^!} \, a \equiv \frac{1}{|W_{F^!}|} \int dt \, \left[\prod_\alpha  2 \sinh \pi (\alpha, t) \right]\,\Tr_{T,t} \, a,
\end{equation} 
where the integral is over real FI parameters $t$. 

In the original setting for the localization formulae, $T$ consists of twisted hypermultiplets and has a trivial Higgs branch. Then $a$ denotes an invariant polynomial on $\ff^!$ and the trace $\Tr_{T,t} \, a$ evaluates $a$ at $t$. 
This expression makes sense for general $T$, though, and can be verified in situations where $T^!$ is itself a gauge theory. 

More general elements of $A[T \rtimes F^!]$ can be ``Abelianized'' \cite{Bullimore:2015lsa} by embedding $A[T \rtimes F^!]$ into an extended version of the Coulomb branch algebra for $A[T \rtimes H^!]$, where $H^!$ is the Cartan subgroup of $F^!$. The extension inverts certain elements in the center of $A_{H^!}$.\footnote{A mathematical version of these statements can be likely demonstrated with the help of the Iwahori Coulomb brach construction \cite{Kamnitzer:2022zkv}.} The trace of a general element in $A[T \rtimes F^!]$ is computed by embedding it into the extension of $A[T \rtimes H^!]$, projecting it to $A_{H^!}$ and using the same integral expression as in (\ref{eq:coust}) \cite{Dedushenko:2017avn,Dedushenko:2018icp}.

The algebra $A[T \rtimes H^!]$ can be decomposed into a sum of bimodules $HC_{b,H^!}$ for $A_{H^!}$, with $b$ being the integral charge under the central generators in $A_{H^!}$ and $HC_{0,H^!} \equiv A_{H^!}$. Each bi-module can be localized to a family of ``Harish-Chandra'' bimodules $HC_{b,t}$ \cite{2018arXiv181007625L}, such that the central generators act as $b/2 + i t$ from the left and $-b/2 + i t$ from the right.
These are $A_{t - i\,b/2}$-$A_{t + i\,b/2}$ bimodules equipped with a multiplication compatible with the bi-module structure. 
They should be thought as part of the data of $T$, associated to sphere correlation functions decorated by certain ``vortex'' line defects. 

In particular, we will have an anti-linear map $\rho: HC_{b,t} \to HC_{-b,t}$ and the trace $\Tr_t$ on $A_t$ gives a bi-module trace pairing $HC_{-b,t}$ and $HC_{b,t}$:
\begin{equation}
	\Tr_{b,t} \, a_1 a_2 \equiv \Tr_{t + i\,b/2}\, a_1 \,a_2 = \Tr_{t - i\,b/2} \,\rho^2(a_2) \,a_1 \qquad \qquad \Tr_{b,t}  \,\rho(a)\, a >0 \, ,
\end{equation}
where the last positivity relation should hold for real $t$. Thus the bi-module trace defines an Hermitean inner product and a collection of Hilbert spaces $\cH_{b,t}$ which complete the $HC_{b,t}$
and carry an $A_{t - i\,b/2}\times A^{\op}_{t + i\,b/2}$ action \cite{Etingof:2020fls}. This should be thought of as an extended version of  sphere quantization for $\cM[T]$, with quantum FI parameters $t - i\,b/2$. 

We can now give a sphere quantization interpretation to the localization formulae for the Abelian gauge theory $T \rtimes H^!$. Essentially by construction, 
the trace $\Tr_{T \rtimes H^!} a_1 a_2$ gives the natural inner product on a direct sum/integral 
\begin{equation}
	\int^\oplus_{(t,b) \in \fh \times \Lambda_{H^!}}dt\, \cH_{b,t} 
\end{equation}
of the $\cH_{b,t}$ Hilbert spaces, built from the individual inner products with a natural 
measure. Here we denoted as $ \Lambda_{H^!}$ the charge lattice $b$ is valued in. 

The resulting direct sum/integral can be identified immediately with $\cH[T \rtimes H^!]$. The only subtlety is that the localization formulae and the direct integral are related by a shift of the integration contour by $i\,b/2$, which in turns requires the trace pairing on $HC_{-b,t} \times HC_{b,t}$ to be analytic on a 
large enough strip around the real $t$ axis. This is expected on physical grounds and can be demonstrated in examples. 

The individual $\cH_{b,t}$ spaces are the eigenspaces for the center of $A_{H^!}$ acting on $\cH[T \rtimes H^!]$, with eigenvalues $b/2 + i t$. The spherical vector $|1\rangle$ is simply the direct integral of the spherical vectors in $\cH_{0,t} \equiv \cH_{t}$.

The sphere quantization for $T \rtimes F^!$ can be understood in a similar manner. An important observation is that the $HC_{b,t}$ bi-modules are invariant under the simultaneous Weyl group $W_{F^!}$ action on the pair $(b,t)$. Then the trace is again identified as the natural inner product on a direct sum/integral of $\cH_{b,t}$ with an appropriate $\sinh$-Vandermonde measure:
\begin{equation}
	\int^\oplus_{(t,b) \in (\fh \times \Lambda_{H^!})/W_{F^!}} \left[\prod_\alpha \sinh^2 \pi \alpha \cdot t\right] \,\cH_{b,t}  \, .
\end{equation}
A potential subtlety is that the 
contour deformations involved in this identification could be obstructed by the denominators which appear in the Abelianized formulae. The $\sinh$-Vandermonde measure is expected to cancel all such dangerous poles. The direct integral/sum is again a spectral decomposition for the center of $A_{F^!}$.

We refer to Section \ref{sec:gaucoulomb} for a detailed discussion of Coulomb branch gauging and associated spectral descriptions of $\cH$. 

\subsection{Non-conformal case}
The availability of FI deformations makes if possible to associate some useful structures to theories which are not conformal by taking a large FI limit of conformal 
theories, or large mass limit in a mirror description. 

For example, the Coulomb branch description of the Hilbert space $\cH[T]$ does not depend strongly on the twisted hypermultiplets present in the theory. If we assign a mass to the twisted hypermultiplets,  a large mass limit simply removes some factors from 
the cyclic vector, possibly making it non-normalizable. This costs us the dense basis of states identifying $\cH[T]$ as the closure of $A[T]$, but $\cH[T]$ itself and the $A \times A^\op$ action appear to survive.  Notice that the algebra for the the simplified theory is actually a subalgebra of the 
original $A$. 

An alternative perspective on $\cH[T]$ is that one can consider a supersymmetric $S^2 \times \bR$ geometry, defined by treating the 3d ${\cal N}=4$ theory as a 2d $(2,2)$ theory 
with fields valued in maps from $\bR$ to the target of the 3d theory. An unbroken $U(1)_C$ subgroup of the R-symmetry acting on the Coulomb branch,
which is always present in a standard gauge theory as long as eventual mass parameters are real, is sufficient for the SUSY compactification to work 
\cite{Festuccia:2011ws,Gomis:2012wy,Doroud:2012xw,Benini:2012ui,Doroud:2013pka,Gerchkovitz:2014gta}. 

The 3d path integral reduces to a 1d path integral for a quantum mechanical system. If $T$ is a standard gauge theory, this gives the path integral for a gauged 1d quantum mechanics with $\cM$ target space, whose space of states is a natural $\cH[T]$ candidate. 
An unbroken $U(1)_H$ subgroup of the R-symmetry endows the answer with positivity properties and should suffice to guarantee that $\cH[T]$ 
is an Hilbert space. A typical case would be a 3d sigma model with target $\cM = T^*X$, giving $\cH[T] = L^2(X)$. 

Mirror symmetry can thus still give some predictions for the spectral decomposition of certain functional spaces, even in the non-conformal case.

\subsection{Line defects}

It is possible to enrich the story by various supersymmetric defects to generate a larger variety of quantizations. This includes
\begin{itemize}
	\item Line defects wrapping a circle linking the great circle supporting local operators \cite{Guerrini:2023rdw}. These line defect insertions do not affect the 
	local operator algebra or $\rho$, but change the twisted trace itself. Line defects invariant under reflections will give alternative 
	choices of positive traces \cite{Etingof:2020fls}. 
	\item Line defects wrapping the circle supporting local operators or sections thereof. These defects modify the local algebra and produce bi-modules 
	at junctions between different defects. A special case of ``vortex defects'' \cite{Bullimore:2016nji} only change the value of quantum FI parameters by integral amounts and produces the
	$HC_{b,t}$ bimodules. Sphere correlation functions equip them with positive-definite inner products \cite{Etingof:2020fls}.
\end{itemize}

\subsection{Boundary conditions and ``Real'' quantizations}
The correlation functions of protected local operators on a three-sphere are part of a larger collection of protected correlation functions which enjoy (reflection) positivity. 

A particularly interesting example are ``hemisphere'' correlation functions, which take as an input a 3d SCFT $T$ as well as an half-BPS super-conformal boundary condition $B$ for $T$. 
Classically, such a boundary condition restricts the space of Higgs branch vacua to a complex Lagrangian cone $\cL \subset \cM$.

One useful application of such boundary conditions is to define interesting (possibly distributional) ``boundary states'' $|B\rangle$ in $\cH$, through the one-point functions of bulk operators on the hemisphere. Some of the presentations of $\cH$ involve bases of states of such kind.  

In order to describe a second, less obvious application, we need some geometric considerations. If we orient our choice of complex structure so that $\cL$ is Lagrangian for the real and imaginary parts of $\Omega$, boundary local operators give a 
quantization of $\cL$ in the sense of deformation quantization, i.e. promote the space $M_\cl$ of holomorphic functions on $\cL$ to a left module $M$ for $A$ or, dually, 
to a right module $\wt M$. The hemisphere correlation functions then give $A$ bi-module maps $M \times \wt M \to \cH$ with images $|B;m,\tilde m\rangle$.

Reflection positivity on the hemisphere implies that the correlation functions $\langle a|B;m,\tilde m\rangle$ can be 
also interpreted as giving a pairing 
\begin{equation}
(\cdot, \cdot): \wt M \otimes_A M \to \bC
\end{equation}
compatible with the action of $A$, as well as an anti-linear invertible map $\rho: M \to \wt M$ such that the inner product $\left( \, \rho(m),m'\, \right)$ 
is positive-definite. 

The definition of $\rho$ involves a non-trivial Grahm-Schmidt procedure \cite{Gerchkovitz:2016gxx} applies to a semiclassical filtration of $M$ and $\wt M$. 
Hermitian conjugation under this inner product gives an action of $A^\op$ on $M$ as well:
\begin{equation}
	m \wt a \equiv \rho^{-1}\left(\rho(m) \, \rho(a) \right)
\end{equation}
This action does {\it not} commute with the $A$ action. 

We give the following geometric interpretation to these structures. We can use an hyper-Kahler rotation on $\cL$ to produce a submanifold $\cL_\bR$ which is Lagrangian for the real part of $\Omega$ and for the third K\"ahler form $\omega_\bR$. Then $\mathrm{Im}\, \Omega$ can play the role of a Kahler form on $\cL_\bR$. 
It may thus be possible to ``quantize'' $\cL_\bR$ in the sense of geometric quantization,
i.e. promote $M_\cl$ to an Hilbert space $\cH_B$. 

We interpret the $L^2$ completion of $M$ precisely as a candidate for $\cH_B$ and the resulting actions of $A$ and $A^\op$ on $\cH_B$ as the quantization of functions on $\cM$ restricted to $\cL_\bR$ in the geometric quantization setting.

In favourable situations, $\cL_\bR$ may be the fixed point $\cM_\bR$ of an anti-holomorphic involution of $\cM$, which extends to a relation between the $A$ and $A^\op$ actions on $M$,
i.e. to a Hermiticity property of the $A$ representation on  $\cH_B$. This situation is the typical context where brane quantization is applied \cite{Gukov:2008ve}. We will employ the setup to discuss the quantization of $\cM_\bR$. In particular, we will encounter real versions of both Higgs- and Coulomb- branch localization formulae, as well as tantalizing hints of 
number-theoretic generalizations. 

We will only touch this subject in passing in Section \ref{sec:real}, leaving a full treatment to future work. 

\section{Higgs branch operators and positivity} \label{sec:extended}
Any  3d ${\cal N}=4$ SCFT is equipped with an $SU(2)_H \times SU(2)_C$ R-symmetry group. Higgs branch operators are defined as belonging to certain 
half-BPS multiplets for the superconformal algebra \cite{Dolan:2008vc,Beem:2016cbd}. They always have non-negative (half)integral dimension $\ell$ and transform as spin $\ell$ irreducible representations of $SU(2)_H$.\footnote{They are automatically Lorentz scalars and $SU(2)_C$ singlets due to BPS bounds.} 
The only operator of dimension $0$ is the identity. The space of operators with any given $\ell$ is finite-dimensional.

The $SU(2)_H$ symmetry acts as hyper-k\"ahler rotations on the Higgs branch of vacua. For each choice of a Cartan generator in $SU(2)_H$ we can consider highest weight Higgs branch operators, which have non-singular OPE and map to holomorphic functions on the Higgs branch $\cM$ in the corresponding complex structure.
The OPE equips the space of highest weight Higgs branch operators with the structure of an algebra $A_\cl$ graded by $\ell$, with finite-dimensional graded pieces. 

The algebra is the same for every choice of highest weight direction. Working twistorially, a multiplet associated to an element $a \in A_\cl$ can be collected into a 
generating function 
\begin{equation}
	O_a(\zeta) = O_a^{(\ell_a)}+ \zeta  O_a^{(\ell_a-1)}+ \cdots + \zeta^{2 \ell_a} O_a^{(-\ell_a)}   \, ,
\end{equation}
and the generating functions multiply accordingly:
\begin{equation}
	O_a(\zeta) O_b(\zeta) =  O_{a b}(\zeta) \, .
\end{equation}
The expectation values of the $O_a(\zeta)$ are holomorphic functions on $\cM$ in the complex structure parameterized by the twistor parameter $\zeta \in \mathbb{C}P^1$. 
The opposite complex structure corresponds to the antipodal point on the twistor sphere. We can correspondingly obtain an new generating function as 
\begin{equation}
	O_{\rho_\cl(a)}(\zeta) \equiv \zeta^{2\ell_a} \overline{O_a(-1/\bar \zeta)}  = (-1)^{2 \ell_a} \overline{O_a^{(-\ell_a)}} + \cdots + \zeta^{2 \ell_a} \overline{O_a^{(\ell_a)}}  \, .
\end{equation}
The map $\rho_\cl: A_\cl \to A_\cl$ is manifestly anti-linear. At the level of functions, it corresponds to the action of a specific hyper-Kahler rotation by an angle of $\pi$ followed by complex conjugation. Furthermore, $\rho_\cl^2 = (-1)^{2\ell}$. 

We should also discuss reality conditions on multiplets. A multiplet with integral spin $\ell$ can be real, in the sense that $O_a^{(-\ell_a)} = \overline{O_a^{(\ell_a)}}$
and thus $a = \rho_\cl(a)$. This is not an option for half-integral spin $\ell$, where we can only employ a pseudo-reality condition
$O_a^{(-\ell_a)} = \overline{O_b^{(\ell_b)}}$ and thus $a = \rho_\cl(b)$, which implies $b = - \rho_\cl(a)$. 

A classical example is the free hypermultiplet, which is valued in $\mathbb{R}^4$: we have elementary generators
\begin{equation}
	P(\zeta) = P + \zeta \bar X \qquad \qquad X(\zeta) = X - \zeta \bar P
\end{equation}
and thus $\rho_\cl(X)=P$ and $\rho_\cl(P) = -X$. 

Quadratic functions such as 
\begin{equation}
X(\zeta)\, P(\zeta) = X P + \zeta (|X|^2 - |P|^2) - \zeta^2 \bar X \bar P
\end{equation}
are pure imaginary. 

In order to define protected correlation functions, one places highest weight Higgs branch operators along a great circle in $S^3$, correlating the highest weight direction and overall normalization to the position along the circle in an appropriate manner:
\begin{equation}
	\cos^{2 \ell_a} \frac{\varphi}{2} \,  O_a\left(\tan \frac{\varphi}{2}\right)\Big|_\varphi
\end{equation}
where $\varphi$ is the angle along the circle. The resulting correlation functions 
\begin{equation}
	\Tr \, a_1 \cdots a_n \equiv \left\langle \prod_i \left[ \cos^{2 \ell_{a_i}} \frac{\varphi_i}{2} \,  O_{a_i}\left(\tan \frac{\varphi_i}{2}\right)_{\varphi_i}\right]\right\rangle_{S^3} 
\end{equation}
only depend on the relative order of the insertions along the circle, up to an overall $(-1)^{2 \ell}$ sign accrued when an operator is brought around the circle. 

When computing the (non-singular) OPE between two operators in the correlation function, the naive multiplication is deformed by the contribution of
singular terms in the OPE of operators which are highest weight along different directions \cite{Beem:2016cbd}. 
This is the same as the deformation which occurs in an $\Omega$-background \cite{Yagi:2014toa,Beem:2018fng}.
In particular, the leading deformation is the Poisson bracket on $A_\cl$ inherited from the complex symplectic form $\Omega$ on $\cM$, 
i.e. the deformed OPE is a star product for $A_\cl$ \footnote{We apologize to the reader for using the symbol $\Omega$ with two distinct meanings in the same paragraph.}

The OPE of protected operators along the circle thus defines a deformation $A$ of the algebra $A_\cl$: the ``quantized Higgs branch algebra'' 
The deformation breaks the $\ell$ grading in a controlled manner, as the deformation parameter (which we set to $1$) has scaling dimension $1$: $A$ is filtered by $2\ell$, with associated graded $\mathrm{gr} A = A_\cl$. The $(-1)^{2 \ell}$ grading is preserved by the deformation. 
By construction, the sphere one-point functions give a trace $\Tr$ on $A$ twisted by $(-1)^{2 \ell}$. 

It is important to observe that the definition of the correlation functions gives a specific ``quantization map'' $A_\cl \to A$, mapping $a$ to the twisted $O_a$ insertion. 
For example, the only operators in the CFT which can have non-zero one-point functions on the three-sphere must have scaling dimension $0$, 
i.e. $\Tr \, a = 0$ if $a\in A$ is the image of an element of non-zero degree in $A_\cl$. Other selection rules apply to the deformed OPE, but can be all subsumed in the 
next property we discuss: positivity \cite{Etingof:2019guc}. 

We should also observe that the manipulations involved in localization computations of the sphere correlation functions may lead to a finite operator renormalization, so that the quantization map $A_\cl \to A$ is obfuscated \cite{Gerchkovitz:2014gta}. We will also review how to recover it from positivity. 

As a preliminary observation, define an anti-linear map $\rho: A \to A$ to be the same as $\rho_\cl$ under the identification $A_\cl \to A$.
We claim that this is still an antilinear algebra morphism. Indeed, $O_{\rho_\cl(a)}$ inserted at $\varphi$ is equivalent to $\overline{O_a}$ inserted at   
$\varphi - \pi$. 

A special case of the sphere correlation functions are two-point functions with operators at antipodal points on the sphere:
\begin{equation}
	\Tr \,a \, b = \langle O^{-\ell_a}_a O^{\ell_b}_b\rangle
\end{equation}
Such sphere correlation functions satisfy reflection positivity and give a positive-definite sesqui-linear inner product on the SCFT local operators: 
\begin{equation}
	\langle a|b\rangle \equiv \Tr \rho(a) b = \langle O^{-\ell_a}_{\rho_\cl(a)} O^{\ell_b}_b\rangle = \langle \overline{O^{\ell_a}_{a}} O^{\ell_b}_b\rangle \, .
\end{equation}
e.g. 
\begin{equation}
	\Tr \rho(a)\,a >0
\end{equation}
for all $a$. We will denote a twisted trace with this property as {\it positive}.

The fact that the pairing $\Tr \, a\,b$ is non-degenerate on $A$ allows one to recover the quantization map if we are only given 
the identification $A_\cl = \mathrm{gr} A$. Indeed, we can map an element $a_\cl \in A_\cl$ to an element $a \in A$ 
with associated graded $a_\cl$ and orthogonal to all elements of degree smaller than the degree of $a_\cl$. This is a simple example of a 
general strategy used to recover positive-definite inner products from localization calculus \cite{Gerchkovitz:2014gta}.

The positive-definite inner product $\Tr \rho(a)\,b$ allows us to define an Hilbert space $\cH$ as the $L^2$ completion of $A$, as anticipated in the introduction. 
We denote the image of $a \in A$ in $\cH$ as $|a \rangle$, and the inner product as
\begin{equation}
	\langle a|b \rangle \equiv \Tr \rho(a)\,b
\end{equation}
By definition, $A$ forms a dense subspace of $\cH$.

The left- and right- multiplication by elements of $A$ 
\begin{equation}
	a |b\rangle \equiv |a\,b\rangle \qquad \qquad \wt a |b\rangle \equiv |b\,a\rangle
\end{equation}
defines (possibly unbounded) operators on $\cH$, with domain $A$. The two action commute with each other and give an action of $A\otimes A^\op$ on $\cH$. We have 
\begin{equation}
	\langle b\, a |c\rangle = \Tr \,\rho(b\,a) \,c  = \langle b| \rho(a)|c\rangle
\end{equation}
This implies that the domain of $\wt a^\dagger$ includes $A$ and 
\begin{equation}
\wt a^\dagger = \rho(a)
\end{equation}
on $A$. 

The  image of the identity is a special ``spherical'' vector $|1 \rangle$, which is cyclic and separating and intertwines the two actions:
\begin{equation}
	a |1 \rangle = \wt a |1 \rangle
\end{equation}

We have thus produced an Hilbert space $\cH$ with an action of the quantized algebra $A$ of holomorphic functions on $\cM$, commuting with and adjoint to an action 
by $A^\op$, which we identify with the quantized algebra $A$ of anti-holomorphic functions on $\cM$. More precisely, $a$ and $\wt a$ quantize holomorphic and anti-holomorphic functions related by an appropriate hyper-K\"ahler rotation. The opposite operator ordering in $A$ and $A^\op$ means that the quantization involves the imaginary part
$\mathrm{Im} \, \Omega$ of the complex symplectic form on $\cM$.

The data of the Hilbert space $\cH$, the dense subspace $A\to \cH$ and the $A\otimes A^\op$ action with spherical vector $|1\rangle$ is the output of the ``sphere quantization'' of $\cM$. 

Three-dimensional ${\cal N}=4$ SCFTs may be equipped with two types of global symmetries: Higgs- and Coulomb- branch flavour symmetries.  These are compact symmetry groups $F$ and $F^!$ acting respectively on the Higgs or Coulomb branch operators. Both are associated to (a triple of) relevant deformation parameters of the theory: ``masses'' are associated to 
generators of $F$ and ``Fayet-Iliopoulos'' (FI) parameters are associated to generators of $F^!$.  

Up to a symmetry rotation, the masses and FI parameters which contribute to a localization calculation can be taken to be valued in the (imaginary) Cartan subalgebras of $F$ and $F^!$, modulo the action of the Weyl group. They are permuted by mirror symmetry. 
 
More precisely, the triple of FI parameters $(t, t_\bR, \bar t)$ can be organized into a combination
\begin{equation}
	t(\zeta) \equiv t + \zeta  t_\bR + \zeta^2 \bar t
\end{equation}
which deforms the Higgs branch algebra in complex structure $\zeta$. Consistency of the sphere correlation functions requires the algebra to be independent of $\varphi$,
i.e.
\begin{equation}
	t \cos^2 \varphi/2 + t_\bR \sin  \varphi/2  \cos \varphi/2  + \bar t\sin^2  \varphi/2 
\end{equation} 
to be $\varphi$-independent, requiring $t$ to be real and $t_\bR=0$. Analogously, sphere correlation functions are compatible with a single real 
mass parameter $m$ within each triple $(m, m_\bR, \bar m)$. 

Both $t$ and $m$ can be formally analytically continued in the sphere correlation functions, but the complexified parameters should not be identified as 
complex FI or mass parameters. We will discuss this point further below. 

In a standard gauge theory description of the SCFT, $F$ is typically already manifest in the UV and generic masses can thus be included in localization computations. On the other hand, only some Abelian subgroup of $F^!$ can appear in the UV and the rest of $F^!$ is emergent. As long as the full Cartan is visible in the UV, generic FI parameters can be included in localization computations. 

Finally, a small point of notation: a theory with symmetry group $F$ can also be always treated as a theory with symmetry group $F' \subset F$ 
by forgetting part of the symmetry. The same is true for $F^!$. 

\subsection{Coulomb branch global symmetries}
We discuss the role of Coulomb branch symmetries first. Most of this information has already been presented in the Introduction. 

Real quantum FI parameters $t$ can be turned on in sphere correlation functions without spoiling their properties, including reflection positivity. The FI parameters deform the Higgs branch $\cM$ and modify accordingly the $A_\cl$ and $A$ algebras to some families $A_{t,\cl}$ and $A_t$ parameterized by $t$. Sphere correlation functions give a family of Hilbert spaces $\cH_t$, etc.  

We can organize this data in terms of a single algebra $A_{F^!}$ with a center consisting of Weyl-invariant polynomials in $t$, such that $A_t$ emerges as a central quotient of $A_{F^!}$ at $t$. 
Then the $\cH_t$ Hilbert spaces give representations of $A_{F^!}\times A^{\op}_{F^!}$ with specific eigenvalues for the center. 

Analytic continuation of sphere correlation functions away from real $t$ is possible, but only by a finite amount controlled by the flavour charges of Coulomb branch operators. Poles should appear beyond that, with interesting features. The ``leading'' poles will appear for values of $t$ such that a non-zero locus on the Coulomb branch is fixed by the combination of 
the corresponding $F^!$ generator and the scaling transformation. 

Physically, a non-zero Coulomb branch vev $c$ restricts the Higgs branch expectation values to some sub-manifold 
$\cM(c) \subset \cM$. This sub-manifold is the Higgs branch of an effective low energy theory describing fluctuations in the $c$ vacuum.
The construction of \cite{Gaiotto:2012xa} suggests that the sphere correlation functions should have a simple pole at the corresponding value $t_c$ with residue 
given by the sphere correlation functions for the low energy theory. 

We conjecture that the construction can be extended to protected Higgs branch correlation functions, which indeed also have a simple pole at $t_c$. 
Then the residue of the trace gives a new trace on $A_{t_c}$ which should encode the trace for the algebra $A(c)$ associated to the low energy theory. 
The UV and IR algebras should be related by a map $A_{t_c} \to A_c$, as every UV operator must have an image in the IR theory. We thus conjecture that the 
residue of the trace on $A_{t_c}$ is the image of the trace on $A_c$ under such map. More precisely, the residue as a trace on $A_t$ should be 
semi-positive definite, with a kernel which can be quotiented away to get a positive trace on $A_c$.

Coulomb branch symmetries can be gauged to produce a new theory. We will discuss details of Coulomb gauging in a separate Section \ref{sec:gaucoulomb}. 
We anticipate here a useful fact. If we gauge the Cartan subgroup $H^!$ of $F^!$, the resulting theory $T \rtimes H^!$ will be an SCFT, 
with an algebra $A[T \rtimes H^!]$ which contains an $A_{H^!}$ sub-algebra. The FI parameter $t$ is promoted to the moment maps for an Abelian Higgs branch symmetry 
and we can decompose $A[T \rtimes H^!]$ into a collection of $A_{H^!}$-bimodules $HC_b$ labelled by the charge $b$ under the symmetry. 

The $HC_b$ bimodules can be localized to $A_{t - i\,b/2}$-$A_{t + i\,b/2}$ bimodules $HC_{b,t}$. 
The positive trace on $A[T \rtimes H^!]$ then gives a positive trace on the $HC_{b,t}$, i.e. pairing $HC_{-b,t} \otimes HC_{b,t}\to \bC$ together with an antilinear map $\rho$
used in defining cyclicity and positivity of the trace \cite{Etingof:2020fls}. The completion of $H_{b,t}$ under the inner product 
gives Hilbert spaces $\cH_{b,t}$. 

This structure can also be defined without invoking $H^!$ gauging, with the help of supersymmetric ``vortex'' line defects \cite{Bullimore:2016nji}. In particular, 
if the $H^!$ global symmetry is associated to Abelian gauge fields in a gauge-theory description for $T$, the vortex defects are just Abelian Wilson lines 
and $T \rtimes H^!$ coincides with the theory where the Abelian gauge fields are removed. The $HC_{b,t}$  and associated trace are readily recovered 
with the tools described in Section \ref{sec:gauhiggs}. We will discuss line defect in some detail at the end of this Section.

\subsection{Higgs branch global symmetries}
Next, we discuss Higgs branch symmetries. 

The $F$ symmetry group acts tri-holomorphically on the Higgs branch. The complexification $F_\bC$ acts holomorphically on $\cM$. The algebras $A_\cl$ and $A$ 
have an $F$ symmetry which preserves the grading and the trace. In particular, the algebra $A$ decomposes into a direct sum of finite-dimensional representations of $F$. 
The infinitesimal action of $F$ is generated by conjugation by complex moment map operators $\mu$ with $\ell=1$. In a natural normalization, 
 $\rho(\mu) = -\mu$.  

As we map $A$ to a dense subset in $\cH$, the $F$ action on $A$ can be promoted to an unitary action on $\cH$. The infinitesimal $F$ action $a \to [\mu,a]$ maps to the action of the Hermitean generators $\mu - \wt \mu = \mu + \mu^\dagger$. The cyclic vector $|1\rangle$ is $F$-invariant and the $|a\rangle$ are organized in finite-dimensional representations of $F$. 

The Hilbert space $\cH$ actually carries an unitary representation of $F_\bC$, as
\begin{equation}
	|a \rangle \to   g\left|a \right\rangle \equiv = \left| e^{2 \pi \beta \cdot \mu} \,a \,e^{2 \pi \bar \beta \cdot \mu} \right\rangle 
\end{equation} 
with $g = e^{2 \pi \beta} \in F_\bC$ preserves the inner product. The images $\left| g g^\dagger  \right\rangle \equiv   g\left|1 \right\rangle$ of the cyclic vector only depend on the combination $g g^\dagger \in F_\bC/F$. 

Each Higgs branch symmetry generator is associated to a ``mass'' relevant deformation of the 3d SCFT. Real mass parameters $m$ can be turned on in sphere correlation functions, but spoil reflection positivity. The mass parameters do not affect the Higgs branch $\cM$ or the $A_\cl$ and $A$ algebras. 
The mass deformations of the sphere partition function effectively ``twist'' the trace by inserting a factor of $e^{2 \pi m \cdot \mu}$:
\begin{equation}
	\Tr  \,e^{2 \pi m \cdot \mu}\,a\,=\Tr \,a \,e^{2 \pi m \cdot \mu}\, .
\end{equation} 
The twisted trace is expected to be well defined/finite for all Hermitean $m$. Indeed, it should decrease exponentially fast at large $m$. It  can be interpreted as the inner product between a state in $A$ and the $F_\bC$ image $|e^{2 \pi m}\rangle$. \footnote{Analytic continuation away from Hermitean $m$ should be possible, but only by a finite amount controlled by the flavour charges of Higgs branch operators. Poles should appear beyond that. The ``leading'' poles 
will appear for values of $m$ such that a non-zero point on the Higgs branch is fixed by the combination of 
the corresponding $F$ generator and the scaling transformation. Differently from the case of FI parameters, the degree of the  poles depends on the operators in the trace. }

The properties of the twisted trace extend to the matrix elements for the $F_\bC$ action: 
\begin{equation}
	\langle a| g |b\rangle = \Tr \,\rho(a) \,g b g^{-1} \,e^{2 \pi m \mu}
\end{equation}
which depends on $g$ as the combination of the action on $b$, which is an holomorphic polynomial, and the dependence of the twisted trace on $g g^\dagger = e^{2 \pi m}\in F_\bC/F$. In particular, this is a smooth function on $F_\bC$. In Section \ref{sec:gauhiggs} we will gain some extra physical insights about the properties of such matrix elements when integrated over $F_\bC$.

\subsection{Spectral decomposition and S-duality}
The Hilbert space $\cH$ provides an unitary representation of $F_\bC$. It is natural to 
consider the decomposition of $\cH$ into irreducible unitary representations. 
For example, we can ask which such representations appear in the decomposition of $\cH$. This is sometimes called the 
``unitary dual'' of $\cH$. 

Recall that $F$-invariant polynomials $\cP(\mu)$ in the moment maps $\mu$ are central and commute with their adjoint $\cP(-\tilde \mu)$. We expect them to be normal operators on $\cH$ \cite{Segal}. As a first step of the decomposition,  we can ask about their joint spectrum and decompose $\cH$ into (possibly distributional) eigenspaces. 

The spherical vector $|1\rangle$ is invariant under the action of the compact group $F$. As we decompose $|1\rangle$ into components in each irreducible representation, we can only have a non-zero image in ``spherical'' unitary representations, which 
admit an unique such $F$-invariant ``spherical'' vector. Once we pick an independent normalization for these spherical vectors, we can ask 
which multiples appear in the decomposition of $|1\rangle$. With a bit more work, we can ask similar questions about the decomposition of general 
states $|a \rangle$. 

As an extension of these questions, one may consider the $F$-invariant part $A^F$ of $A$, which contains the $\cP(\mu)$ generators as central elements
and will act separately on each summand of the decomposition of $\cH$ into irreducible unitary representations. 

We will find a general physical answer to these questions with the help of S-duality for ${\cal N}=4$ four-dimensional $F$ gauge theory and its boundary conditions \cite{Gaiotto:2008sa,Gaiotto:2008ak}. We refer for details to Section \ref{sec:spher}, but we can anticipate here the general idea, at least for non-anomalous $F$: 
we can use S-duality to produce a mirror description of $T$ involving (a subgroup of) $F^\vee$ gauge fields, where $F^\vee$ is the Langlands dual group to $F$.  
The resulting Coulomb branch presentation of $A$, $\cH$ and $|0\rangle$ predicts the detailed form of the spectral data. The theory of (hyper)spherical varieties and their quantization will be an important example \cite{spher}. 

\subsection{Local operators on line defects}
An ${\cal N}=4$ SCFT admits two classes of half-BPS super-conformal line defects, exchanged by mirror symmetry.\footnote{These line defects preserve an $U(2|1,1)$ subgroup of the $OSp(4|4)$ superconformal group in three dimensions.} 

The class we consider here preserves the full $SU(2)_H$ R-symmetry group and breaks $SU(2)_C$ to a Cartan subgroup. There are protected local operators living 
on such line defects or interpolating between two defects, which have properties very similar to bulk Higgs branch operators: they transform in $SU(2)_H$ irreps  
with spin and scaling dimension $\ell$ and the highest weight components have non-singular OPE. We can thus organize them into a category, with graded spaces of
endomorphisms $\mathrm{Hom}_\cl(L_1,L_2)$ consisting of operators interpolating between line defects $L_1$ and $L_2$. Notice that conformal line defects 
only admit dimension $0$ intertwiners if they have identical direct summands. 

Loosely speaking, the category of line defects is analogous to a category of hyper-holomorphic sheaves on $\cM$. 

Protected sphere correlation functions can be defined as before, with some sequence of line defects placed along the same greater circle of the sphere 
protected operators would be placed at. We thus get a modified composition operation, which we encode in a deformation $\mathrm{Hom}(L_i,L_j)$ of the spaces of morphisms, together with a collection of compatible traces on $\mathrm{Hom}(L_i,L_i)$. There is also an anti-linear 
map $\rho: \mathrm{Hom}(L_i,L_j) \to \mathrm{Hom}(L_j,L_i)$, such that $\Tr \rho(a) b$ gives a positive-definite inner product on $\mathrm{Hom}(L_i,L_j)$.

\section{Examples}\label{sec:ex}
In this section we will work through a somewhat idiosyncratic sequence of detailed examples. The main purpose of this section is to demonstrate the sort of representation theoretic 
results which can emerge from mirror symmetry or S-duality. We will borrow results or conjectures from later sections when strictly needed, but employ abstract algebraic considerations as much as possible. 

\subsection{The Weyl algebra.} \label{sec:weyl}
The simplest possible non-trivial Higgs branch is $\cM = \bC^2$ and arises from the theory of a single free hypermultiplet, which we can also denote as $T = \bC^2$. 
The construction is readily generalized to the theory $T = \bC^{2n}$ of $n$ hypermultiplets.

The scalar fields in the hypermultiplet give two elementary Higgs branch operators with $\ell=\frac12$, which we can denote as 
$(X,\wt X)$ and $(P, \wt P)$. The reality structure is 
\begin{equation}
	\wt X^* = P \equiv \rho_\cl(X) \qquad \qquad \wt P^* = - X \equiv \rho_\cl(P) \,.
\end{equation}
The map $\rho_\cl$ extends anti-linearly to more general highest weight Higgs branch operators, which are 
degree $2\ell$ polynomials in $X$ and $P$.

The algebra $A$ here is the Weyl algebra. We denote the generators again as $X$ and $P$, with 
\begin{equation}
[P,X]=1 \,.
\end{equation} 
We also define $\rho$ as 
\begin{equation}
	\rho(X) = P  \qquad \qquad \rho(P) = - X  \, ,
\end{equation}
extended anti-linearly. 

The Weyl algebra does not admit any untwisted trace, but it admits a trace twisted by 
\begin{equation}
	\rho^2: X \to -X \qquad \qquad P \to -P
\end{equation}
The trace is unique once we normalize $\Tr \, 1 = 1$. E.g. 
\begin{equation}
	\Tr \, P \, X = - \Tr \, X \,  P = \frac12 \Tr[P,X] =\frac12
\end{equation}
This is the first example of positivity: we can rewrite the relations as 
\begin{equation}
	\Tr\, \rho(X) \,X = \Tr \,\rho(P) \,P = \frac12
\end{equation}
The trace of a general polynomial in $X$ and $P$ can be obtained by Wick contractions. 

Positivity is not obvious, though it can be demonstrated recursively. We will instead demonstrate it momentarily 
by giving a functional description of $\cH$ and of the spherical vector reproducing the trace. 

As the twisted trace is unique, this must coincide with the trace defined by the sphere correlation functions. 
It gives an abstract Hilbert space $\cH$ with the dense basis of states $|X^n\, P^m\rangle$
and generators of $A \times A^\op$ acting as 
\begin{align}
	X |X^n\, P^m\rangle &= |X^{n+1}\, P^m\rangle \cr P |X^n\, P^m\rangle &= |X^{n}\, P^{m+1}\rangle + n |X^{n-1}\, P^m\rangle \cr
	\wt X |X^n\, P^m\rangle &= |X^{n+1}\, P^m\rangle + m |X^{n}\, P^{m-1} \rangle \cr \wt P |X^n\, P^m\rangle &= |X^{n}\, P^{m+1}\rangle
\end{align}
with $\wt X^\dagger = \rho(X)$ and $\wt P^\dagger = \rho(P)$.

We can readily identify this with a more familiar quantization of the $\bC^2$ phase space: we pick a polarization 
and consider the Hilbert space $L^2(\bC)$, with generators of the Weyl algebra acting as  
\begin{align}
	X&= z  \cr
	P& = \partial_{z} \cr 
	\wt X &= - \partial_{\bar z} \cr
	\wt P&=- \bar z
\end{align}
where $z$ and $\bar z$ are coordinates on $\bC$. 

The spherical vector is identified with the Gaussian vector  
\begin{equation}
	|1\rangle \equiv e^{- |z|^2} 
\end{equation}
in $L^2(\bC)$, which satisfies the intertwining relations
\begin{equation}
	X |1\rangle  = \wt X |1\rangle \qquad \qquad P |1\rangle  = \wt P |1\rangle
\end{equation}
We normalize the $L^2$ measure on $\bC$ so that $|1\rangle$ has norm $1$. 

Positivity of the twisted trace is now manifest:
\begin{equation}
	 \Tr \, \rho(a) \, a = \left| a|1 \rangle \right|^2
\end{equation}

The identification of the abstract $\cH$ with $L^2(\bC)$ follows from the observation that the collection of vectors of the form 
\begin{equation}
	|X^n P^m \rangle = z^n (-\bar z)^m e^{- |z|^2} 
\end{equation}
give a dense image of $A$ in $L^2(\bC)$. Notice that the abstract definition of $\cH$ treated $X$ and $P$ democratically, whereas 
the $L^2(\bC)$ presentation picks a polarization of $\cM$. 

The free hyper-multiplet has an $F = SU(2)$ Higgs branch global symmetry, rotating $X$ and $P$ as a doublet. This becomes $USp(2n)$ for $n$ hypermultiplets. The symmetry has a discrete anomaly, though the implications of that are not immediately visible here. 

\subsubsection{Abelian Higgs branch symmetry}
We will often focus on the (non-anomalous) $U(1)$ (or $U(n)$) subgroup, acting on $X$ with charge $1$ and on $P$ with charge $-1$. It has moment map $\mu = :X P: \equiv X P +\frac12$, with $\rho(\mu) = - \mu$. The unitary $\bC^*$ action on $\cH$ corresponds to the action of dilatations on wavefunctions (half-densities) in $L^2(\bC)$. 

We can compute
\begin{equation}
	\Tr  \, e^{2 \pi m \mu} = \frac{1}{\cosh \pi m} \, ,
\end{equation}
e.g. by working in $L^2(\bC)$ and contracting 
\begin{equation}
	e^{2 \pi m \mu} |1\rangle = e^{\pi m} e^{- e^{2 \pi  m} |z|^2}
\end{equation}
with $\langle 1|$. The answer is decreasing exponentially fast along the real axis, as expected. It has poles along the imaginary axis at $m = i (2n+1)$. 
\footnote{The leading poles at $m= \pm i$ are associated to the existence of loci in $\cM$ where either $X$ or $P$ get a vev, which preserves a diagonal combination of $GL(1)$ scale transformations generated by $\mu$ and scale transformations in space-time \cite{Gaiotto:2012xa}. Poles as a function of the masses do not play an important role in this paper.}

The most basic mirror symmetry statement is that the free hypermultiplet is mirror dual to a $U(1)$ gauge theory coupled to a single hypermultiplet of charge $1$. In our language, 
$\bC^2 = (\bC^2)^! \rtimes U(1)$. Thus the Weyl algebra must admit a presentation as a Coulomb branch algebra for an $U(1)$ gauge theory, the positive trace must be reproduced by a Coulomb branch localization formula and the Hilbert space $\cH$ must admit a corresponding spectral decomposition. We will now make these facts manifest. 

The algebra $A[(\bC^2)^!]$ is trivial, as twisted hypermultiplets have no Higgs branch operators. The theory has an $U(1)$ (subgroup of an anomalous $SU(2)$) Coulomb branch symmetry and a corresponding $FI$ parameter $t$. We can identify $A[(\bC^2)^!]_{U(1)}$ with the polynomial algebra with a single generator $\mu$, anticipating the mirror map 
 $A_{U(1)} [(\bC^2)^!]=  A^{U(1)}[\bC^2]$. The rest of $A[\bC^2]$ is organized into bi-modules $HC_{b,U(1)}[(\bC^2)^!]$ for $A_{U(1)} [(\bC^2)^!]$ consisting of elements of the form $X^b \mu^n$ or $P^{-b} \mu^n$. 
 
Correspondingly, the Coulomb branch presentation of $\cH[\bC^2]$ must coincide with the spectral decomposition of $\cH[\bC^2]$ under the action of $\mu$ and $\tilde \mu$. 

We can write a complete set of distributional eigenvectors of the $U(1)$ moment map $\mu$:
\begin{equation}
	|t;b \rangle \equiv z^{-\frac12 + i t+ \frac12 b}\bar z^{-\frac12 + i t- \frac12 b} \, ,
\end{equation}
with eigenvalue $i t+ \frac12 b$ for $\mu$. These allow one to identify $L^2(\bC) = L^2(\bR \times \bZ)$ via Mellin transform.
The spherical vector can be described in the new basis as:  
\begin{equation}
	|1\rangle= \int d^2z \,z^{-\frac12 + i t+ \frac12 b}\bar z^{-\frac12 + i t- \frac12 b} e^{- |z|^2}  = \delta_{b,0}\Gamma\left(\frac12 + i t\right) 
\end{equation}
in $L^2(\bR \times \bZ)$. 

We can also state the spectral decomposition as a direct sum/integral 
\begin{equation}
	L^2(\bC) = \int^{\oplus}_{(b,t) \in \bZ \times \bR} d t \, \cH_{b,t}
\end{equation}
of one-dimensional eigenspaces $\cH_{b,t}$. This is precisely the description of the Hilbert space which emerges from Coulomb branch localization applied to the mirror description 
$\bC^2 = (\bC^2)^! \rtimes U(1)$ of the theory. See Section \ref{sec:gaucoulomb}. 

Mirror symmetry at the level of the sphere partition function is expressed by the equality
\begin{equation}
	\Tr_{\bC^2} \, e^{2 \pi m \mu}  = \frac{1}{\cosh \pi m} = \int_{-\infty}^\infty dt \,e^{2 \pi i m t} \frac{1}{\cosh \pi t}  =  \int_{-\infty}^\infty dt \, \Tr_{t,(\bC^2)^!} e^{2 \pi m \mu} \, .
\end{equation}
Recall that $A[(\bC^2)^!]_{U(1)}$ is localized here at $\mu = i t$. The trace $\Tr_{t,(\bC^2)^!}$ on the trivial algebra $A[(\bC^2)^!]$ is defined as 
\begin{equation}
	\Tr_{t,(\bC^2)^!} 1 =  \frac{1}{\cosh \pi t} \, ,
\end{equation}
i.e. the sphere partition function of a free twisted hypermultiplet. 
 
Similarly, we have relations such as
\begin{equation}
	\Tr_{\bC^2} \, P \, e^{2 \pi m \mu} \,X  =  \int_{-\infty}^\infty dt \,e^{2 \pi m (1/2+i t)} \frac{t}{\sinh \pi t} \, ,
\end{equation}
which contains the trace on $HC_{-1}[(\bC^2)^!] \times HC_1[(\bC^2)^!]$. Notice the extended width of the strip around the real axis where the integrand is analytic. This allows the integral over the real $t$ axis with $\mu = 1/2+i t$ to match the answer for different cyclic order of the operators e.g. 
\begin{equation}
	- \Tr_{\bC^2} \, X P \, e^{2 \pi m \mu}   =  \int_{-\infty}^\infty dt \,e^{2 \pi m t} \frac{\frac{1}{2}- i t}{\cosh \pi t} \, .
\end{equation}

A final observation is that pure $U(1)$ gauge theory, without the hypermultiplet matter, is not an SCFT and is mirror to $T^* \bC^*$. We can still associate the theory to 
a natural Hilbert space $L^2(\bC^*) = L^2(\bZ \times \bR)$, which is actually indistinguishable from $L^2(\bC)$: the $(\bC^2)^!$ matter fields only affect the form of the spherical vector, 
not of the ambient Hilbert space. We will use analogous statements in some interesting examples below. 

The localization formulae for the sphere correlation functions of the pure gauge theory do not converge: they take the form of expectation values for a non-normalizable spherical vector $\delta_{b,0}$.

\subsubsection{Full Higgs branch symmetry}

The $SU(2)$ moment maps are 
\begin{equation}
	E = \frac12 X^2 \qquad \qquad H =X P+\frac12 \qquad \qquad  F = -\frac12 P^2 \, ,
\end{equation}
with $\rho(E) = - F$, $\rho(H)=-H$ and $\rho(F) = -E$.

 For later reference, we can compute the $SL(2,\bC)$ image of the cyclic vector 
\begin{equation}
	\left| \begin{pmatrix} a & b \cr c & d \end{pmatrix} \right\rangle = \frac{1}{\sqrt{d}} e^{- \frac{1}{d} |z|^2 - \frac{c}{2 d} z^2 + \frac{b}{2 d} \bar z^2} \,.\end{equation}

The $SU(2)$ moment maps generate a sub-algebra of $A$ which coincides with the central quotient of $U(\fsl(2))$, with quadratic Casimir $j(j+1) = -\frac{3}{16}$.

The $\bZ_2$ quotient of the Weyl algebra, generated by the above moment maps, is one possible quantization of the $A_1$ singularity and arises from an 
SCFT defined as the $\bZ_2$ quotient of a free hypermultiplet. We will momentarily encounter a different quantization of the $A_1$ singularity, associated to a different 
SCFT. This illustrates the fact that sphere quantization depends on the full theory, not just on the Higgs branch geometry. 

The Hilbert space $\cH$ decomposes into two irreducible representations of $SL(2,\bC)$, generated by even and odd polynomials in the generators acting on the cyclic vector. Because of the anomaly for the $SU(2)$ symmetry, we cannot explain this decomposition via the simplest form of the S-duality strategy we employ in later examples. At the very end of this Section we will discuss 
S-duality for anomalous theories.

\subsection{Example: $T[SU(2)]$ and the spherical principal series representation of $SL(2,\bC)$.} \label{sec:pri}
An important example of 3d ${\cal N}=4$ SCFT is a theory denoted as $T[SU(2)]$ \cite{Gaiotto:2008ak}.  
This theory is self-mirror and has an Higgs branch $\cM$ which is an $A_1$ singularity. The Coulomb branch is also an $A_1$ singularity. There is an $SU(2)$ Higgs branch 
global symmetry and an $SU(2)$ Coulomb branch global symmetry. The latter implies the presence of an FI parameter $t$, defined up to $t \to -t$. 

The theory has a gauge-theory description as a $U(1)$ gauge theory coupled to hypermultiplets valued in $T^*\bC^2$. In Section \ref{sec:gauhiggs} we will review how tho derive 
the quantized Higgs branch algebra $A_t$ from that description. 
For now we will just state that the quantized Higgs branch algebra $A_t$ 
coincides with the central quotient of the universal enveloping algebra $U(\fsl(2))$ with quadratic Casimir $-\frac14(t^2+1)$ and proceed algebraically. 

We can express the Casimir as $j(j+1)$ with 
\begin{equation}
	j = -\frac12 + \frac{i}{2} t \,.
\end{equation}
Recall that $t$ is real in the sphere quantization. The enlarged algebra $A_{SU(2)}$ coincides with $U(\fsl(2))$. This statement can also be derived directly from the definition of $T[SU(2)]$
as S-dual to Dirichlet boundary conditions for 4d $SU(2)$ SYM \cite{Dedushenko:2020vgd}. 

The algebra $A_t$ is generated by $SL(2)$ moment maps $E,H,F$, with relations 
\begin{align}
	[H,E] &= 2 E \cr [H,F]&=- 2 F \cr 4 E\, F +H(H-2) &= -t^2-1\cr 4 F\, E +H(H+2)&= -t^2-1 \, , 
\end{align} 
and anti-holomorphic involution 
\begin{equation}
	\rho(E) = -F \qquad \qquad \rho(H)= -H \qquad \qquad \rho(F) = -E \, .
\end{equation}
This is an algebra map only if $j(j+1)$ is real, which is the case here. 

We have $\rho^2=1$ so the trace will be untwisted. The symmetrized, traceless degree $\ell$ polynomials in the generators form a single irreducible representation of $\fsl(2)$ of dimension $2\ell+1$ and give the graded pieces of $A_t$. The trace relations imply that the trace annihilates all operators which transform non-trivially under $\fsl(2)$. The trace is thus unique and must coincide with the sphere correlation functions. 

The simplest test of positivity is 
\begin{equation}
	2 \Tr \rho(E) E +2 \Tr \rho(F) F+  \Tr \rho(H) H = t^2+1
\end{equation}
which is indeed positive for the physically relevant range of FI parameters. 

We will now verify that $\cH_t$ is a well-known ``spherical principal series'' representation of $SL(2,\bC)$. 

\subsubsection{An Hilbert space of twisted half-densities}
In order to demonstrate positivity we can again introduce an auxiliary Hilbert space, which is a spherical principal series unitary representation of $SL(2,\bC)$. 

Consider the space of twisted half-densities $L^2(\bC P^1,|K|^{1-i\,t})$. These are expressions of the form 
\begin{equation}
	\psi(x) |dx|^{1-i\,t} 
\end{equation}
Under inversion $x = 1/x'$ they transform as 
\begin{equation}
	\psi(1/x')|x'|^{-2 +2 i t} |dx'|^{1-i\,t} 
\end{equation}
i.e. 
\begin{equation}
	\psi'(x')=\psi(1/x')|x'|^{-2 +2 i t}
\end{equation}

We have an action of $A_t$ by holomorphic vectorfields
\begin{equation}
	E = \partial_x \qquad \qquad H = -2 x \partial_x -1 + i t \qquad \qquad  F = - x^2 \partial_x+ (-1+i \, t) x = \partial_{x'}
\end{equation}
and of $A^\op_t$ by anti-holomorphic vectorfields
\begin{equation}
	\wt F = -E^\dagger = \partial_{\bar x} \qquad  \wt H =  -H^\dagger =-2 \bar x \partial_{\bar x} -1 + i t \qquad  \wt E =  -F^\dagger=-\bar x^2 \partial_{\bar x}+ (-1+i \, t) \bar x
\end{equation}
defined e.g. on the space $C^\infty(\bC P^1,|K|^{1-i\,t})$ of smooth twisted half-densities in $L^2(\bC P^1,|K|^{1-i\,t})$. 

The combinations $E - \wt E$, etc. generate an unitary $SU(2)$ action with an unique invariant vector 
\begin{equation}
	|1 \rangle \simeq (1+ |x|^2)^{-1+i\,t} |dx|^{1-i\,t} 
\end{equation}
Because of these relations, the expectation values $\langle 1|a|1\rangle$ define a positive trace which must coincide with $\Tr_t \, a$, as the trace is unique. 

More generally, we can decompose $L^2(\bC P^1,|K|^{1-i\,t})$ into irreps of spin $\ell$. The lowering generator annihilate primary 
wavefunctions 
\begin{equation}
	|\ell,-\ell \rangle = \bar x^\ell (1+ |x|^2)^{-1+i\,t-\ell} |dx|^{1-i\,t}  \sim E^\ell |1 \rangle
\end{equation}
which generate irreducible representations of dimension $2\ell+1$ with basis elements $|\ell,m \rangle$. 

These wavefunctions give a dense image of $A_t$ in $L^2(\bC P^1,|K|^{1-i\,t})$, which is thus identified with the completion $\cH_t$ of $A_t$ under the trace inner product. Conversely, we learn that the sphere quantization of the $A_1$ singularity as the Higgs branch of $T[SU(2)]$ produces the spherical principal series unitary representation of $SL(2,\bC)$ \cite{Etingof:2020fls}.

The reader may wonder why one would define the Hilbert space as $L^2(\bC P^1,|K|^{1-i\,t})$ rather than just taking $x$ to be a coordinate in $\bC$ and defining the Hilbert space as $L^2(\bC)$. Indeed, there is no difference at the level of Hilbert spaces. The main challenge is to promote the vector-fields representing $A$ to unbounded operators on the Hilbert space, which requires a choice of dense domain. Smooth twisted half-densities on $\bC P^1$ can provide a nice candidate. 

On the other hand, the $|1 \rangle \simeq (1+ |x|^2)^{-1+i\,t}$ wave-function in $L^2(\bC)$ and its images under the action of $A$ also provide a natural domain of definition for the action of the vectorfields. Indeed, one can rewrite any element $a$ in $A$ as a combination of $E$, $H$ and $\wt F$, which map $|1\rangle$ to a normalizable wave-function $|a\rangle$. This obviates the need of an alternative definition of 
a domain for the $A \otimes A^\op$ operators. 

We can compute a trace twisted by a Cartan generator
\begin{equation}
	\Tr_t \,e^{\pi m H} \sim \frac{\sin \pi m t}{t \sinh \pi m} \,. 
\end{equation}
using the above arbitrary choice of normalization of $|1 \rangle$. The correct normalization of the sphere partition function is instead 
\begin{equation}
	\Tr_t \, e^{\pi m H} =\frac{\pi \sin \pi m t}{\sinh \pi t \sinh \pi m} \, 
\end{equation}
and in particular 
\begin{equation}
	\Tr_t \, 1 =\frac{\pi t}{\sinh \pi t} \, 
\end{equation}
The partition function manifests self-mirror symmetry: it is symmetric under the exchange of $t$ and $m$. 
We will also derive it momentarily from the gauge theory description, together with the correct physical normalization for the spherical vector
\begin{equation} \label{eq:sphtsu2}
	|1;t \rangle =\Gamma(1-i\,t)(1+ |x|^2)^{-1+i\,t} |dx|^{1-i\,t} \, .
\end{equation}

The residue of the trace at the leading pole at $t = \pm i$, i.e. $j(j+1)=0$, is the trace induced from the trivial representation of  
$SL(2,\bC)$. Physically, that should compute sphere correlation functions for the theory obtained by an RG flow triggered by generic Coulomb branch vev.
That theory is indeed trivial. We thus see an example of the general phenomenon we described in the introduction: 
the residue of the trace at the leading pole has a kernel $I$, which here is the whole of $A_t$, and the algebra 
for the new theory is the quotient of $A_t/I$, which here is trivial. 

We can present the $SL(2,\bC)$ image of the spherical vector:
\begin{equation}
	\left| \begin{pmatrix} a & b \cr c & d \end{pmatrix} \right\rangle =  \Gamma(1-i\,t)(d + b\, x + c\, \bar x + a\, |x|^2)^{-1+i\,t} |dx|^{1-i\,t}  \, .
\end{equation}

As the sphere quantization only depends on $t^2$, there should be an unitary transformation inter-twining the actions on $L^2(\bC P^1,|K|^{1-i\,t})$
and $L^2(\bC P^1,|K|^{1+i\,t})$. This can simply be defined by matching the spherical vectors, but there is also a well-known integral transformation
$L^2(\bC P^1,|K|^{1+i\,t})\to L^2(\bC P^1,|K|^{1-i\,t})$:
\begin{equation}
	\psi(x,\bar x) \to \int_{\bC P^1} d^2 x' |x-x'|^{-2+2 i\,t} \psi(x',\bar x')\, .
\end{equation}

For later reference, it is also interesting to Fourier-transform wavefunctions with respect to $x$ and $\bar x$. The Hilbert space is thus 
identified with $L^2$-normalizable functions of $p$ and $\bar p$, with 
\begin{equation}
	E = p \qquad \qquad H = 2 p \, \partial_p +1 + i\, t \qquad \qquad  F = - p \, \partial_p^2- (1 + i t)\, \partial_p \, .
\end{equation}
The Fourier transform of the spherical vector is 
\begin{equation}
	|1\rangle = |p|^{- i t} K_{- i t}(|p|)\, .
\end{equation}
This function and the functions obtained by the action of $A$ decay exponentially fast at infinity but behave as a smooth linear combination of 
$1$ and $|p|^{- 2 i t}$ near the origin. 

If we rescale wavefunctions by a factor of  $|p|^{- i t}$ we can actually make the $t \to -t$ equivalence manifest. The moment map operators become
\begin{equation}
	E = p \qquad \qquad H = 2 \, p \, \partial_p +1 \qquad \qquad  F = - p \, \partial_p^2- \partial_p +\frac{t^2}{4 p}\, .
\end{equation}

\subsubsection{Anticipating vortex line defects}
The representation on $L^2(\bC P^1,|K|^{1 - i\, t})$ has a nice generalization to an action of $A_{t-\frac{i b}{2}} \times A^\op_{t+\frac{i b}{2}}$ on 
\begin{equation}
\cH_{n,t}\equiv L^2(\bC P^1,K^{\frac12 -\frac{b}{4} -\frac{i}{2} t } \otimes \bar K^{\frac12 +\frac{b}{4} -\frac{i}{2} t })\, .
\end{equation}
For simplicity, we can take $b \leq 0$. The case $b \geq 0$ can be treated in an  analogous manner. 

This Hilbert space defines a generic unitary principal series representation of $SL(2,\bC)$. It has a natural collection of special states of the form 
\begin{equation}
	|1;n \rangle \equiv x^n (1+ |x|^2)^{-1-\frac{b}{2} + i t} dx^{\frac12 -\frac{b}{4} -\frac{i}{2} t} d\bar x^{\frac12 +\frac{b}{4} -\frac{i}{2} t} \qquad \qquad   0 \leq n \leq b
\end{equation}
These have the property that the combinations $E-\tilde E$, etc. act on the space of $|1;a \rangle$ as generators of an irreducible representation of $\mathfrak{sl}_2$ of dimension $b+1$. 

We can generate a dense basis of the Hilbert space by acting with $A_{t-\frac{i b}{2}}$ on the $|1;a \rangle$. Each finite-dimensional irreducible representation of $SU(2)$ of dimension greater than $b+1$ will appear exactly once in the basis. These states define an $A_{t-\frac{i b}{2}} \times A_{t+\frac{i b}{2}}$ ``Harish-Chandra'' bimodule $HC_{b,t}$. 

Inner products between states in the dense basis give a pairing between $HC_{b,t}$ and $HC_{-b,t}$ together with an anti-linear map $\rho:HC_{b,t} \to HC_{-b,t}$. This data generalizes to bi-modules the notion of a positive trace and $\cH_{b,t}$ can be recovered as the $L^2$ completion of $HC_{b,t}$ \cite{Etingof:2020fls}. Both $HC_{b,t}$ and $\cH_{b,t}$ 
are invariant under a Weyl reflection acting on $t$ and $b$ simultaneously. 

These inner products can be identified as sphere correlation functions in the presence of a background vortex line defect stretching along half of the 
equator, with elements in the bimodules representing the defect endpoints. These sphere correlation functions thus give a physical meaning to all of the unitary principal series representations of $SL(2,\bC)$.

We should elaborate on the abstract algebraic properties of $HC_{b,t}$. We will do so momentarily.

A final observation is that $SL(2,\bC)$ has another ``complementary'' series of unitary representations, which could arise in our context from situations where the 
trace on $A$ or $HC_b$ is still positive definite even though $t$ is not pure imaginary. It should be possible to give a physical meaning to these representations by 
twisting the reflection positivity constraint by a Weyl reflection. We will not pursue this idea further in this paper, but we should remark that 
complementary series representations for general complex groups are an important subject in representation theory and a QFT perspective on the problem would be very interesting. 
 
\subsection{$T^* \bC^2$ as $T[SU(2)] \rtimes U(1)$}
We can now illustrate the role the bimodules $HC_{b,t}$ can play in the context of Coulomb gauging. 

Consider the theory $T[SU(2)] \rtimes U(1)$, obtained by gauging a $U(1)$ subgroup of the $SU(2)$ Coulomb branch symmetry of $T[SU(2)]$.  
This operation is known to produce (up to RG flow) a theory of free hypermultiplets with target $T^* \bC^2$, with the original Higgs branch $SU(2)$ enlarged by 
a ``magnetic' $U(1)$ Coulomb branch symmetry to $U(2)$ rotations of $\bC^2$. 

Sphere quantization of $T^* \bC^2$ naturally gives $L^2(\bC^2)$, equipped with an unitary $GL(2,\bC)$ action. 
We can decompose $T^* \bC^2$ into eigenspaces for the diagonal $GL(1,\bC)$ action. The eigenspaces will consist of 
distributional wavefunctions with specific weight under the $GL(1,\bC)$ action rotating both coordinates in the same way. 

Such a wavefunction can be identified with a twisted half-density on $\bC P^1$. Conversely, we can take a twisted half-density in $L^2(\bC P^1,K^{\frac12 -\frac{b}{4} -\frac{i}{2} t } \otimes \bar K^{\frac12 +\frac{b}{4} -\frac{i}{2} t })$ and promote it to a (distributional) half-density on $\bC^2$ with scaling weights controlled by $b$ and $t$. 

We will review this construction in greater detail in the next section. For now, we can present the basic statement: 
there is a spectral decomposition 
\begin{equation}
	L^2(\bC^2) = \int^{\oplus}_{(b,t) \in \bZ \times \bR} dt \, \cH_{b,t} \, . 
\end{equation}
If a wavefunction $|\psi\rangle$ in $L^2(\bC^2)$ is decomposed to a family $|\psi;b,t\rangle \in \cH_{b,t}$ the inner product is written as 
\begin{equation}
	\langle \psi|\chi\rangle= \sum_{b=-\infty}^\infty \int_{-\infty}^\infty dt \, \langle \psi;b,t|\chi;b,t\rangle \, .
\end{equation}
The spherical vector $|1;T^* \bC^2\rangle$ decomposes into a collection of spherical vectors $|1;t\rangle$ 
in $\cH_{0,t}$ with the specific physical normalization in (\ref{eq:sphtsu2}). 

For example, the partition function of the $T^* \bC^2$ theory twisted by a $U(1)$ mass $\beta$ is 
\begin{equation}
	\frac{1}{\cosh^2 \pi \beta} = \int_{-\infty}^\infty dt \, e^{2 \pi i t \beta} \frac{2 t}{\sinh \pi t}
\end{equation}
and the integrand contains the correct physical normalization for $\Tr_t \,1$ in $T[SU(2)]$. If we add an $SU(2)$ mass $\alpha$, we get
\begin{equation}
	\frac{1}{\cosh \pi (\beta+\alpha/2)\cosh \pi (\beta-\alpha/2)} = \int_{-\infty}^\infty dt \, e^{2 \pi i t \beta} \frac{2 \sin \pi \alpha t}{\sinh \pi t \sinh \pi \alpha}
\end{equation}

The mirror symmetry statement also provide a natural relative normalization for specific vectors in $\cH_{b,t}$ as we vary $t$. For example, 
the relation
\begin{equation}
	\langle \rho(X_1^n X_2^{b-n})|e^{2 \pi \beta \mu} |X_1^n X_2^{b-n} \rangle =  \int_{-\infty}^\infty dt \, e^{\pi (b+ 2 i t) \beta} \langle 1;n|1;n \rangle
\end{equation}
gives a natural physical normalization for $|1;n \rangle \in \cH_{b,t}$, which will appear in Coulomb branch calculations for any other theory
where twisted gauge fields are coupled to $T[SU(2)]$. We will sketch some examples momentarily. 

\subsection{Abelian and non-Abelian S-duality}
One basic property of Abelian gauging is that 
\begin{equation} \label{eq:u1s}
(T/U(1))\rtimes U(1) = T \, .
\end{equation}
 As a consequence, any theory $T$ with an $U(1)$ Higgs branch symmetry can be recast as the result 
of gauging an $U(1)$ global symmetry in a different theory $T/U(1)$.

The Hilbert space $\cH$ can accordingly always be decomposed into eigenspaces for the moment map $\mu$:
\begin{equation}
	\cH[T] = \int^{\oplus}_{(b,t) \in \bZ \times \bR} dt \, \cH_{b,t}[T/U(1)] \, .
\end{equation}
and the spherical vector $|1;T\rangle$ analogously decomposed to $|1;T/U(1);t\rangle \in \cH_{0,t}[T/U(1)]$. The inner product can be decomposed as 
\begin{equation}
	\langle f|g\rangle = \sum_{b\in \bZ} \int_{-\infty}^\infty dt \langle f;b,t|g;b,t \rangle \, .
\end{equation}

If we apply our averaging definition for the inner product on co-invariants to define $\cH_{0,\sigma}[T/U(1)]$, we restrict to states supported at $b=0$ and we get the correct answer as a repeated Fourier transform:
\begin{equation}
	\int_{-\infty}^\infty d \beta \langle f|e^{2 \pi \beta (\mu- i \sigma)} |g\rangle = \int_{-\infty}^\infty d \beta \int_{-\infty}^\infty dt e^{2 \pi i \beta (t- \sigma)} \langle f;0,t|g;0,t \rangle = \langle f;0,\sigma|g;0,\sigma \rangle \, .
\end{equation}

We will next focus on examples of theories with $SU(2)$ global symmetries. For a sufficiently ``generic'' theory $T$ with non-anomalous $SU(2)$ Higgs branch symmetry, we have a non-Abelian analogue of the relation (\ref{eq:u1s})  
\begin{equation}
	T = \left(\left(\left(T \times T[SU(2)] \right)/SU(2)\right)\times T[SU(2)]\right) \rtimes SU(2)
\end{equation}
Here we are not terribly careful with the global form of the gauge groups. Strictly speaking, one of the gauge groups should be $SU(2)$ and the other its Langlands dual to $SO(3)$.
The ``generic'' constraint essentially means that we should have an SCFT at all intermediate steps of the calculation. 

There are generalizations which hold when 
this constraint fails as well as in the presence of an anomaly and for other groups. 
We will discuss some examples were $\left(T \times T[SU(2)] \right)/SU(2)$ has an alternative known description and give further details on S-duality in Section \ref{sec:spher}.

This identity holds at the level of sphere partition functions thanks to the fact that the partition function of $T[SU(2)]$
\begin{equation}
	\Tr_{t,T[SU(2)]} e^{2 \pi m \mu} = \frac{\pi \sin \pi m t}{\sinh \pi m \sinh \pi t}
\end{equation}
is essentially a $\sin$-Fourier transform kernel for the measures $\sinh^2 \pi m\, dm$ and $\sinh^2 \pi t \,dt$.

The Coulomb branch decomposition of $\cH[T]$ is thus 
\begin{equation}
	\cH[T] = \int^{\oplus}_{(b,t) \in (\bZ \times \bR)/\bZ_2} dt \, \cH_{b,t}\left[[\left(T \times T[SU(2)] \right)/SU(2)\right] \otimes \cH_{b,t}[T[SU(2)]]\, .
\end{equation}
%If we apply our averaging definition for the inner product on co-invariants to define $\cH_{0,\sigma}\left[[\left(T \times T[SU(2)] \right)/SU(2)\right]$,
%we will encounter $SL(2,\bC)$ averages of products of $SL(2,\bC)$ matrix elements in two principal series representations, reproducing the standard ortho-normality 
%properties of such matrix elements as functions on $SL(2,\bC)$.
This is a decomposition of $\cH[T]$ into principal series representations of $SL(2,\bC)$, together with a decomposition of the spherical vector 
on the left hand side into tensor products of canonically-normalized spherical vectors on the right hand side. 

\subsection{$\bC^8$ as $T[SU(2)]^3 \rtimes SU(2)$}
Consider the theory of hypermultiplets valued in $\bC^8$, equipped with a ``tri-fundamental'' Higgs branch action of $SU(2)^3$. Concretely, we can denote the 
generators of the $A$ Weyl algebra as $Z_{abc}$ with commutator
\begin{equation}
\{Z_{abc}, Z_{a'b'c'}\} = \epsilon_{aa'}\epsilon_{bb'}\epsilon_{cc'}
\end{equation}
Here the Greek indices run from $1$ to $2$ and $\epsilon$ is the elementary antisymmetric tensor. The three copies of $SU(2)$ act respectively on 
$a$, $b$ and $c$ in $Z_{abc}$.

Remarkably, the quadratic Casimirs built from the three sets of $SL(2)$ moment maps coincide,
essentially because there is an unique $SU(2)^3$-invariant quartic polynomial $\cP$ in the $Z$'s. 

We have available a non-trivial mirror symmetry statement: $\bC^8 = T[SU(2)]^3 \rtimes SU(2)$. This means $\bC^8$ can be identified as the Coulomb branch of the theory 
obtained by gauging the diagonal $SU(2)$ symmetry of three copies of $T[SU(2)]$. Correspondingly, $\cH$ should be decomposed as 
\begin{equation}
	\cH =  \int^\oplus_{(b,t) \in \frac{\bZ \times \bR}{\bZ_2}} \left[dt \sinh^2 \pi t\right] \, \cH_{b,t}^{(1)} \otimes \cH_{b,t}^{(2)} \otimes \cH_{b,t}^{(3)} \, ,
\end{equation}
where the notation for the measure means that we define inner products as 
\begin{equation}
	\langle \psi|\chi\rangle= \frac12 \sum_{n=-\infty}^\infty \int_{-\infty}^\infty dt \, \sinh^2 \pi t\, \langle \psi;1;n,t|\chi;1;n,t\rangle\langle \psi;2;n,t|\chi;2;n,t\rangle\langle \psi;3;n,t|\chi;3;n,t\rangle
\end{equation}
with the same normalization for the $\cH_{b,t}$ inner products as in the previous subsection. 

For example, the partition functions match as 
\begin{align}
	&\frac{1}{\cosh \pi/2 (\alpha + \beta + \gamma)\cosh \pi/2 (\alpha - \beta + \gamma)\cosh \pi/2 (\alpha + \beta - \gamma)\cosh \pi/2 (\alpha - \beta - \gamma)} = \cr &= \frac12 \int_{-\infty}^\infty dt \,  \sinh^2 \pi t \frac{2 \sin \pi \alpha t}{\sinh \pi t \sinh \pi \alpha} \frac{2 \sin \pi \beta t}{\sinh \pi t \sinh \pi \beta}\frac{2 \sin \pi \gamma t}{\sinh \pi t \sinh \pi \gamma}
\end{align}
The spherical vector $|1\rangle$ in $\cH$ should have components $|1;i;0,t\rangle$ equal to the canonically normalized spherical vectors in $\cH_{0,t}^{(i)}$.

A richer check would involve comparing expressions such as $\Tr \rho(Z_{\alpha \beta \gamma}) Z_{\alpha \beta \gamma}$ in the mirror theories, 
using the relative normalization for $\langle 1;a|1;a \rangle$ derived above. 

In order to test these assertions further, we can break the symmetry among the three $SU(2)$ groups, presenting $\cH$ as $L^2(\bC^4)$ and organizing the 4 coordinates as a $2 \times 2$ matrix $z$
so that two $SL(2)$'s act on $z$ by multiplication on the left and on the right. The third has moment maps $\det z$, $: \Tr z \partial_z:$ and $\det \partial_z$. 
We can then pass from $\bC^4$ to the open set $\bC^4- \{ \det z =0 \}$  and use a ``polar'' parameterization $z = x g$ with $\det g = 1$. 
That presents $\cH$ as 
\begin{equation}
L^2\left[\left(\bC^* \times SL(2,\bC)\right)/\bZ_2 \right]
\end{equation}
It is well known that 
\begin{equation}
	L^2\left[SL(2,\bC)\right] =  \int^\oplus_{(b,t) \in \frac{\bZ \times \bR}{\bZ_2}} \left[dt \sinh^2 \pi t\right] \, \cH_{b,t}^{(1)} \otimes \cH_{b,t}^{(2)} \, .
\end{equation}

The $L^2[\bC^*]$ factor can be identified with the Fourier transform of twisted half-densities on $\bC P^1$ by a $x \to \sqrt{p}$ coordinate re-definition, reproducing 
the third factor of $\cH_{n,t}^{(3)}$.  

We can compare the dual descriptions of the spherical vector. We have
\begin{equation}
	|1\rangle = e^{- |z|^2} = e^{- |x|^2 \Tr g^\dagger g} \, ,
\end{equation}
in $L^2\left[\left(\bC^* \times SL(2,\bC)\right)/\bZ_2 \right]$
which should be compared with the direct integral of the expected product of three factors
\begin{equation}
	\Gamma(1-i \, t)(1+ |x_1|^2)^{-1+i\,t} \Gamma(1-i \, t)(1+ |x_2|^2)^{-1+i\,t}K_{- i t}(|x|^2)\, .
\end{equation}

\subsection{$\bC^8/U(1)$ as $\left(T[SU(2)]^2\times (\bC^4)^!\right) \rtimes SU(2)$}
If we gauge the Cartan of one $SU(2)$ flavour groups acting on $\bC^8$ to get $\bC^8/U(1)$, we obtain a $U(1)$ gauge theory coupled to four flavours.
On the mirror side, one of the three $T[SU(2)]$ factors is converted to $(\bC^4)^!$. 

The $U(1)$ gauge theory has Higgs branch $\cM = \bC^8/\!\!/GL(1,\bC)$, which is a minimal orbit for 
$\fsl_4$. The associated algebra is a certain quotient of $U(\fsl_4)$ and has a unique trace. We can describe the sphere quantization much as for the $T[SU(2)]$ example: 
realize $\cH$ as a space of twisted half-densities on $\bC P^3$ with a spherical vector 
\begin{equation}
	|1 \rangle = \Gamma(2-i s) \left(\Tr z^\dagger z\right)^{-2 + i s} \, .
\end{equation}
Here we represented the homogeneous coordinates as a $2 \times 2$ matrix $z$. The parameter $s$ is the quantum FI parameter for the $U(1)$ gauge group. This is just the Mellin transform of the $e^{- |z|^2}$ Gaussian vector. 

The mirror description would now give a presentation 
\begin{equation}
	\cH =  \int^\oplus_{\frac{\bZ \times \bR}{\bZ_2}} \left[dt \sinh^2 \pi t\right] \cH_{n,t}^{(1)} \otimes \cH_{n,t}^{(2)}  \, ,
\end{equation}
with spherical vector 
\begin{equation}
	\Gamma\left(\frac12 + i s + \frac{i t}{2}\right)\Gamma\left(\frac12 + i s - \frac{i t}{2}\right)\Gamma(1-i t)(1+ |x_1|^2)^{-1+i t} \Gamma(1-i t)(1+ |x_2|^2)^{-1+i t} \, .
\end{equation}
The first factor is the contribution from $(\bC^4)^!$. 

Notice that we can map $T^*SL(2,\bC)$ to an open patch in $T^*\bC P^3$ by mapping $g \to z=g$, so the Hilbert space $\cH$ can be identified with $L^2[SL(2,\bC)]$.  
This has a physical meaning: if we remove the matter fields $(\bC^4)^!$ in the mirror description, we get $T[SU(2)]^2\rtimes SU(2)$ which is the same as a sigma model 
with target $T^*SL(2,\bC)$. \footnote{These and similar statements below are obtained as an application of S-duality for $SU(2)$ ${\cal N}=4$ SYM, using the relation between $T[SU(2)]$ and the S-dual of Dirichlet boundary conditions and of $(\bC^4)^!$ to a boundary condition reducing $SU(2)$ to $U(1)$ \cite{Gaiotto:2008ak}.} 

\subsection{$\bC^8/(U(1)\times U(1)$ as $\left(T[SU(2)]\times (\bC^8)^!\right) \rtimes SU(2)$}
If we gauge the Cartan of two $SU(2)$ flavour groups acting on $\bC^8$, we obtain a $U(1)\times U(1)$ gauge theory coupled to four flavours, which is essentially the same as $T[SU(2)]^2$, up to a $\bZ_2$ subgroup which acts trivially on the matter fields and is not visible in the Higgs branch. We thus get a Hilbert space $\cH_{s_1,s_2}$ which is just the tensor product of two spherical principal series representations:
\begin{equation}
	\cH_{0,s_1} \otimes  \cH_{0,s_2} 
\end{equation}
We have FI parameters $(s_1,s_2)$ for an $SO(4)$ Coulomb branch symmetry. 

On the mirror side, two of the three $T[SU(2)]$ factors are converted to $(\bC^4)^!$. The mirror description gives a presentation 
\begin{equation}
	\cH = \int^\oplus_{(b,t)\in \frac{\bZ \times \bR}{\bZ_2}} \left[dt \sinh^2 \pi t\right] \cH_{b,t}
\end{equation}
with spherical vector 
\begin{align}
	&\Gamma\left(\frac12 + \frac{i s_1}{2} + \frac{i s_2}{2} + \frac{i t}{2}\right)\Gamma\left(\frac12 + \frac{i s_1}{2} + \frac{i s_2}{2} - \frac{i t}{2}\right)\Gamma\left(\frac12 + \frac{i s_1}{2} -\frac{i s_2}{2} + \frac{i t}{2}\right)\Gamma\left(\frac12 + \frac{i s_1}{2} - \frac{i s_2}{2} - \frac{i t}{2}\right)\cr &\cdot\Gamma(1-i t)(1+ |x|^2)^{-1+i t} 
\end{align}
We thus learn how to decompose the tensor products of two spherical principal series representations into principal series representations. With a bit more work we could 
add Wilson lines and decompose $\cH_{b_1,s_1} \otimes  \cH_{b_2,s_2}$.

If we remove a $(\bC^4)^!$ set of the matter fields in the mirror description, we get $\left(T[SU(2)]\times (\bC^4)^!\right)\rtimes SU(2)$,
 which is the same as a sigma model with target $T^*\left[SL(2,\bC)/GL(1,\bC)\right]$. The Hilbert space $\cH$ can thus be identified with a space of $L^2$-normalizable twisted half-densities on $SL(2,\bC)/GL(1,\bC)$. 
 
 We thus obtain the spectral decomposition of this functional space as a direct integral of $\cH_{b,t}$, each appearing once modulo Weyl. This construction generalizes to other Lie algebras. We will give some details momentarily. 

\subsection{Sicilian theories for $SU(2)$ and the minimal orbit of $SO(8)$.}
The series of examples based on $\bC^8$ can be generalized to the mirror of Sicilian theories $T[SU(2)]^n \rtimes SU(2)$ for $n \geq 4$. 
These theories have a direct gauge theory description as $SU(2)^{n-3}$ gauge theories with $n-2$ copies of the tri-fundamental $\bC^8$. 
Here we will specialize to $n=4$, where the theory has special properties. 

The Higgs branch $\cM$ of the theory is $O_4$, the minimal nilpotent orbit of $SO(8)$. The quantized algebra $A$ is a well-known quotient of $U(\mathfrak{so}_8)$.
As for other quantizations of minimal orbits, the associated graded take the form $\oplus_n \fso_8^{(n)}$ where
%({\bf Harvest citations from $0512296$}), 
$\fso_8^{(n)}$ is the irreducible representation of $\fso_8$ whose weight is $n$ times the weight of the adjoint. The unique trace on $A$ is positive, but it is challenging to make that manifest in an $SO(8)$-covariant way. If we use the description as $\bC^{16}/SU(2)$ and  we break $SO(8)$ to $U(4)$, we could present $\cH$ as $L^2\left[\bC^8/SL(2,\bC)\right]$.

The mirror description $T[SU(2)]^4 \rtimes SU(2)$ presents $\cH$ as
\begin{equation}
	\cH = \int^\oplus_{(b,t)\in \frac{\bZ \times \bR}{\bZ_2}} \left[dt \sinh^2 \pi t\right] \cH_{b,t}^{(1)} \otimes \cH_{b,t}^{(2)} \otimes \cH_{b,t}^{(3)}\otimes \cH_{b,t}^{(4)}
\end{equation}
 with a spherical vector presented as the tensor product of spherical vectors in each $b=0$ representation, with canonical normalization.
 
  \subsection{The theory $T[SU(3)]$ and the relation with $O_4$.}
 The theory $T[SU(3)]$ is the direct analogue to $T[SU(2)]$. It is self-mirror, with a gauge theory description involving a triangular quiver with $U(1)$ and $U(2)$ gauge groups and 
 $3$ flavours at the latter node. Symbolically, 
 \begin{equation}
 T[SU(3)] = (T[SU(2)] \times \bC^{12})/U(2)
 \end{equation}
 The algebra $A$ here is expected to be the central quotient of $U(\fsl_3)$, with Casimirs controlled by the FI parameters $t$ at the two nodes. We would also 
 expect $\cH$ to consist of an irreducible spherical principal series representation for $SL(3,\bC)$.
 
 This makes sense geometrically: the Higgs branch of $T[SU(3)]$ is the closure of the maximal nilpotent orbit of $SL(3,\bC)$,
 which is the affine closure of the cotangent bundle $T^* \left[SU(3,\bC)/B_\bC\right]$ to the flag manifold. This leads to a realization of 
 $\cH_t$ as a space of twisted half-densities on the flag manifold, which is a standard way to describe the principal series representations.
 
 We can elaborate further on the $\cH_{b,t}$ representations associated to vortex defects by 
gauging the $U(1)^2$ Cartan sub-algebra of the Coulomb branch global symmetry, which results in 
  \begin{equation}
 T[SU(3)] \rtimes U(1)^2= \bC^{16}/SU(2) = O_4
 \end{equation}
We should thus look at the $SL(3,\bC)$ action on $\cH[O_4]$. The manifold $\bC^{16}/\!\!/SU(2)$ can actually be identified with the 
affine closure of $T^* \left[SU(3,\bC)/U_\bC\right]$ where $U_\bC$ is the maximal unipotent subgroup of $SL(3,\bC)$. We thus tentatively identify 
\begin{equation}
	\cH[O_4] = L^2\left[SU(3,\bC)/U_\bC\right]
\end{equation}
This agrees with the decomposition
\begin{equation}
	 L^2\left[SU(3,\bC)/U_\bC\right] = \sum_{b} \int dt \, \cH^{SL(3)}_{b,t}
\end{equation}
and the spherical vector on the left hand side is decomposed into spherical vectors with a physical relative normalization as a function of $t$. 

The self-mirror property of $T[SU(3)]$ follows from the mirror descriptions of $O_4$ by gauging/ungauging $U(1)$ symmetries. 
The presentation 
 \begin{equation}
 T[SU(3)] = (T[SU(2)] \times (\bC^{12})^!) \rtimes U(2)
 \end{equation}
 makes manifest an $A_{U(2)}[T[SU(2)] \times (\bC^{12})^!]$ subalgebra which coincides with $U(\fgl_2)$. This is just the block-diagonal subalgebra in $U(\fsl_3)$,
 with the central elements playing the role of the Gelfand - Zeitlin sub-algebra of $U(\fsl_3)$.

The sphere partition function of $T[SU(N)]$ computed from the gauge-theory description (see e.g. \cite{Chang:2019dzt,Gaiotto:2019mmf} for details) is 
\begin{equation}
	\Tr_{t,T[SU(N)]} e^{2 \pi m \mu} = \frac{\sum_{\sigma \in S_N} (-1)^\sigma e^{2 \pi i \sum_a m_{\sigma(a)} t_a}}{\prod_{i<j} \sinh \pi (m_i-m_j)\sinh \pi (t_i-t_j)}
\end{equation}
with $\sum_i m_i=0$ and $\sum_i t_i=0$. The FI parameters in the gauge theory are differences between consecutive $t$'s, but the expression has the full Weyl symmetry of the 
$SU(N)^!$ Coulomb branch symmetry.

This displays clearly the simple poles associated to Coulomb branch vevs, at $t_j = t_i+1$. For $N=3$, the simple pole at $t_2 = t_1+1$ has a residue which 
reproduces the partition function of a $U(1)$ gauge theory with three flavours, e.g. $\bC^6/U(1)$, which gives a low energy effective description of the theory in the presence of the vev. 
The identification holds up to a function of $t$ only, associated to an extra free twisted hypermultiplet found at low energy. 

The algebra for the simplified theory is a truncation of $A[T[SU(3)]]$, which quantizes the minimal nilpotent orbit of $SL(3)$. 

  \subsection{The interplay between $T[SU(3)]$ and $T[SU(2)]$. }
 Define now 
 \begin{equation}
 T =\bC^{12}/U(1)
 \end{equation}
i.e. a $U(1)$ gauge theory with $6$ flavours. As long as the S-duality construction for $SU(2)$ applies, we have a mirror symmetry 
\begin{equation}
	T=\left(T[SU(3)]\times T[SU(2)] \right) \rtimes SU(2)
\end{equation}
and thus a decomposition 
\begin{equation}
	\cH_{n,s}[T] = \int^{\oplus}_{(b,t) \in (\bZ \times \bR)/\bZ_2} dt \, \cH_{b,n;t,s}\left[T[SU(3)]\right] \otimes \cH_{b,t}[T[SU(2)]]\, .
\end{equation}
in principal series representations of $SL(3, \bC) \times SL(2,\bC)$, with $\bC^6$ transforming in a bi-fundamental representation. 

This is an example of a general phenomenon which applies to the bifundamental action of $SL(N+1, \bC) \times SL(N,\bC)$ on $\bC^{N(N+1)}$, 
which follows from a basic family of S-duality statements concerning 3d interfaces. 

\subsection{$\bC^{18}/U(1)$ as $\left(T[SU(3)]^2\times (\bC^6)^!\right) \rtimes SU(3)$}
There is another family of S-duality statements which control the bifundamental action of $SL(N, \bC) \times SL(N,\bC)$ on $\bC^{N^2}$. 
It e.g. leads to 
\begin{equation}
	T= \bC^{18}/U(1)=\left(T[SU(3)]^2\times (\bC^6)^!\right) \rtimes SU(3)
\end{equation}
and thus to a decomposition 
\begin{equation}
	\cH_{n,s}[T] = \int^{\oplus}_{(b,t) \in (\bZ^2 \times \bR^2)/S_3} dt \, \cH_{b;t}\left[T[SU(3)]\right] \otimes \cH_{b,t}[T[SU(3)]]\, .
\end{equation}
with the spherical vector involving an extra factor of $\prod_{a=1}^3 \Gamma\left(\frac12 + i s + \frac{i t_a}{2}\right)$ from $(\bC^6)^!$. 

Embedding $SL(3,\bC) \subset \bC^9/GL(1)$, the decomposition of the Hilbert space is the standard decomposition of $L^2(SL(3,\bC) )$ in principal series representations. 

\subsection{The $SU(3)$ trinion}
Our final $SU(3)$-related example is the trinion theory $T[SU(3)]^3 \rtimes SU(3)$. This theory lacks a gauge theory description, but is expected to possess an $E_6$ 
global symmetry extending the naive $SU(3)^3$. We expect the associated algebra $A$ to be the quantization of the minimal nilpotent orbit of $E_6$. 
It would be nice to verify the expected decomposition 
\begin{equation}
	\cH_{n,s}[T] = \int^{\oplus}_{(b,t) \in (\bZ^2 \times \bR^2)/S_3} dt \, \cH_{b;t}\left[T[SU(3)]\right] \otimes \cH_{b,t}[T[SU(3)]]\otimes \cH_{b,t}[T[SU(3)]]\, .
\end{equation}
under the action of the $SU(3)^3$ subgroup of $E_6$. 

\subsection{Example: $T[G]$ and principal series representations of $G_\bC$.}
To each reductive group $G_\bC$ we can associate a theory $T[G]$ which has a $G$ Higgs branch symmetry and a $G^\vee$ Coulomb branch symmetry, where $G^\vee$ is the 
Langlands dual group to $G$. It is mirror to $T[G^\vee]$. 

The Higgs branch of this theory is the regular nilpotent orbit of $G$, possibly deformed to other regular orbits by 
FI parameters. With a real FI parameter turned on, this can be identified with the cotangent bundle to the 
complete flag variety $G_\bC/B_\bC$, with $B_\bC$ being the Borel subgroup of $G_\bC$. 

The algebra $A$ is the central quotient of $U(\fg)$, with Casimirs determined by the FI parameters.
The trace on $A$ is again unique. The extended algebra $A_{G^\vee}$ coincides with $U(\fg)$.
The quantum FI parameters can be identified as imaginary elements $i t$ in the Cartan algebra $\fh^\vee$ for $G^\vee$,
modulo the action of the Weyl group. 

The positivity of the trace can be demonstrated by identifying a cyclic/spherical vector in a space of $L^2$-normalizable twisted half-densities on 
$G_\bC/B_\bC$, giving a concrete realization for $\cH$ as a spherical principal series representation 
of $G_\bC$. We have seen this happen for both $SU(2)$ and $SU(3)$. 

We expect vortex line defects to allow access to unitary representations of 
$G_\bC$ and associated Harish-Chandra bi-modules built from $L^2$-normalizable twisted half-densities on 
the complete flag variety for $G$ and all possible twists, i.e. general principal series representations of $G$. 
These are associated to quantum FI parameters $b/2 + i t$ for any integer weight $b$, modulo the action of the Weyl group.   

There is a construction analogue to the ones given for $T[SU(2)]$ and $T[SU(3)]$, involving the Coulomb gauging 
$T[G] \rtimes H^\vee$, with $H^\vee$ being the Cartan subgroup of $G^\vee$. This gives a theory whose Higgs branch 
is the cotangent bundle to $G_\bC/N_\bC$  with $N_\bC$ being the unipotent subgroup of $G_\bC$. 
This has an $G_\bC \times H_\bC$ global symmetry acting on the Higgs branch, 
with $H$ being the Cartan subgroup of $G$.

The Hilbert space associated to this extended theory can be identified with $L^2(G_\bC/N_\bC)$. We have a decomposition 
\begin{equation}
	L^2(G_\bC/N_\bC) = \sum_b \int dt \, \cH_{b, t}
\end{equation}
and if we normalize inner products so that the spectral measure is ``1'':
\begin{equation}
	\langle \psi|\chi\rangle= \sum_b \int dt \, \langle \psi;b,t|\chi;b,t\rangle
\end{equation}
we should recover the physical normalization of sphere correlation functions:
\begin{equation}
	\Tr_{t,T[G]} e^{2 \pi m \mu} = \frac{\sum_{\sigma \in W_G} (-1)^\sigma e^{2 \pi i \sigma(m) \cdot t}}{\prod_\alpha \sinh \pi \alpha \cdot m\prod_{\alpha^\vee}\sinh \pi \alpha^\vee \cdot t}
\end{equation}
where the product in the denominator runs over positive roots. This appears to be a somewhat non-trivial combinatorial statement. 
  
\subsection{Example: $T_{\rho^\vee}[G]$ and unitary representations of $G_\bC$.}
There is a further collection of theories  $T_{\rho^\vee}[G]$ labelled by an $\fsl_2$ embedding $\rho^\vee$ in $G^\vee$. 
 They can be engineered with the help of S-duality and Nahm pole boundary conditions in 4d SYM \cite{Gaiotto:2008sa,Gaiotto:2008ak}. We denote their mirror as 
 $T^{\rho^\vee}[G^\vee]$. For $A$ type, these theories have an explicit gauge theory description as certain A-type quivers.

The Higgs branches of  $T_{\rho^\vee}[G]$ have $G$ symmetry and are expected to roughly consist of some union of 
nilpotent orbits of $G$, conjecturally controlled by the Spaltenstein map \cite{Chacaltana:2012zy}.
We thus expect to have quantized algebras $A^{\rho^\vee}_{t}$ as well as families of $HC^{\rho^\vee}_{b,t}$, 
all equipped with positive traces and completed to unitary representations of $G_\bC$. 

A mathematical construction of algebras, modules and unitary structures which appear to match the physical expectations 
about $T_{\rho^\vee}[G]$ was recently described in \cite{2021arXiv210803453L}. It should be possible to explore the dictionary in depth. 

The $T_{\rho^\vee}[G]$ theories should emerge as a description of $T[G]$ near a point in the Coulomb branch  
described by the raising operator in $\rho^\vee$. Correspondingly, $A^{\rho^\vee}_t$ and their traces 
should emerge from the residue of the sphere correlation functions of $T[G]$ at the pole determined by 
the Cartan generator of $\rho^\vee$. It would be nice to verify this statement in detail as well. 

The Coulomb branch of $T_{\rho^\vee}[G]$ is expected to be given by the Drinfeld-Sokolov reduction of the regular nilpotent orbit of $G^\vee$, i.e. the Coulomb branch of $T[G]$ \cite{Gaiotto:2008ak}. The data which goes into the Drinfeld-Sokolov reduction is precisely $\rho^\vee$. We expect the Coulomb branch algebra $A^![T_{\rho^\vee}[G]]$ to also be given by the corresponding quantum DS reduction, i.e. to be the finite W-algebra $W^{\rho^\vee}_{G^\vee}$ \cite{2001math......5225G}.  

A natural question is how to derive the correct physical trace on $A^!$ in a manner uniform in $G$ and $\rho^\vee$. This is a special case of a more general 
problem, as the Drinfeld-Sokolov reduction has a precise physical meaning and can be executed on general theories with a Higgs branch $G^\vee$ symmetry. 

The Drinfeld-Sokolov reduction is very similar to a gauging operation, involving the parabolic subgroup $P^\vee$ of $G_\bC^\vee$ associated to $\rho^\vee$. 
An important difference is that the moment map $\mu[f]$ for the $f$ generator of $\fsl_2$ is set to $1$ instead of $0$. A natural conjecture would thus be 
a $P^\vee$ averaging formula:
\begin{equation}
	\Tr_{T^{\rho^\vee} } \, a \equiv \oint_{h \in \gamma \subset P^\vee} d\Vol_h \, \chi(h)\Tr_T  \, h \, a \,
\end{equation} 
on some middle-dimensional cycle in $P^\vee$. Here $\chi(h)$ is a character such that the integral imposes the $\mu[f]=1$ condition. We leave a full definition of such a formula to future work. 

All of these considerations should extend to more general class of theories $T^\rho_{\rho^\vee}[G]$, which should give unitary representations 
of $W^{\rho}_{G}$.

\subsection{S-duality and theories with anomalous $SU(2)$ symmetry}
The $SU(2)$ global symmetry for a 3d ${\cal N}=4$ theory can have a $\bZ_2$-valued anomaly. For example, 
$n$ free hypermultiplets, each forming a doublet for $SU(2)$, have a non-zero anomaly if $n$ is odd. 

These theories admit a coupling to 4d ${\cal N}=4$ SYM with a topological discrete $\theta$ angle,  
which changes the S-duality properties of the theory so that it is self-dual, instead of being dual to an $SO(3)$ 
gauge theory. Relatedly, we will discuss a theory $\widehat{T}[SU(2)]$ which appears to play a role analogous to that of $T[SU(2)]$ in the non-anomalous case. 

Consider a 3d $U(2)$ gauge theory coupled to an adjoint hypermultiplet and a fundamental one, e.g. the ADHM quiver whose Higgs branch is the Hilbert scheme of two points in $\bC^2$. 
This theory is self-mirror and has both an $SU(2)$ global symmetry acting on the adjoint hypermultiplet and an $SU(2)^!$ Coulomb branch symmetry. 

It includes a decoupled hypermultiplet: the $U(1)$ part of the adjoint corresponding to the center-of-mass of the two points in $\bC^2$. By the self-mirror property, it must also have a decoupled twisted hypermultiplet. We strip both off to define $\widehat{T}[SU(2)]$. Notice that the Higgs branch is now $\bC^2/\bZ_2$ and so is the Coulomb branch, just as for $T[SU(2)]$. On the other hand, the $SU(2)$ global symmetry acts on an odd number ($3$) doublets in the gauge theory description, and thus has an anomaly. The same must be true for $SU(2)^!$.

We can compute the sphere partition function with some effort. As the $U(1)$ subgroup of the gauge group acts only on the fundamental hypermultiplets, 
we can combine them into a copy of $T[SU(2)]$, coupled by $SO(3)$ gauge fields  to the adjoint hypermultiplet:
\begin{equation}
	\int_{-\infty}^\infty d\beta \frac{\sinh^2 \pi \beta}{\cosh \pi m/2 \cosh \pi(m/2 - \beta)\cosh \pi(m/2 + \beta)}\frac{\sin \pi \beta t}{\sinh \pi \beta \sinh \pi t} \sim \frac{ \pi \sin \pi m t/2}{\sinh \pi m \sinh \pi t}
\end{equation}
We expect the Higgs branch algebra to still be the central quotient of $U(\fsl_2)$. 
The $m$ dependence of the partition function characterizes this as the trace in a spherical principal series representation with Casimir $- \frac14 - \frac{t^2}{16}$, 
with a normalization $\Tr_t 1 = \frac{\pi t}{2 \sinh \pi t}$ which is different from the one we found for $T[SU(2)]$. 

Here $t$ is the FI parameter in the standard normalization for the Cartan on $SU(2)^!$. So the relation between FI parameter and Casimir has a factor of $2$ in $\widehat{T}[SU(2)]$
compared to $T[SU(2)]$. This is related to the fact that $T[SU(2)]$ expresses a Langland duality relation between $SU(2)$ and $SO(3)$ while $\widehat{T}[SU(2)]$ 
between two $SU(2)$ gauge groups. 

The relation 
\begin{equation}
	\widehat{T}[SU(2)] \times (\bC^2)^! = (T[SU(2)] \times \bC^6)/SU(2)
\end{equation}
implies 
\begin{equation}
	\bC^6 = (\widehat{T}[SU(2)] \times (\bC^2)^!  \times T[SU(2)] ) \rtimes SU(2) \, , 
\end{equation}
which gives a spectral decomposition of $L^2(\bC^3)$ under the action of $SO(3) \times SU(2)$, with 
$SU(2)$ acting on each of the three doublets $(X_i, P_i)$ of coordinates in $T^* \bC^3$:
\begin{equation}
	L^2(\bC^3) = \int^\oplus_{(b,t)\in \frac{2 \bZ \times \bR}{\bZ_2}} dt \, \sinh^2 \pi t \, \cH^{SO(3)}_{b,t} \otimes \cH^{SU(2)}_{b/2,t/2} \,.
\end{equation}
Indeed, we can easily verify that the quadratic Casimirs in $T^* \bC^3$ are related as $\frac14 + \frac{t^2}{4}$ and $\frac14 + \frac{t^2}{16}$.

We can also see that, for example, $(X_i, P_i)$ has spins $1$ and $1/2$ respectively for the two groups and is the leading element in $\cH^{SO(3)}_{1,t} \otimes \cH^{SU(2)}_{1/2,t/2}$, etcetera. 

The decomposition comes with a decomposition of the spherical vector into canonically normalized spherical vectors in $\cH^{SO(3)}_{b,t} \otimes \cH^{SU(2)}_{b/2,t/2}$,
the latter with the physical $\widehat{T}[SU(2)]$ normalization, weighed by an extra $\Gamma(\frac12 - i t)$ from the factor of $(\bC^2)^!$.

A final interesting observation is that the reduction of the $\widehat{T}[SU(2)]$ theory associated to a Coulomb branch vev is expected to yield 
$\bC^2$, or perhaps $\bC^2/\bZ_2$. This precisely occurs thanks to the normalization of $t$: the reduction should occur as we set $t=i$, which in $T[SU(2)]$ would give zero Casimir and the trivial theory, 
but in $\widehat{T}[SU(2)]$ gives a Casimir $\frac{3}{16}$, precisely consistent with the Weyl algebra of $\bC^2$. 

We thus expect that the 
residue at $t=i$ of the (appropriately normalized) trace on the central quotient of $U(\fsl_2)$ will give a trace with a kernel $I$ such that $U(\fsl_2)/I$ is the $\bZ_2$ quotient of the 
Weyl algebra. 

\section{Gauging Higgs branch symmetries} \label{sec:gauhiggs}
Gauge theories provide a large class of examples of 3d ${\cal N}=4$ SCFTs. A gauge theory is defined by a choice of reductive gauge group $G$
and of some ``matter theory'' $T$ with an action of $G$ on the Higgs branch.\footnote{Some discrete anomalies may obstruct the existence of the gauge theory. 
We have not seen any manifestation of that constraint in this paper.} We will denote as $G_\bC$ the complexification of $G$ and as $T/G$ the new theory.

The Higgs branch itself, the quantized Higgs branch algebra and the protected sphere correlation functions all transform in a predictable way under 
gauging:
\begin{itemize}
	\item The Higgs branch $\cM[T]$ has a tri-holomorphic $G$ symmetry and $\cM[T/G]$ coincides with the complex symplectic quotient $\cM[T]/\!\!/G_\bC$. 
	Correspondingly, $A_\cl[T/G]$ consists of $G_\bC$-invariant elements in $A_\cl[T]$ modulo the ideal generated by the moment maps $\mu$ for the $G_\bC$ action.
	\item The quantized algebra $A[T]$ has a $G_\bC$ action implemented by quantum moment maps, which we also denote as $\mu$. The algebra $A[T/G]$ is a quantum Hamiltonian reduction of $A[T]$, i.e. it consists of $G_\bC$-invariant elements in $A[T]$ modulo the ideal generated by the quantum moment maps.
	\item The sphere correlation functions $\Tr_{T/G}$ are computed by an integral average \cite{companion}
	\begin{equation}
	 \Tr_{T/G} \,a \equiv \int_{G_\bC/G \subset G_\bC} d\mathrm{Vol}_h\, \Tr_{T} \,a \,h
	\end{equation}
	Here $h$ is integrated over a middle-dimensional integration contour consisting of positive-definite Hermitean elements of $G_\bC$, i.e. elements of the form $e^{2 \pi \beta}$, with $\beta = \beta^\dagger$. This is identified with the $G_\bC/G$ coset by a $h = g g^\dagger$ parameterization. The measure $d\mathrm{Vol}_h$ is the invariant holomorphic top form on $G_\bC$. The $\Tr_{T} \,a \,h$ expression is a $G_\bC$-twisted trace defined as $\Tr_T \,a \,e^{2 \pi \beta \cdot \mu}$. 
\end{itemize}
The integral formula will converge if $\Tr_T \,a \,e^{2 \pi \beta \cdot \mu}$ decreases sufficiently fast at large $\beta$. This is expected to be the case if $T/G$ is an SCFT. 

In order to see that the integral above defines a trace on $A[T/G]$, notice that $G_{\bC}$-invariant operators can be commuted across the $h$ insertion and that 
if $a = a_i \mu^i$ we can write the trace as a total derivative in $\beta$ which integrates to $0$ up to an infinitesimal contour deformation. 

The integral formula can be reduced to a more standard form by observing that $\Tr_{T} \,a \,h$ for $G_\bC$ invariant $a$ only depends on the conjugacy class of $h$.  We can thus diagonalize $\beta$ up to a unitary transformation, much as we would do for an integral over the compact group $G$. In the latter case, $d\mathrm{Vol}_h$ reduces to a Vandermonde measure over anti-Hermitean diagonal $\beta$. In the current case, we obtain instead an integral over 
Hermitean diagonal $\beta$ and the Vandermonde measure is analytically continued, replacing $\sin$ with $\sinh$ functions:
\begin{equation}
	 \Tr_{T/G} \,a = \int_{\beta \in \fh|\beta = \beta^\dagger} d^{\mathrm{rk} \fg} \beta \left[\prod_{\alpha} \sinh^2 \pi \alpha \cdot \beta \right]\, \Tr_{T} \,a \, e^{2 \pi \beta}
\end{equation}
where $\fh$ is the Cartan and $\alpha$ runs over the positive roots. 

Positivity of the trace appears to be a remarkable property which is not manifest from the integral formula. It is thus interesting to compare the 
Hilbert space $\cH_[{T/G}]$ for the gauged theory with the Hilbert space $\cH[T]$ for the original theory. From that perspective, the $G$ action is generated by 
unitary operators on $\cH[T]$ and thus the space $A[T]^{G_\bC}$ of $G_\bC$-invariants in $A[T]$ is the intersection of $A[T]$ with the $G$-invariant subspace $\cH[T]^G$ of $\cH[T]$. Then $A[T/G]$ can be identified with ``co-invariants'' of $A[T]^{G_\bC}$, i.e. the quotient of $A[T]^{G_\bC}$ by the image of $\mu$ in $A[T]^{G_\bC}$. 

The above integral formulae define an inner product to the space of co-invariants, defined by lifting states to $A[T]^{G_\bC}$, pairing them with 
$\Tr_{T} \,\rho(a) b \,h$ and averaging over positive-definite Hermitean $h$. The formula can be derived in a BRST formalism.  
Leaving aside physics considerations, it is not obvious that this averaging operation should be well-defined or positive definite. 

A naive way to implement the quantum Hamiltonian reduction on an Hilbert space would be to average vectors over the unitary action of $G_\bC$. 
In particular, we could take a $G$-invariant vector such as $|1\rangle$ and average over $G_\bC/G$:
 \begin{equation}
	|\tilde 1\rangle =\int_{G_\bC/G \subset G_\bC} d\mathrm{Vol}_h\,h \,|1\rangle
\end{equation}
Unfortunately, this will be a distributional state, with an infinite norm which is morally proportional to the volume of $G_\bC/G$. If we can regularize the 
volume divergence, though, the expectation value of $a$ on $|\tilde 1\rangle$ is formally proportional to $\Tr_{T/G} \,a$. Indeed, 
we can write
 \begin{equation}
	\langle \tilde 1| a |\tilde 1\rangle =  \int_{G_\bC/G \subset G_\bC} d\mathrm{Vol}_{h'} \int_{G_\bC/G \subset G_\bC} d\mathrm{Vol}_h\,\langle 1| a (h')^{-1} h \,|1\rangle
\end{equation}
Changing the integration variables from $(h',h)$ to $(h',(h')^{-1} h)$ and deforming the integration contour for the internal integral, we recover the integral over 
$G_\bC/G$ of $\Tr_{T/G} \,a$. It should be possible to refine this argument to a rigorous proof of positivity of $\Tr_{T/G}$. 

\subsection{Functional representations}
We have already encountered several examples where $\cH_T$ can be identified with a space of half-densities $L^2(\cV,|K|\otimes \cL)$ on some auxiliary space $\cV$ and ($A[T]^\op$) $A[T]$ is represented by (anti)holomorphic differential operators on $\cV$. In particular, the action of some flavour symmetry group  $G_\bC$  is induced from a geometric action on $\cV$. 
Furthermore, $A \in \cH[T]$ consists of smooth functions, perhaps with some allowed singularities at special loci.

In such a situation, $A[T/G]$ is represented by $G_\bC$-invariant holomorphic differential operators modulo the vectorfields implementing the $G_\bC$ action. 
Suppose now that $G_\bC$ acts freely outside some locus $D$ of lower codimension. Denote $\cU = \cV-D$. Then $A_{T/G}$ will act naturally on 
some space of twisted half-densities on $\cU/G_\bC$. We would like to identify $\cH_{T/G}$ with the corresponding space $L^2(\cU/G_\bC,|K|\otimes \cL')$ of $L^2$-normalizable half-densities. 

Our strategy is to define a candidate spherical vector $|1;T/G\rangle$ in $L^2(\cU/G_\bC,|K|\otimes \cL')$ as a $G_\bC$ average of the spherical vector $|1;T\rangle$.
Denote as $\psi_\cV(v, v^*)$ the wavefunction representing  $|1;T\rangle$. We can average it as 
\begin{equation}
	 \int_{G_\bC} |d\mathrm{Vol}_{g}|^2 \,\psi_\cV(g v, (gv)^*) 
\end{equation}
We expect this to give a reasonable wavefunction $\psi_{\cU/G_\bC}$ in $L^2(\cU/G_\bC,|K|\otimes \cL')$. Indeed, we can rewrite the integral as 
  \begin{equation}
	\psi_{\cU/G_\bC} \simeq \int_{G_\bC/G \subset G_\bC} d\mathrm{Vol}_{h}\, h \circ \psi_\cV
\end{equation}
up to a factor of volume of $G$, except that we interpret this as a twisted half-density on $\cU/G_\bC$ rather than a $G_\bC$-invariant distributional half-density on $\cV$. 
We take this to define $|1;T/G\rangle$. It is easy to verify that it is annihilated by $a-\wt a$ for $a \in A_{T/G}$, as expected. 

When we compute an expectation value 
\begin{equation}
	\langle 1;T/G| a |1;T/G\rangle
\end{equation}
the integral over $\cU/G_\bC$ can be combined with the $G_\bC$ integral in the definition of $\langle 1;T/G|$, promoting the $L^2(\cU/G_\bC,|K|\otimes \cL')$ pairing to an 
$L^2(\cV,|K|\otimes \cL)$ pairing between $\langle 1;T|a$ and the average of $|1;T\rangle$. This is precisely $\Tr_{T/G}$. 

We conclude that the averaging operation maps $A_T \subset L^2(\cV,|K|\otimes \cL)$ to $A_{T/G} \subset L^2(\cU/G_\bC,|K|\otimes \cL')$ and thus gives us an embedding of
$\cH_{T/G}$ into $L^2(\cU/G_\bC,|K|\otimes \cL')$, which we expect to be an isomorphism.

\subsubsection{Harish-Chandra bimodules from Abelian Wilson lines}
A simple modification of the quantum Hamiltonian reduction is to consider the space $A_T^{G_\bC,\chi}$ 
of operators of charge $\chi$ under the Abelian generators of the gauge group. The action of $\mu -i t- \frac12 b$ on this space from the left
equals the action of $\mu -i t+\frac12 b$ from the right. 

Accordingly, we obtain an $A_{T/G}\left[t- \frac{i}{2} b\right]$-$A_{T/G}\left[t+ \frac{i}{2} b\right]$ bimodule $HC[t, b]$. This represents the space of local operators 
at the junction of two line defects. The sphere correlation functions can be extended in a natural way to a pairing 
\begin{equation}
	HC[t, -b] \times_{A\left[t- \frac{i}{2} b\right] \times A\left[t+ \frac{i}{2}b\right]^\op} HC[t, b] \to \bC
\end{equation}	
which together with the involution $\rho$ inherited from $A_T$ gives a positive-definite inner product on $HC[t,b]$ \cite{Etingof:2020fls}. 

\subsection{Example: $G=U(1)$ acting on $T^*\bC$, i.e. SQED with one flavour.}
In this example the Higgs branch algebra is trivial, but we can at least check that the integral giving the trace and cyclic vectors are 
convergent. We will run our definitions in detail. 

We represent $\cH_T= L^2(\bC)$ as in the example in the previous section. The sphere partition function for the gauge theory is the average
\begin{equation}
	\int_{-\infty}^\infty d \beta \Tr e^{2 \pi \beta (X P + \frac12+ i t)} =\int_{-\infty}^\infty d \beta \frac{e^{2 \pi i t \beta}}{\cosh \pi \beta} = \frac{1}{\cosh \pi t}
\end{equation}
This is a beautiful example of mirror symmetry: the mirror description of the gauge theory is a theory with no gauge group and matter in $T^* \bC$. 

The average of the Gaussian vector is 
\begin{equation}
	\int_{-\infty}^\infty d \beta e^{2 \pi \beta (X P + \frac12+ i t)} |1\rangle = \int_{-\infty}^\infty d \beta e^{\pi \beta+ 2 \pi i t} e^{-  e^{2 \pi \beta}|z|^2}= \Gamma\left(\frac12 + i t\right) |z|^{-1-2 i t} 
\end{equation}
We take $D = {0}$, so that $\cU = \bC^*$ and $\cU/G_\bC$ is a point.  The resulting ``wavefunction'' is 
\begin{equation}
	\Gamma\left(\frac12 + i t\right)
\end{equation}
whose norm $\frac{\pi}{\cosh \pi t}$ agrees with the sphere partition function. 

\subsection{$U(1)$ with two flavours}
This is a gauge theory realization of $T[SU(2)]$. 
In order to get the quantum Higgs branch algebra, we take two copies of the Weyl algebra and do a quantum Hamiltonian reduction by the diagonal $\bC^*$ action. 
The result is the central quotient of $U(\fsl_2)$, at a value of the Casimir controlled as before by the FI parameter as $j =-\frac12 + \frac{i t}{2}$. 

We can take $\cV=\bC^2$ with $G=U(1)$ acting in the same way on both factors.
The invariant combinations $X^\alpha P_\beta$ modulo the $\bC^*$ moment map generate the whole reduction and satisfy the $sl_2$ Lie algebra:
\begin{equation}
E= X^1 P_2 \qquad \qquad H = X^1 P_1- X^2 P_2\qquad \qquad P = X_2 P_1
\end{equation}
The whole algebra is a central quotient of $U(sl_2)$, with a Casimir determined by the FI parameter in the moment map $\mu = X^\alpha P_\alpha +1-i t$. 

We can compute a character 
\begin{equation}
	\Tr e^{\pi m (X^1 P_1 - X^2 P_2)} = 
	\int_{-\infty}^\infty \frac{e^{- 2 \pi i t \beta} d \beta }{\cosh \pi(\beta + m/2)\cosh \pi(\beta + m/2)} = \frac{2 \sin \pi m t}{\sinh \pi t \sinh \pi m}
\end{equation}

The average of the Gaussian $e^{-|z_1|^2 - |z_2|^2}$ is 
\begin{equation}
	\int_0^\infty ds \, s^{-i t} e^{-s |z_1|^2 - s |z_2|^2}= \frac{\Gamma(1-i t)}{\pi^2}  (|z_1|^2 + |z_2|^2)^{-1+i t}
\end{equation}
In inhomogeneous coordinates, this becomes the familiar wavefunction representing $|1\rangle$ in the principal series representation. 
\begin{equation}
	|1 \rangle \equiv \frac{\Gamma(1-i t)}{\pi} (1+ |x|^2)^{-1+i t} |dx|^{1-i t} \,.
\end{equation}

An alternative choice of $\cV$ is $\bC^2$ with $U(1)$ acting in the opposite way on the two factors. 
The averaged cyclic vector becomes 
\begin{equation}
	\int_0^\infty ds \, s^{1-i t} e^{- s |z_1|^2 - 1/s |p_2|^2} \sim |z_1 p_2|^{-i t} K_{-i t}(|z_1 p_2|)\,,
\end{equation}
giving the Fourier-transformed version of $|1\rangle$ in the principal series representation.

The normalization factor $\Gamma(1-i t)$ is an interesting piece of information, which does not immediately emerge from a direct algebraic analysis. 
In certain contexts \cite{Teschner:1997fv}, the wavefunction is normalized as 
\begin{equation}
	\Psi(j;x,\bar x|h) = \frac{(2j+1)}{\pi} \left((x\,\,1)\, h\, {\bar x\choose 1}\right)^{2j}\,,
\end{equation}
which has useful orthonormality properties with a measure $dj$:
\begin{align}
	\int_{H_3^+} d\Vol_h \Psi(j;x,\bar x|h)^* \Psi(j';x',\bar x'|h) &= 2 \pi i \delta(j-j') \delta^{(2)}(x-x') \cr
	\int_{j \in -\frac12+i \bR} dj \int_\bC d^2x \Psi(j;x,\bar x|h)^* \Psi(j;x,\bar x|h') &= 2 \pi i \delta_h \,,
\end{align}
expressing the decomposition of $L^2(H_3^+)$ into principal series representations of $SL(2,\bC)$. 

Our cyclic vector is thus 
\begin{equation}
	|h \rangle = \Gamma(-2j-1) \Psi(j;x,\bar x|h)  |dx|^{-2j} 
\end{equation}
and has orthogonality properties with measure $(2j+1) \sinh \pi (2j+1) dj$.

If we insert a background vortex defect of charge $k$, we could look at the average of something like $z_1^{k-a} z_2^a e^{- |z_1|^2 - |z_2|^2}$, giving the collection of vectors
\begin{equation}
	|k,a \rangle= \frac{\Gamma(k+1-i t)}{\pi^2} x^a (1 + |x|^2)^{1-i t-k}dx^{\frac12-\frac{k+i t}{2}} d\bar x^{\frac12+\frac{k- i t}{2}} 
\end{equation}
in the principal series representation labelled by $t$ and $k$. 

Analogously, we can define 
\begin{equation}
	\Psi(j,k,a;x,\bar x|h) = \frac{1}{\pi} x^a \left((x\,\,1)\, h\, {\bar x\choose 1}\right)^{2j-k}
\end{equation}
These also have orthonormality properties as elements of $L^2(H_3^+,\bC^{k+1})$, where $\bC^{k+1}$ is interpreted as 
a bundle associated to the principal $SU(2)$ bundle on $H_3^+$, expressing the decomposition of the Hilbert space into 
principal series representations of $SL(2,\bC)$.

\section{Coulomb branch traces} \label{sec:gaucoulomb}
As mentioned in the introduction, the protected sphere correlation functions of Coulomb branch operators in a Lagrangian gauge theory of gauge group $G$ are already 
written upon localization as an expectation value in some auxiliary functional Hilbert space. The auxiliary space takes the form  $L^2\left[\frac{\fh_\bR \times \Lambda_w}{W_G}\right]$, where $\fh_\bR$ is a real version of the Cartan Lie algebra of $G$, $\Lambda_w$ the magnetic weight lattice and the measure is a $\sinh$-Vandermonde. The Coulomb branch algebra and its opposite act as Weyl-invariant meromorphic difference operators \cite{Dedushenko:2017avn,Dedushenko:2018icp} and an explicit spherical vector is defined as a product of $\Gamma$ functions:
\begin{equation}
	|1 \rangle = \delta_{b,0} \prod_{(w,w_f)} \Gamma\left(\frac12 - i \sigma\cdot w- i m \cdot w_f \right)
\end{equation}
where the product runs over the gauge and flavour weights for all the matter hypermultiplets. Here $\sigma \in \fh_\bR$ and $b \in  \Lambda_w$. 
The symbol $m$ denotes the mass parameters.

A clarification is in order here. Hypermultiplets transform in an symplectic representation $M$ of $G$. If we can write $M = T^* N$ for a representation $N$, the above product runs over the weights of $N$. Different choices of $N$ are related by unitary transformations: 
the ratio of the $\Gamma$ functions which enter in the spherical vector for two different choices is an overall $\sigma$-dependent phase. 
The different presentations of the algebra $A$ are conjugated into each other by conjugation by that ratio.  
If no $N$ exists, we can still split the weights of $M$ into opposite pairs and include only one of the two in the product. 

The actual Hilbert space $\cH$ should then be identified with the closure of the image of $|1 \rangle$ under the action of the Coulomb branch algebra. 
Notice that the difference operators shift $\sigma$ in the imaginary direction by integral amounts and thus have domains which consist of functions which admit 
holomorphic extensions to certain regions of the complexified $\fh_\bC\times \Lambda_w$. 

Geometrically, the Coulomb branch of a Lagrangian gauge theory is a fibration over a middle-dimensional base space, parameterized by 
Poisson-commuting expectation values of gauge-invariant polynomials ${\cal P}(\varphi)$ of an adjoint-valued complex scalar field $\varphi$ of weight $2$,
with $\rho_\cl(\varphi) = \varphi$. This integrability structure persists after quantization, with commuting generators $H_{\cal P}$ in $A$ which form the same commutative subring. 

In the localization description, $H_{\cal P}$ act as multiplication operators by 
\begin{equation}
	{\cal P}(\sigma- i b/2)
\end{equation}
In particular, the Coulomb branch description of the abstract Hilbert space $\cH$ diagonalizes action of the the normal operators $H_{\cal P}$ and predicts the quantization of $b$

If the matter in the gauge theory is not free, the fibration will still be available but the base may not be middle-dimensional. If we denote the matter theory as $T$
and use mirror conventions where the $G$ gauge group acts on the Coulomb branch of $T$, then we can denote the result as $T \rtimes G$ and discuss the Higgs branch of 
that. Then the statement is that $A[T \rtimes G]$ contains an $A[T]_G$ sub-algebra, i.e. the combination of $A[T]_\varphi$ and of the ${\cal P}(\varphi)$ operators. 

As we diagonalize the $H_{\cal P}$ operators, $\cH[T \rtimes G]$ will decompose into eigenspaces which carry an action of 
$A_{\sigma- i b/2} \times A^\op_{\sigma+ i b/2}$. This is closely related to describing $A[T \rtimes G]$ as an $A[T]_G-A[T]_G$
bi-module and localizing it to $A[T]_{\sigma- i b/2}-A[T]_{\sigma+ i b/2}$ bi-modules. 

The remaining operators in $T \rtimes G$ are ``monopole operators''. If $T$ is the mirror of free hyper-multiplets, the algebra of monopole operators is defined by the BFN construction
\cite{Nakajima:2015txa,Braverman:2016wma,Braverman:2022zei}. It is constructed as a space of global sections of a certain ring object in the ``equivariant derived Satake'' category for $G$.
For general $T$, the expectation if that one can attach such a ring object to any $T$ with Coulomb branch symmetry $G$ \cite{Braverman:2017ofm}, whose global sections give the Coulomb branch algebra of $T \rtimes G$. Physically, the basic idea is to promote $T$ to a boundary condition for a  four-dimensional ${\cal N}=4$ SYM with gauge group $G$ and look at spaces of local operators living at the end of a 't Hooft line, keeping track of what happens when 
't Hooft lines are fused. Taking the space of global sections corresponds to ending the 't Hooft lines on a second, pure Neumann, boundary condition to get 3d local operators. 

The interplay between sphere correlation functions and the BFN construction is mostly unexplored. Localization calculations involve the ``Abelianized'' description of the Coulomb branch, 
which approximates the Coulomb branch by gauging only the Cartan subgroup of $G$. As mentioned in the introduction, a mathematical version of Abelianization can be likely formulated with the help of the Iwahori Coulomb brach construction \cite{Kamnitzer:2022zkv}.

The Abelianization formulae should thus have as a natural ambient space $\cH[T \rtimes H]$, decomposed into a direct sum/integral of $A[T]_{\sigma- i b/2}-A[T]_{\sigma+ i b/2}$ bimodules
we denote as $\cH_{b,\sigma}[T]$. The monopole operators will lie in a ``localized'' version of $A[T\rtimes H]$. 

\subsection{A bad example: free $U(1)$ gauge theory}
The Coulomb branch of free $U(1)$ gauge theory is $T^*\bC^*$, which is not a cone. A natural quantization of the Coulomb branch 
is $L^2(\bR \times S^1)$. In a convenient parameterization, denoting as $y$ and $\phi$ coordinates on $\bR \times S^1$, holomorphic functions on the base are exponentials such as $v^n \equiv e^{n(y/2 + i \phi)}$ and $\tilde v^n \equiv e^{n(-y/2 + i \phi)}$. 
Fibre coordinates are $u = \partial_y - \frac{i}{2} \partial_\phi$ and $\wt u = \partial_y + \frac{i}{2} \partial_\phi$. 

A natural domain of definition for these operators is the space of $C^\infty$ functions which decay faster than any exponential 
along $\bR$. On that domain, $v$ is adjoint to $\wt v^{-1}$ and $u$ to $-\wt u$.

The distribution $|1\rangle = \delta(y)$ interpolates the action of holomorphic and anti-holomorphic differentials, 
but it is not normalizable. Acting on it with holomorphic functions, we get a collection of distributions of the form 
$\partial^n \delta(y) \, e^{i m \phi}$ which gives a non-normalizable image of $A$ in $L^2(\bR \times S^1)$. 

For general $U(1)$ gauge theories, the Coulomb branch operators will be built from $u$ and $v$.
The resulting embedding of $A$ and $A^\op$ will be such that the intertwining states are normalizable.

In preparation, we can Fourier transform $L^2(\bR \times S^1) \to L^2(\bR \times \bZ)$ with coordinates $\sigma$ and $b$ 
and accordingly multiplication operators $u = i\sigma +\frac12 b$ and $\wt u = i \sigma-\frac12 b$ and $v$, $\tilde v$  
acting as 
\begin{equation}
	v \,\psi(\sigma,b) = \psi\left(\sigma + \frac{i}{2}, b-1\right) \qquad \qquad \tilde v \,\psi(\sigma,b) = \psi\left(\sigma - \frac{i}{2}, b-1\right)
\end{equation}
so that 
\begin{equation}
	u v = v (u+1) \qquad \qquad \wt u \wt v = \wt v(\wt u -1) 
\end{equation}
and $|1\rangle = \delta_{b,0}$. 

The domain of $A$ is now functions which can be extended holomorphically in the complex $\sigma$ plane and decay faster than polynomials in $b$ and $\sigma$. 
Notice that in order to prove that $\wt v^n$ and $v^{-n}$ are adjoint to each other in an inner product, the $\sigma$ integral needs to be shifted by 
$i n/2$. This is possible precisely because the functions in the domain can be analytically continued to the complex $\sigma$ plane. 

\subsection{Mirror of free hypermultiplet}
The Coulomb branch of a $U(1)$ gauge theory coupled to a single hypermultiplet is $\bC^2$, i.e. $\bC^2 = (\bC^2)^! \rtimes U(1)$. 

The Coulomb branch algebra $A$ is embedded in the above-defined shift algebra as 
\begin{equation}
	X = v \qquad \qquad P = \left(\frac12 +u\right)v^{-1}
\end{equation}
and 
\begin{equation}
	\wt X  = \left(\frac12 + \wt u \right) \wt v \qquad \qquad  \wt P  = \wt v^{-1}
\end{equation}

These operators are defined on a larger functional space than in the case of $T^* \bC^*$. The action of $X^n$ and $\wt P^n$ still require the wavefunctions to admit holomorphic analytic 
continuation to the upper half plane. On the other hand, $P$ could act reasonably on a function whose analytic continuation to the lower half plane has a simple pole at 
$\sigma = -\frac{i}{2} + \frac{i}{2} b$. Similarly, the action of $\wt X$ allows a simple pole at $\sigma = -\frac{i}{2} - \frac{i}{2} b$ in the lower half plane. More generally, the algebra acts 
well on the space of functions of $\sigma$ which are holomorphic in the upper half plane and are allowed simple poles at $\sigma =  -\frac{i}{2} (|b|+n)$ for integer $n>0$. 

The spherical vector 
\begin{equation}
	|1 \rangle =  \delta_{b,0} \Gamma\left(\frac12 - i \sigma \right) 
\end{equation}
intertwines the two actions and defines the trace on $A$. It lies in the correct functional space and so do 
\begin{equation}
	X^n |1 \rangle =  \delta_{b,n} \Gamma\left(\frac12 +\frac{n}{2}- i \sigma \right) \qquad \qquad  \wt P^n |1 \rangle =  \delta_{b,-n} \Gamma\left(\frac12 +\frac{n}{2}- i \sigma \right)\, ,
\end{equation}
as well as states obtained from these by multiplication by polynomials in $\sigma$. These states are all normalizable and are a dense basis for the desired Hilbert space $\cH$. 

The cancellations of poles by the prefactors in $P$ and $\wt X$ are instrumental for the $X^\dagger = \wt P$ and $P^\dagger = - \wt X$ relations to hold in 
inner products, as they require a shift of the $\sigma$ integration contour which would fail if the contour is deformed across an uncancelled pole.

We can make contact between this presentation of the trace and Hilbert space and the mirror trace on $L^2(\bC)$ described in Section \ref{sec:weyl}. Recall the complete set of distributional eigenvectors of the moment map $\mu$ for the $\bC^*$ action on $L^2(\bC)$:
\begin{equation}
	|\sigma;n \rangle \equiv z^{-\frac12 + i \sigma+ \frac12 b}\bar z^{-\frac12 + i \sigma- \frac12 b} \, ,
\end{equation}
with eigenvalue $i \sigma+ \frac12 b$ for $\mu$. These allow one to map $L^2(\bC)$ to $L^2(\bR \times \bZ)$.
Conversely, the distribution 
\begin{equation}
	|x, \bar x \rangle = x^{-\frac12 - i \sigma+ \frac12 b}\bar x^{-\frac12 - i \sigma- \frac12 b}
\end{equation}
in $L^2(\bR \times \bZ)$ represents a delta-function distribution in $L^2(\bC)$. It is easy to see that these states intertwine the action of the Weyl algebras on the two 
descriptions of $\cH$. 

Finally, the alternative embedding
\begin{equation}
	X = v \left(\frac12 +u\right) \qquad \qquad P = v^{-1}
\end{equation}
and 
\begin{equation}
	\wt X  =  \wt v \qquad \qquad  \wt P  = \wt v^{-1} \left(\frac12 + \wt u \right) 
\end{equation}
arises in conventions where the hypermultiplet has charge $-1$. It can be related to the previous description by conjugation by the phase
\begin{equation}
	\frac{\Gamma\left(\frac12 - i \sigma \right)}{\Gamma\left(\frac12 + i \sigma \right)}\, .
\end{equation}

\subsection{Mirror of $T[SU(2)]$}
We can analyze $T[SU(2)] = (\bC^4)^! \rtimes U(1)$ in the same fashion. There are different equivalent ways to present the algebra and spherical vector, depending on our charge conventions. 

One of the presentations emerge naturally if we diagonalize the action of the $H$ generator 
in the principal series representation of $SL(2,\bC)$, as we discussed in Section \ref{sec:pri}.  

The eigenvectors become 
\begin{equation}
	|\sigma;n \rangle \equiv x^{-\frac12+\frac{im}{2} + i \sigma+ \frac12 b}(x^*)^{-\frac12+\frac{im}{2} + i \sigma- \frac12 b} |dx|^{1 - i m}\, ,
\end{equation}
with eigenvalues $i \sigma+ \frac12 b$ for $H$ and $i \sigma- \frac12 b$ for $\wt H$. 

Now the cyclic vector becomes 
\begin{align}
	|1 \rangle &= \int |dx|^2 x^{-\frac12 -  \frac{im}{2} - i \sigma- \frac12 b}(x^*)^{-\frac12 -  \frac{im}{2}  - i \sigma+ \frac12 b} \Gamma(1-i m) (1+ |x|^2)^{-1+i m} = \cr &= \delta_{b,0} \Gamma\left(\frac12-\frac{im}{2}- i \sigma \right)\Gamma\left(\frac12-\frac{im}{2}+ i \sigma \right)\, .
\end{align}
Each factor comes from one hypermultiplet, in a convention where they have opposite gauge charge and the same flavour charge.
The $E$ and $F$ operators act as difference operators
\begin{equation}
	E =\left(\frac12+\frac{i m}{2}+ u\right) v \qquad \qquad F = \left(\frac12+\frac{i m}{2} -u\right)v^{-1}
\end{equation}
and analogously 
\begin{equation}
	\wt E =\left(\frac12+\frac{i m}{2}- \wt u\right) \wt v^{-1} \qquad \qquad \wt F= \left(\frac12+\frac{i m}{2} +\wt u\right) \wt v \, .
\end{equation}

Notice that the unitary transformation defined by conjugation by 
\begin{equation}
	\frac{\Gamma(\frac12-\frac{i m}{2} + i \sigma)}{\Gamma(\frac12+\frac{i m}{2}- i \sigma)} \, ,
\end{equation}
maps the cyclic vector to 
\begin{equation}
	|1 \rangle =\delta_{b,0} \Gamma(-\frac12-\frac{i m}{2} - i \sigma)\Gamma(\frac12+\frac{i m}{2}- i \sigma)
\end{equation}
and 
\begin{equation}
	E = v \qquad \qquad F  = \left(\frac12+\frac{i m}{2} -u\right)\left(\frac12-\frac{i m}{2}+ u\right) v^{-1}
\end{equation}
and analogously 
\begin{equation}
	\wt E =\left(\frac12+\frac{i m}{2}- \wt u\right) \left(\frac12-\frac{i m}{2}+\wt u\right) \wt v^{-1} \qquad \qquad \wt F= \wt v \, ,
\end{equation}
which make manifest the $m \to -m$ symmetry of the system. These formulae arises from the general 
definition of Coulomb branch algebra in a convention where both hypermultiplets have the same gauge charge and opposite flavour charge. 

\subsection{SQED}
Consider now an $U(1)$ gauge theory with $n$ flavours. The Abelianized expressions are 
\begin{equation}
	E = v \qquad \qquad F  =\left[ \prod_a \left(\frac12 + i s_a -u\right) \right]v^{-1}
\end{equation}
and analogously 
\begin{equation}
	\wt E =\left[ \prod_a \left(\frac12 + i s_a - \wt u\right) \right]\wt v^{-1} \qquad \qquad \wt F= \wt v \, .
\end{equation}

They generate the quantized $A_{n-1}$ singularity, as 
\begin{equation}
	F E = \prod_a \left(\frac12 + i s_a -u\right) \qquad \qquad E F = \prod_a \left(-\frac12 + i s_a -u\right) \, .
\end{equation}

The cyclic vector is 
\begin{equation}
	|1 \rangle =\delta_{b,0} \prod_a \Gamma\left(\frac12 + i s_a - i \sigma\right)
\end{equation}
and the image of more general elements in $A$ is
\begin{equation}
	E^n |1 \rangle =\delta_{b,n} \prod_a \Gamma\left(\frac12 + n + i s_a - i \sigma\right) \qquad \qquad \wt F^n |1 \rangle =\delta_{b,-n} \prod_a \Gamma\left(\frac12 + n + i s_a - i \sigma\right) \, ,
\end{equation}
as well as the product of these with polynomials in $b$ and $\sigma$.

A novel phenomenon occurs in these examples. Consider vectors 
\begin{equation}
	|1;k \rangle =e^{\pi k \sigma} \delta_{b,0} \prod_a \Gamma\left(\frac12 + i s_a - i \sigma \right) \, .
\end{equation}
As long as $n>2k$, these are still normalizable and define alternative positive twisted traces on $A$, with the same twist. Linear combinations of these with $k$ which differs by an even amount 
would have the same property. We thus get a whole collection of positive twisted traces beyond the standard one \cite{Etingof:2020fls}. 

\subsection{Wilson line defect insertions}
The new traces can also be understood in terms of the characters $\Tr e^{2 \pi m \mu}$. The character is analytic on a strip of width $n/2$ in the complexified $m$ 
plane, and the extra twist is not visible if $m$ has integral imaginary values. This has a physical interpretation in terms of Wilson line defects linking the 
great circle where local operators are placed \cite{Etingof:2020fls,Guerrini:2023rdw}.  

This generalizes to other Coulomb branch algebras as the insertion of $W_R \equiv Tr_R e^{2 \pi (\sigma- \frac{i}{2} b)} = Tr_R e^{2 \pi (\sigma+ \frac{i}{2} b)}$ 
in the localization integral. This will only converge if the representation has sufficiently small weight. If the representation is a square, so that the
trace in $W_R$ is a positive-definite function of $\sigma$ on the integration cycle, this gives alternative positive traces on the algebra. 

%\section{Intermission: positive traces and short star products}
%As a preparation to the study of boundary condition, it is useful to review the relationship between positive traces and star products. 
%
%The commutative algebra $A_\cl$ is graded by the Cartan generator in $SU(2)_H$ or equivalently by twice the scaling dimension, with the symplectic form having weight $2$. 
%The quantized algebra $A$, instead, is a priori only filtered. This is related to the fact that quantization does not immediately provide an identification between elements of $A$ and 
%of $A_\cl$. It provides, though, an identification of $A_\cl$ as the associated graded to $A$. 
%
%The Gram-Schmidt procedure applied to the non-degenerate pairing $\Tr \rho(a) b$ offers a way to sharpen this identification to an actual invertible map $A_\cl \to A$. We simply map $\alpha \in A_\cl$ to an element $a_\alpha \in A$ which has $\alpha$ as the highest weight component and which is orthogonal to all elements of $A$ of lower weight. 
%
%Such a quantization map defines a star product on $A_\cl$ by
%\begin{equation}
%	a_{\alpha * \beta} \equiv a_\alpha a_\beta  \, .
%\end{equation}
%This has remarkable properties \cite{...}.
%
%An important aspect of this construction, which will generalize below, is that the quantum $\rho$ map 
%could actually be recovered from the classical $\rho$ map together with the Gram-Schmidt procedure.
%Indeed, we can define $a_\alpha$ as being orthogonal to lower weight $b$'s in $\Tr b \, a$ and 
%$\rho(a_\alpha)$ as having leading term $\rho(\alpha)$ and being orthogonal to all $b$'s in $\Tr a\, b$.

\section{S-duality and spherical varieties} \label{sec:spher}

In previous sections, we have encountered the idea that a (conformal) theory $T$ with non-anomalous Higgs branch symmetry $G$ can be promoted to a (conformal) boundary 
condition for four-dimensional ${\cal N}=4$ Yang-Mills theory with gauge group $G$. If an anomaly is present, the promotion is still possible, but 
the four-dimensional action has to be modified by a topological term which changes the properties of the theory under dualities. We will assume that is not the case unless otherwise specified. 

The result is not the most general possible conformal boundary condition \cite{Gaiotto:2008sa}. General conformal boundary conditions are expected to be labelled by an $\fsl_2$ embedding $\rho$ into $G$, characterizing a Nahm pole, an unbroken subgroup $H$ of the centralizer $L_\rho$ of $\rho$ in $G$ and a 3d ${\cal N}=4$ conformal theory $T$ with Higgs branch symmetry $H$. 
We can denote that as $[\rho,H,T]$. 

It is sometimes possible to replace $T$ in the construction with a theory which is not conformal to build a boundary condition which still flows to a conformal one, possibly with different $\rho$ and $H$. For example, if we denote with $T^*G_\bC$ the sigma model with the same target and consider the left $G$ action, $[0,G,T^*G_\bC]$ flows to the Dirichlet boundary condition $[0,0,0]$. 
In other words, the labelling $[\rho,H,T]$ is expected to be unique if $T$ is restricted to be conformal, but not otherwise. 

Two boundary conditions for the same group $G$ can be combined via a slab geometry $[0,L] \times \bR^3$ to produce a 3d theory, which we could denote as $[\rho_1,H_1,T_1]\times_G [\rho_2,H_2,T_2]$. This theory is not conformal to start with, as $L$ provides a scale, but has unbroken R-symmetry and could flow to a conformal field theory. The theory can be described effectively as 
\begin{equation}
	T_1 \times_{H_1} B_{\rho_1, \rho_2} \times_{H_2} T_2
\end{equation}
Here $B_{\rho_1, \rho_2}[G]$ is a sigma model with target the space $B_{\rho_1, \rho_2}[G]$ of solutions of $G$ Nahm equations on a segment with Nahm poles $\rho_1$ and $\rho_2$ at the endpoints. It can also be described as the DS reduction of $T^* G_{\bC}$ by $\rho_1$ on the left and by $\rho_2$ on the right. It has symmetry given by $L_{\rho_1}$ acting from the left and $L_{\rho_2}$ acting from the right. The $H_1$ and $H_2$ subgroups act as subgroups of that. 

It is also possible to consider interfaces between four-dimensional gauge theories. These are essentially boundary conditions for $G_\ell \times G_r$ of the form 
$[\rho_\ell \times \rho_r,H,T]$, with $\rho_\ell$ and $\rho_r$ embeddings in the respective groups. Interfaces can be composed. If we have interfaces 
$[\rho_1 \times \rho_2,H_{12},T_{12}]$ between $G_1$ and $G_2$ and $[\rho'_2 \times \rho_3,H_{23},T_{23}]$ between $G_2$ and $G_3$,
the composition will have $\rho_1 \times \rho_3$ embedding, unbroken subgroup $H_{12} \times H_{23}$ and theory 
\begin{equation}
	T_{12} \times B_{\rho_2, \rho'_2}[G_2] \times T_{23}
\end{equation}
The composition may flow in the IR to a conformal interface with different-looking data.

Upon S-duality, the boundary condition will map to a boundary condition $[\rho, H, T]^\vee$ for a four-dimensional gauge theory with gauge group $G^\vee$, labelled by S-dual data we sometimes denote as $[\rho^\vee, H^\vee, T^\vee]$. This is an involution. The main claim relating S-duality and mirror symmetry \cite{Gaiotto:2008ak} is that 
\begin{equation}
	[\rho_1,H_1,T_1]\times_G [\rho_2,H_2,T_2] = \left([\rho_1,H_1,T_1]^\vee \times_{G^\vee} [\rho_2,H_2,T_2]^\vee\right)^! 
\end{equation}
Analogous statements hold for interfaces composition, which commutes with S-duality.

An important ingredient of the above mirror symmetry statement is that it allows one to match on the two sides the Higgs (Coulomb) branch local operators 
which arise from the four-dimensional theory, which are labelled by $G$-invariant polynomials on $\fg$ and can be identified with $G^\vee$-invariant polynomials on $\fg^\vee$.
On one side of the duality, they appear as canonical holomorphic functions on  $B_{\rho_1, \rho_2}[G]$. On the other side, they can be expressed in terms of the commuting 
Hamiltonians built from the $H_1^\vee$ and $H_2^\vee$ gauge fields. This match can be extended to local operators built from Wilson lines of the 4d gauge theory on one side of the duality, and 't Hooft lines on the other side. 

These statements can be combined with ``elementary'' S-duality statements to predict the mirror or S-dual of a large variety of conformal boundary conditions or interface. 
Intermediate steps, though, involve non-conformal theories and some care may be needed with the RG flow to the desired conformal systems.

The elementary S-duality statements belong roughly to two classes. The first class of S-duality statements involves auxiliary theories which are {\it defined} through S-duality. In particular, one defines 
$T[G]$ as the conformal limit of $[0,0,0] \times_G [0,0,0]^\vee$, which is a theory with a $G$ Higgs branch symmetry and $G^\vee$ Coulomb branch symmetry. 
The Higgs and Coulomb branches are expected to be regular nilpotent orbits for $G$ and $G^\vee$ respectively. 

More generally, $T_{\rho^\vee}[G]$ is defined as the conformal limit of $[0,0,0] \times_G [\rho^\vee,0,0]^\vee$. The Coulomb branch is expected to be the Slodowy 
slice $S_{\rho^\vee}$ to $\rho^\vee$.  One can also define a mirror $T^{\rho}[G]$ as the conformal limit of $[\rho,0,0] \times_G [0,0,0]^\vee$
and $T_{\rho^\vee}^{\rho}[G]$ as the conformal limit of $[\rho,0,0] \times_G [\rho^\vee,0,0]^\vee$. The latter theory may be empty for sufficiently ``large'' $\rho$ and $\rho^\vee$. 

The importance of $T[G]$ follows from the observation that a boundary condition can be reconstructed from the slab with $[0,0,0]$: 
\begin{equation}
	[0,0,0] \times_G [\rho, H, T] = B_{0,\rho} \times_H T
\end{equation}
and $[0,G, B_{0,\rho} \times_H T]$ is expected to flow back to $[\rho, H, T]$. In particular, $[0,0,0]^\vee$ can be recovered from $[0,G,T[G]]$.

Applying S-duality and some optimism about RG flows, we find that $[\rho, H, T]^\vee$ can be recovered from 
\begin{equation}
	\left( [0,0,0]^\vee \times_{G} [\rho, H, T] \right)^! =  \left([0,G,T[G]]\times_{G} [\rho, H, T] \right)^!=  \left(T^{\rho}[G] \times_H T \right)^!
\end{equation}

A second class of S-duality statements is inherited from string theory constructions. It typically involves classical groups and 
$T$'s consisting of free hypermultiplets transforming in representations with up to two (anti)fundamental indices. Notice that the S-dual description of a boundary condition 
$[\rho,H,T]$ where $T$ is a theory of free hypermultiplets typically involves strongly-coupled SCFTs. 

Examples where both sides involve hypermultiplets only should be rare. A boundary condition $[\rho, H,T]$ could be dubbed ``spherical'' if $T$ is a theory of free hypermultiplets and $\left(T^{\rho}[G] \times_H T \right)^!$
flows to a theory of free hypermultiplets. In particular, the manifold $(S_\rho \times T)/\!\!/G$ should consist of a single point. 
This condition can be matched to the mathematical definition of hyper-spherical variety $X = B_{0,\rho}[G]\times_H T$ \cite{spher}, which has recently played an interesting role in the Geometric Langlands program. The natural conjecture is that the BFN Coulomb branch of $T^{\rho}[G] \times_H T$ will be the dual hyper-spherical variety.

We expect that sphere quantization applied to these examples should recover classical results on the harmonic analysis of spherical varieties. We have already anticipated many examples for small gauge groups in Section \ref{sec:ex}.

\subsection{Example: $G \times G$ acting on $T^*G_{\bC}$}
We begin with the simplest general example: the diagonal interface $[0,G,0]$ for $G \times G$, with $G$ being the diagonal subgroup. 
This is self-dual, i.e. $[0,G,0]^\vee =[0,G^\vee,0]$ for $G^\vee \times G^\vee$. 

We can associate to this interface to the non-conformal theory given by the sigma model into $T^*G_\bC$, i.e. the moduli space of solutions of Nahm equations on a segment, with Dirichlet boundary conditions. We then predict the mirror symmetry statement $T^*G_\bC=  (T[G] \times T[G]) \rtimes G^\vee$. 

Notice that the $G$-invariant polynomials of left and right moment maps (``Casimirs'') in $T^*G_\bC$ coincide. Part of the S-duality dictionary 
is the identification of these with the Hamiltonians in the $G^\vee$ BFN construction. The moment maps for the $G_\bC$ action commute with the Casimirs 
and are identified with the moment maps in $T[G]$. 

The Hilbert space which is naturally associated to $T^*G_{\bC}$ is $L^2[G_\bC]$ with the invariant measure. This space carries a natural unitary action of $G_\bC \times G_\bC$ by 
left and right multiplication. The spectrum of Casimir operators for this action and the decomposition of $L^2[G_\bC]$ into eigenspaces are classic results in representation theory. 

The mirror description presents the Hilbert space as a direct integral over all of the principal series representations $\cH_{b,t}$ of $G_\bC$, realized as the 
closure of the HC modules for $G_\bC$ associated to background vortex lines. The Casimirs of the $G_\bC$ actions coincide and 
are expressed as $G^\vee$ invariant polynomials of $b/2 + i \sigma \in \fh^\vee$. This reproduces the expected structure of $L^2[G_\bC]$!

As $T^*G_{\bC}$ is non-conformal, we do not have a normalizable spherical vector. 

%\subsection{Example: $U(1)$ acting on $T^*\bC$}
%The boundary condition $[0,U(1), \bC^2]$ for $G=U(1)$ is expected to be self-dual. We have already effectively examined this example 
%when discussing $\bC^2$ and $(\bC^2)^! = \bC^2/U(1)$. 
%
%We have the Hilbert space $L^2(\bC)$ with an unitary $GL(1,\bC)$ action and 1-dimensional eigenspaces 
%generated by 
%\begin{equation}
%	|\sigma;b \rangle \equiv z^{-\frac12 + i \sigma+ \frac12 b}\bar z^{-\frac12 + i \sigma- \frac12 b}
%\end{equation}
%with the cyclic vector being 
%\begin{equation}
%	\delta_{b,0} \Gamma\left(\frac12 - i \sigma \right)
%\end{equation}
%in that basis. The single factor of $\Gamma$ arises from the single factor of $\bC^2$ with $U(1)$ weight $1$ in $\bC^2/U(1)$.
%
%\subsection{Example: $U(2)$ acting on $T^*\bC^2$}
%Here we can also piggyback on a previous calculation for the gauge theory description of $T[SU(2)]$,
%where we decomposed $L^2(\bC^2)$ into eigenspaces of the diagonal $GL(1,\bC)$ moment map with eigenvalues $j$, which turned out to 
%coincide with the principal series representations of $SL(2,\bC)$. Equivalently, $L^2(\bC^2)$ decomposes into principal series 
%representations of $GL(2,\bC)$ labelled by a two-dimensional $( j + j, j - j= 0)$ weight. 
%
%This agrees with the dual boundary condition, which is $[0,GL(1), 0]$, with $GL(1) \in GL(2)$ embedded as the upper left entry of the $2 \times 2$ matrix. 
%

\subsection{Example: $U(n) \times U(n)$ acting on $T^*\bC^{n^2}$}
This interface $[0,U(n) \times U(n), T^*\bC^{n^2}]$ is expected to be dual to a diagonal interface with an extra fundamental hypermultiplet, i.e. 
$[0,U(n),T^*\bC^n]$. 

We can study the Higgs branch of both and recover either information about the $GL(n,\bC) \times GL(n,\bC)$ action on $L^2[\bC^{n^2}]$ 
or the $GL(n,\bC) \times GL(n,\bC)$ action on $L^2[GL(n,\bC) \times \bC^{n}]$.

In the former case, $L^2[\bC^{n^2}] \sim L^2[GL(n,\bC)]$, which is the direct integral of principal series 
representations. The extra information we gain is the decomposition of the spherical vector in $L^2[\bC^{n^2}]$ as a direct integral,
with a normalization coefficient which is the product of $\Gamma$ functions associated to twisted hypermultiplets valued in $T^*\bC^{n}$. 

In the latter case, we learn about the decomposition of the tensor product of $L^2(\bC^n)$ and a principal series 
representation into a direct integral of other principal series.

\subsection{Example: $U(n) \times U(n-1)$ acting on $T^*\bC^{n(n-1)}$}
This is an S-duality between $[0,U(n) \times U(n-1), T^*\bC^{n(n-1)}]$ and $[0,U(n-1),0]$ for $U(n) \times U(n-1)$ gauge theories, 
with the $U(n-1)$ embedded diagonally as a block of $U(n)$. 

This example contains information about the decomposition of $L^2[\bC^{n(n-1)}]$ in principal series representations 
or about the decomposition of a principal series representation of $GL(n)$ into $GL(n-1)$ representations. 
\subsection{Example: $U(n) \times U(m)$ acting on $T^*\bC^{n\, m}$}
This is an S-duality between $[0,U(n) \times U(n-k), T^*\bC^{n(n-k)}]$ and $[(k),U(n-k),0]$ for $U(n) \times U(n-k)$ gauge theories, 
with the $U(n-k)$ embedded diagonally as a block of $U(n)$ commuting with an $\mathfrak{sl}_2$ principal embedding in a $k \times k$ block. 

This example contains information about the decomposition of $L^2[\bC^{n(n-k)}]$ in principal series representations 
or about the decomposition of the DS reduction of a principal series representation of $GL(n)$ into $GL(n-k)$ representations.
 
\subsection{Example: $SO(n) \times Sp(2m)$ acting on $T^*\bC^{n\, m}$}
These examples are completely analogous to the unitary ones, but involve Hilbert spaces which cannot be presented as $L^2(\cdots)$ 
without breaking some symmetries. 

\section{Boundary conditions} \label{sec:real}
Three-dimensional ${\cal N}=4$ SQFTs admit half-BPS boundary conditions preserving a $(2,2)$ two-dimensional super-symmetry algebra. 
Such boundary conditions will typicaly preserve the Cartan subgroup of the $SU(2)_C\times SU(2)_H$ R-symmetry group, if present. It plays the same role as the $U(1)_V \times U(1)_A$ R-symmetry group of a 2d $(2,2)$ SQFT. In particular, three-dimensional mirror symmetry 
for he bulk theory can act like a two-dimensional mirror symmetry on boundary conditions, permuting the two $U(1)$'s. 

These boundary conditions are equipped with an chiral ring of boundary local operators $M_\cl$, a module for $A_\cl$ which is akin to holomorphic functions on a complex Lagrangian sub-manifold $\cL[B]$ of $\cM[T]$. The boundary conditions are compatible with the sort of $\Omega$-deformation employed to define non-commutative algebras of local operators and thus $M_\cl$ can be promoted either to a left $A$-module
$M$ or to a right $A$-module $\wt M$. The same boundary condition also gives sub-manifolds and modules for the 
Coulomb branch. 

Some aspects of boundary conditions and associated Lagrangian submanifolds and modules were described in \cite{Bullimore:2016nji}. 
Two-categorical aspects of twisted $(2,2)$ boundary conditions have recently been analyzed in \cite{2022arXiv221006548G,Gammage:2022caj}. 

We now consider half-BPS, super-conformal $(2,2)$ boundary conditions for 3d ${\cal N}=4$ SCFTs.  For super-conformal boundary conditions the Lagrangian $\cL[B]$ must be a cone invariant under the Cartan of $SU(2)_H$. In particular, boundary local operators will be graded, but the grading could be fractional compared to the grading of $A_\cl$.

Boundary conditions which are half-BPS but not conformal can also be employed to define hemisphere partition functions \cite{Dedushenko:2018tgx}. 
Positivity conditions, though, are not generically available away from the conformal case. We refer to Appendix \ref{app:posi} for a discussion of positivity of 
sphere correlators for $(2,2)$ theories, which is generalized in the present setting. 

Given an half-BPS boundary condition $B$ for $T$, we can define protected hemisphere correlation functions. 
The great circle along which bulk operators are placed 
intersects the boundary at two points, where one can insert special ``boundary local operators''.
The hemisphere correlation function can thus be written as $\left(\wt m, a_1 \cdots \,a_n \,m\right)$, 
where we denoted the boundary local operators  at the two intersection points as $\wt m$ and $m$ 
and we included a sequence of bulk local operators $a_i$ along the half-circle. 

Boundary local operators have the structure of a module for the bulk operators, compatible with the correlation functions. More precisely, we get a left $A$-module $M$ and a right $A$-module $\wt M$. The $A$-module structure of $M$ is implied in the above notation. Furthermore, we have 
\begin{equation}
\left(\wt m\, a_1 \cdots a_k, a_{k+1} \cdots a_n \,m\right) = \left(\wt m, a_1 \cdots \,a_n \,m\right)
\end{equation}
In other words, the hemisphere correlation functions factor through a linear map $M_R \otimes_A M_L \to \bC$ as the bulk local operators can be collided with either intersection points. 

The pairing $(\wt m,a m)$ defines implicitly a collection of distributional states 
$|\wt m;m\rangle$ in $\cH$ as
\begin{equation}
(\wt m,a m) =\langle a |m; \tilde m\rangle
\end{equation}
which gives a map of $A$-bimodules from $\wt M \otimes M$ to the space of distributions in $\cH$. This is true for any half-BPS boundary condition. 
In some concrete localization formulae, the hemisphere correlation functions does actually take the form of such an inner product.

The spaces $M$ and $\wt M$ are filtered by the grading in a manner compatible to $A$, with associated graded isomorphic to $M_\cl$ or to the conjugate $\bar M_\cl$ respectively. 
At the level of the associated graded, the $A_\cl$ actions on $M_\cl$ or $\bar M_\cl$ are intertwined by $\rho_\cl$. 

We can apply a Gram-Schmidt procedure \cite{Gerchkovitz:2016gxx} to the pairing $(\cdot, \cdot)$ to refine this identification to identifications of $M$ with $M_\cl$ and $\wt M$ with $\bar M_\cl$. To each $u \in M_\cl$ we associate $m_u$ with highest weight $u$ and orthogonal to all lower weight elements in $\wt M$. Analogously, to each 
$u \in M_\cl$ in we associate $\wt m_u$ with highest weight $\bar u$ and orthogonal to all lower weight elements in $\wt M$.

The outcome of the Gram-Schmidt procedure is an anti-linear map $\rho$ from $M$ to $\wt M$, sending $m_u$ to $\wt m_u$. 
The crucial claim is that 
\begin{equation}
	\langle m|m' \rangle \equiv \left( \rho(m), m' \right)
\end{equation}
defines a positive-definite Hermitian inner product on $M$. 

We thus define a new Hilbert space $\cH_B$ as the $L^2$ completion of $M$
under this inner product. The Hilbert space $\cH_B$ inherits an action of $A$ from $M$. It also inherits an action of $A^\op$ from $\wt M$:
\begin{equation}
	m \wt a \equiv \rho^{-1}\left(\rho(m) \, \rho(a) \right)
\end{equation}
i.e. 
\begin{equation}
	\langle m \wt a|m' \rangle = \langle m| \rho(a) m' \rangle
\end{equation}
so that we are essentially defining $\wt a$ as the adjoint to $\rho(a)$. 

At first sight, this construction is rather mysterious. There is not obvious relation between the $A$ and $A^\op$ actions beyond this 
and they do not typically commute. 

In order to understand it better, we jump to an apparently different problem: the quantization of the submanifold $\cL_\bR[B]$ of $\cM$ defined by a hyper-k\"ahler rotation of $\cL[B]$, on which the real part of the complex symplectic form $\Omega$ on $\cM$ vanishes, but the imaginary part defines a real symplectic form. 

The manifold $\cL_\bR[B]$ inherits functions which are the restriction of holomorphic functions in $\cM$. 
The Poisson bracket defined on this restrictions by the imaginary part of $\Omega$ coincides with the restriction of the complex Poisson bracket. 
The same is true for the restriction of anti-holomorphic functions with the opposite sign, but the Poisson bracket between the restriction of holomorphic and anti-holomorphic functions depends on the geometry of $\cL_\bR[B]$. 
A quantization of $\cL_\bR[B]$ could thus naturally include actions of $A$ and $A^\op$, but no obvious relation between them besides 
 $\wt a = \rho(a)^\dagger$, just as what happens for $\cH_B$. 
 
 Furthermore, $\cL_\bR[B]$ inherits a complex structure from $\cL[B]$ via the hyper-k\"ahler rotation. Geometric quantization of 
$\cL_\bR[B]$ would naturally produce an Hilbert space which includes a basis of holomorphic functions, i.e. elements of $M_\cl$,
possibly with a complicated inner product. This also reminds us of the properties of $\cH_B$. 

We thus conjecture that $\cH_B$, equipped with the $A$ and $A^\op$ actions, is a natural quantization of $\cL_\bR[B]$.
We will test this general conjecture in a non-trivial example below. 

There is an important situation where the reality properties of the $A$ action should be more manifest and the conjecture easier to understand. 
This is the setup conventionally employed in brane quantization \cite{Gukov:2008ve}, where $\cM$ is equipped with a complex conjugation $\tau_\cl$ which fixes the imaginary part of $\Omega$ and fixes $\cM_\bR \equiv \cL_\bR$. We will also assume that $\tau_\cl$ changes the sign of the third K\"ahler form and anti-commute with the generator we use for hyper-k\"ahler rotations. Notice that the conjugation $\tau_\cl$ should map $\cL_\bR$ to its image under a full hyper-k\"ahler rotation, i.e. the involution $\tau_\cl \rho_\cl$ should fix $\cL$. 

If we identify $A^\op$ with the quantization of anti-holomorphic functions, 
we may have a quantum version of the complex conjugation: an anti-linear map $\tau: A \to A^\op$. 
We can thus look for a quantization such that $a = \tau(a)^\dagger$ i.e. $\wt a = \tau(\rho(a))$. 

In the context of hemisphere partition functions, we will take $\tau \rho$ to be a symmetry of the full theory $T$ and pick a boundary condition $B$ 
fixed by $\tau \rho$, so that $\cM_\bR = \cL_\bR[B]$. This should insure that $\tau$ persists as a symmetry of the hemisphere correlation functions and thus the $A$ and $A^\op$ actions on $\cH_B$ are indeed intertwined by $\tau \rho$. We will first proceed through some examples and then try to derive some general principles. 

A surprising pattern which emerges in certain example is that all formulae become ``real'' version of the formulae we obtain over the complex numbers from bulk calculations. Such formulae, involving integrals of Gaussian functions, may potentially make sense over other fields as well. This happens in surprising ways in the context of the Langlands program \cite{Gaiotto:2021tzd} and may give a link to representation theory over other fields. See also \cite{spher}.

We can elaborate a bit on the relationship between boundary conditions and symmetries. As we discussed in previous sections, if the theory $T$ has a symmetry $F$, then the anti-linear involution $\rho$ acts as $\mu \to - \mu$ on the moment maps for the $F_\bC$ action, fixing the Lie algebra for the compact group $F$. 

A natural choice of anti-linear involution $\tau$ will fix some other real form $F_\bR$ of $F$. Intuitively, $\tau$ 
will fix the non-compact $\mu$ generators and invert the compact ones. Correspondingly, $\tau \rho$ should fix 
the generators for the complexification $K_\bC$ of the maximal compact subgroup $K \in F_\bR$. 
This would naturally occur if we consider boundary conditions which break $F$ to $K$. 

In particular, the identity state $|1\rangle \in M$ will be spherical in the sense of representation theory: it will be annihilated by 
the moment maps for  the $K_\bC$ action. We could thus call $M$ a spherical Harish-Chandra module, generalizing the representation-theoretic notion. Adding bulk line defects ending at the boundary, such as vortex line defects, may give more general Harish-Chandra modules.

If $T$ has a $F^!$ Coulomb branch symmetry, a boundary condition may break it to some subgroup $K^!$ as well. 
This will accordingly restrict the allowable ranges for the FI parameters $t$, though vortex line defects may shift that by integral amounts. 

In concrete examples, there seem to be often a tension between preserving both $F$ and $F^!$. For example, 
consider the theory $T[G]$ so that $A_t$ is a central quotient of $U(\fg)$. A boundary condition preserving $G$ 
will contain a spherical vector fixed by the whole $G$. This will fix the value of all the Casimirs and thus of $t$, indicating that 
$F^! = G^\vee$ must be completely broken. It would be nice to explore this point further. 

\subsection{The $P=0$ boundary condition for $\bC^2$}
Consider the distributional boundary state 
\begin{equation}
	|1;1\rangle \equiv 1
\end{equation}
in $L^2(\bC)$. It is annihilated by $P$ and $\tilde X$ and generates a collection of distributional states 
\begin{equation}
	|z^m; z^n \rangle \equiv   \bar z^n z^m =(-\tilde P)^n \, X^m |1;1\rangle 
\end{equation}
It represents the $P_\cl=0$ free boundary condition in the theory of a free hypermultiplet. Recall that $\tilde X_\cl$ represents classically the action of $\bar P_\cl$, hence 
the $P_\cl=0=\tilde X_\cl$ classical boundary condition. 

Up to an overall normalization, we have correlation functions 
\begin{equation}
	(1;1) = \int d^2 z e^{-|z|^2} =1
\end{equation}
and more generally 
\begin{equation}
	(\bar z^n ;z^m) = \int d^2 z e^{-|z|^2} \bar z^n z^m  =n! \delta_{n,m}
\end{equation}

Here $M_\cl$ consists of polynomials in $z$. The deformation quantization $M$ is just the 
usual Verma module $\bC[z]$ for the Weyl algebra, with $P$ acting as $\partial_z$
and $X$ as multiplication by $z$. 

We thus find $\rho(z^n) = \bar z^n$, making the inner product positive-definite and recovering the standard unitary structure on the Verma module. 
The adjoints of the Weyl algebra generators under the inner product are $X^\dagger = P$ and $P^\dagger = X$. 
These relations can be written as 
\begin{equation}
	P - X^\dagger = 0 \qquad \qquad P^\dagger - X = 0 \, .
\end{equation}
Their semiclassical limit is the submanifold $P_\cl = \bar X_\cl$ in $\bC^2$, which is the hyper-K\"ahler rotation of $P_\cl=0$,
as expected. 

This is a situation where the reality conditions can be encoded in 
\begin{align}
	\tau(P) &= X \qquad \qquad \tau(X) = P \cr
	\tau\rho(P) &= -P \qquad \qquad \tau\rho(X) = X
\end{align}
The locus $P=0$ is fixed by $\tau \rho$, as expected. 

\subsection{The $P=X$ boundary condition for $\bC^2$ and $L^2(\bR)$.}
We will now rotate the classical boundary condition to $P_\cl + X_\cl=0$. This leads to
 the modified distributional boundary state 
\begin{equation}
	|1;1\rangle \equiv e^{-\frac12 z^2 + \frac12 \bar z^2}
\end{equation}
in $L^2(\bC)$. It is annihilated by $P+X$ and $\tilde X- \tilde P$ and generates a collection of distributional states 
\begin{equation}
	|z^m; \bar z^n \rangle \equiv   \bar z^n z^m e^{-\frac12 z^2 + \frac12 \bar z^2}
\end{equation}
Up to an overall normalization, we have correlation functions 
\begin{equation}
	(\bar z^n ;z^m) = \int d^2 z e^{-|z|^2-\frac12 z^2 + \frac12 \bar z^2} \bar z^n z^m 
\end{equation}

We can analytically continue the Gaussian integration contour to make $z$ and $\bar z$ independent:
\begin{equation}
	(\bar z^n ;z^m) = \int dx dp \,e^{-xp-\frac12 x^2 + \frac12 p^2} p^n x^m =  \int_{-\infty}^\infty dx \,x^m e^{-\frac12 x^2} (-\partial_x)^n e^{- \frac12 x^2} 
\end{equation}
This allows us to identify the Hilbert space as $L^2(\bR)$ with a standard Weyl algebra action and the inner products $(1;a)$ as expectation values of $a$ on the {\it real spherical vector}
\begin{equation}
	|1;\bR \rangle \equiv e^{- \frac12 x^2} 
\end{equation}
Notice that this Gaussian function is the natural real analogue of the complex $e^{- |z|^2}$.

In particular, in this representation we have Hermiticity conditions $X=x=X^\dagger$, $P=\partial_x = -P^\dagger$. 
The underlying reason for this particularly simple result is that the classical limit of these relations is the hyper-K\"ahler rotation of $X_\cl + P_\cl=0$: 
\begin{align}
	\tau(P) &= -P \qquad \qquad \tau(X) = X \cr
	\tau\rho(P) &= -X \qquad \qquad \tau\rho(X) = -P \,.
\end{align}

The Hermitean structure is best expressed in terms of 
\begin{equation}
	|f_m(z); g_n(\bar z) \rangle \equiv   (\tilde P + \tilde X)^n (P-X)^m e^{-\frac12 z^2 + \frac12 \bar z^2} = (\partial_z - z)^m (- \partial_{\bar z} -\bar z)^n e^{-\frac12 z^2 + \frac12 \bar z^2}
\end{equation}
leading to an Harmonic oscillator orthogonal inner product
\begin{equation}
	(g_n(\bar z) ;f_m(z)) = \int dx \,\left[(\partial_{x} -x)^n e^{-\frac12 x^2}\right]\left[(\partial_x - x)^m  e^{-\frac12 x^2}\right] 
\end{equation}

The $P=X$ boundary condition obviously breaks the $U(1)$ global symmetry of the theory. The $\mu = XP + \frac12$ moment map still acts on $L^2(\bR)$, with $\mu^\dagger = - \mu$. It generates dilatations. We can compute the twisted norm
\begin{equation}
	\langle1;\bR| e^{2 \pi m \mu} |1;\bR \rangle = \int_{-\infty}^\infty dx e^{\pi m} e^{- \frac12 (1+e^{4 \pi m}) x^2} = \frac{1}{\sqrt{2\cosh 2 \pi m}}
\end{equation}

As the boundary condition is a symplectic rotation of $P=0$, it preserves a different $U(1)$ subgroup of the $SU(2)$ global symmetry of the hypermultiplet theory. 
Accordingly, $L^2(\bR)$ carries a unitary action of the $\mathfrak{sl}(2,\bR)$ real conformal algebra, which has a one-dimensional compact subgroup given by the Harmonic oscillator Hamiltonian. 

\subsection{Mirror to $P=X$}
Mirror symmetry for $(2,2)$ boundary conditions is only partly understood. In this situation, we can make a simple guess based on the sphere correlation functions. 

A simple shortcut is to Mellin transform the boundary state:
\begin{equation}
	|1,1\rangle = \int |dz|^2 z^{-\frac12 - i \sigma+ \frac12 b}\bar z^{-\frac12 - i \sigma- \frac12 b} e^{-\frac12 z^2 + \frac12 \bar z^2} = (1+(-1)^b) 2^{- i s}\frac{\Gamma\left(\frac{b+1}{4}- \frac{i \sigma}{2}\right)}{\Gamma\left(\frac{b+3}{4}+ \frac{i \sigma}{2}\right)}
\end{equation}
This is identical to the contribution to the Coulomb hemisphere partition function of a 2d free chiral multiplet $\varphi$ of gauge charge $\frac12$ and specific R-charge assignment. Such a fractional charge is possible if the $U(1)$ gauge group is extended by a $\bZ_2$ at the boundary. The above constraint on the magnetic charge $b$ to be even is compatible with that. The 
R-charge assignment seems compatible with a $P^! \varphi^2$ super-potential coupling at the boundary. 

A full discussion of this system goes beyond the scope of this paper. It is interesting to note, though, that the pairing $(1;1)$ can be written as 
\begin{equation}
	\int d\sigma 2^{- i s}\frac{\Gamma\left(\frac{1}{4}- \frac{i \sigma}{2}\right)}{\Gamma\left(\frac{3}{4}+ \frac{i \sigma}{2}\right)} \Gamma\left(\frac12+i \sigma \right) = \int d\sigma 2^{- i s}\Gamma\left(\frac{1}{4}- \frac{i \sigma}{2}\right)\Gamma\left(\frac{1}{4}+ \frac{i \sigma}{2}\right)
\end{equation}
matching the real Mellin transform of the real Gaussian
\begin{equation}
	\int_{-\infty}^\infty dx |x|^{-\frac12 - i t} e^{- x^2/2} = 2^{\frac34+\frac{it}{2}}\Gamma\left(\frac14- \frac{i t}{2}\right)
\end{equation}
in $L^2(\bR)$. 

The hemisphere partition function of the mirror theory thus encodes the spectral decomposition of the real spherical vector. 
This is the sort of representation-theoretic statement we'd like to find in other examples, giving the spectral decomposition of functional representations of 
real forms of flavour symmetry groups. 

\subsection{Boundary conditions for $T[SU(2)]$ and representation theory of $\mathfrak{sl}(2,\bR)$.}
Consider now the Hilbert space $L^2(\bR^2)$, associated to an $X_i + P_i=0$ boundary condition for $T=\bC^2$. 
We can consider two different versions of the $U(1)$ symmetry acting on $\bC^2$:
\begin{enumerate}
	\item An $SO(2)$ symmetry rotating $X^i$. This commutes with an $\mathfrak{sl}(2,\bR)$ action generated by $X^2$, $X\cdot P+1$ and $P^2$. The $X^2 + P^2$ generator is compact. 
	\item An $\bR^+$ scaling symmetry acting on both $X^i$ in the same manner. This commutes with an $\mathfrak{sl}(2,\bR)$ action generated by the traceless part of $X_i P_j$. The $SO(2)$ 
	rotation subgroup is compact. 
\end{enumerate}

In both cases we can decompose $L^2(\bR^2)$ under the action of the Abelian generator to obtain unitary representations of $\mathfrak{sl}(2,\bR)$:
\begin{enumerate}
	\item If we expand $L^2(\bR^2)$ as a Fock space built on the spherical vector $|1,\bR\rangle \equiv e^{- x^2/2}$, with $X_i-P_i$ creation generators,  the charge $b$ part of
	the Fock space gives a Verma module for  $\mathfrak{sl}(2,\bR)$ with highest weight $|n|+1$, which we identify with discrete series representations ${\cal D}_{b}$. 
	\item The dilatation operator has a continuum spectrum. In polar coordinates, we can write a basis of distributional eigenfunctions of the form $r^{-1+i t}e^{i n \theta}$, thus identifying each 
	eigenspace with $L^2(S^1)$. Better, we can split the Hilbert space into even and odd eigenspaces for the $X_i \to -X_i$ symmetry and identify each 
	eigenspace with a space of twisted half-densities on $\bR P^1$. We obtain the two principal series representations ${\cal P}_{t,\pm}$ for $\mathfrak{sl}(2,\bR)$.
\end{enumerate}

We can match these constructions to two boundary conditions for $T[SU(2)]$, in a gauge theory description. We impose the 
$X_\cl + P_\cl=0$ boundary condition on both hypermultiplets, but we have two different choice of boundary condition for the gauge fields:
\begin{enumerate}
	\item If the gauge group acts as $SO(2)$, the matter boundary condition preserved the bulk gauge symmetry. It is natural to impose Neumann boundary conditions for the gauge fields. In localization formulae, the boundary states are projected to eigenspaces for the action of the compact gauge group. We add a bulk Wilson line defect of charge $b$ to shift the charge in the projection to get boundary local operators building up ${\cal D}_b$.
	\item If the gauge group rotates both $X^i$ with charge $1$, it is broken by the matter boundary conditions. This forces us to impose Dirichlet boundary conditions for the gauge fields. In localization formulae, the boundary states are averaged over the action of the complexified gauge group. The FI parameter $t$ controls which direct sum of two principal series representations ${\cal P}_{t,\pm}$ emerge from the boundary local operators.
\end{enumerate}

It is also instructive to discuss the classical geometry of these boundary conditions. 
\begin{enumerate}
	\item In the first case, the reality condition on the Higgs branch is set by the anti-linear map $\tau(E) = E$, $\tau(F) = F$ and $\tau(H) = -H$. The moment map for the gauge action 
	satisfies $\tau(\mu) = \mu$ as long as the FI parameter is set to zero and thus $\tau \rho(\mu) =-\mu$. Indeed, $\mu$ vanishes. The complex symplectic quotient reduces 
	to a GIT quotient of the complex Lagrangian $X_i +P_i=0$. 
	\item In the second case, the reality condition on the Higgs branch is set by the anti-linear map $\tau(E) = -E$, $\tau(F) = -F$ and $\tau(H) = -H$. The moment map for the gauge action 
	satisfies $\tau(\mu) = -\mu$. The complex symplectic quotient smears the the complex Lagrangian $X_i +P_i=0$ along the orbit of the complexified gauge group and imposes the moment map constraint $\mu=0$. 
\end{enumerate}

We expect both boundary conditions to descend to super-conformal boundary conditions for $T[SU(2)]$. It is natural to ask which subgroup of the $SU(2)$ Coulomb branch symmetry group 
will they preserve. This affects how could they appear in Coulomb gauging situations. 
\begin{enumerate}
	\item Neumann b.c break the $U(1)$ part of the Coulomb branch global symmetry visible in the gauge theory description. A reasonable assumption is that the symmetry will not be restored in the IR. 
	\item Dirichlet b.c. preserve the $U(1)$ part of the Coulomb branch global symmetry visible in the gauge theory description. The $t \to -t$ symmetry of principal series representations suggests that the boundary condition will preserve the full $SU(2)$ Coulomb branch symmetry group in the IR. 
\end{enumerate}

There are a few more interesting possibilities which we have not considered above. For example, we can take boundary conditions $P_i=0$ for the matter fields and  
impose Neumann boundary conditions. Now the Hilbert space before gauging is a Fock space generated by the $X_i$. There is no state with charge $0$ under the gauge $U(1)$. 
States with charge $n$ are organized in irreducible unitary representations of $SU(2)$ of dimension $n$. 

We have thus found gauge theoretic constructions of a large variety of unitary representations of real forms of $SL(2,\bC)$.

\subsection{More about unitary principal series for $SL(2,\bR)$.}
Consider the space $\cH_\bR= L^2(\bR)$, with an $SL(2,\bR)$ action generated by 
\begin{equation}
	E = \partial_x \qquad \qquad H = x \partial_x- j \qquad \qquad  F = - x^2 \partial_x+ 2j x = \partial_{x'}
\end{equation}
We take the usual range $j = - \frac12 + i \frac{t}{2}$. Clearly, $H^\dagger= -H$, $E^\dagger = -E$ and $F^\dagger = -F$
Thus $\tau$ acts as $-1$ on the generators. Accordingly, 
\begin{equation}
	\tau \rho(E) = F \qquad \qquad \tau \rho(H)= H \qquad \qquad \tau \rho(F) = E
\end{equation}

Consider now the ``real'' spherical vector 
\begin{equation}
	|1;\bR \rangle \simeq (1+ x^2)^{j} 
\end{equation}
This satisfies 
\begin{equation}
	E |1;\bR \rangle = F |1;\bR \rangle 
\end{equation}

The generator $(E-F)/2$ rotates the $H\pm i(E+F)/2$ generators with charge $\pm 1$. 
Acting with these on the spherical vector we get a series of vectors of the form 
\begin{equation}
	|n\rangle \equiv (i+ x)^{j+n} (-i+ x)^{j-n} 
\end{equation}
for integer $n$. These are a dense basis, related to Fourier modes by mapping $\bR$ to $S^1$ as $\theta = \frac{i+x}{i-x}$, i.e. identifying $\cH_\bR$ as a space of twisted half-densities on the circle. 
The $E-F$ is generator can be exponentiated. 

This unitary representation of $SL(2,\bR)$ fits nicely in our story. The Hilbert space $\cH_\bR$ naturally quantizes a real locus $T^* S^1$ in $T^* \bC P^1$, 
or better the nilpotent $SL(2,\bR)$ orbit inside the nilpotent $SL(2,\bC)$ orbit. If we write $H = \mu_3$, $E = \mu_1 + i \mu_2$, $F= \mu_1 - i \mu_2$, 
then $\rho$ acts as usual as $-1$ on the $\mu_a$, and $\tau$ as $\mu_3 \to - \mu_3$, $\mu_1 \to - \mu_1$, $\mu_2 \to \mu_2$, as appropriate to 
fix $SL(2,\bR)\subset SL(2,\bC)$. 

Working in $\cH$, we could write distributional states
\begin{equation}
	|n;n'\rangle \equiv  (i+ x)^{j+n} (-i+ x)^{j-n} (i+ \bar x)^{j+n'} (-i+ \bar x)^{j-n'} 
\end{equation}
It takes a bit of work/analytic continuation to show that the inner product of these with the complex spherical vector
$\langle 1;\bC|$ reproduces the inner product in $\cH_\bR$. Essentially, $\bar x \to \bar x^{-1}$ maps $\langle 1;\bC|$
to the integral kernel for $t \to -t$, much as what happened with the Fourier kernel in the free hypermultiplet analysis. 

A natural way to produce the correct correlation functions from the gauge theory picture is to do a real version of the averaging procedure: average $e^{-\frac12 x_1^2 -\frac12 x_2^2}$ over real scale transformations 
This reproduces the desired answer:
\begin{equation}
	|1;\bR\rangle = \int_{-\infty}^\infty e^{2 \pi i \beta t}   e^{2 \pi \beta s}  e^{-\frac12 e^{4 \pi \beta s} (x_1^2 +x_2^2)} = \Gamma\left(\frac12-i \frac{t}{2} \right) 2^{i \frac{t}{2}} (x_1^2 +x_2^2)^{-\frac12+i \frac{t}{2}}
\end{equation}
with an interesting normalization factor. It gives a norm
\begin{equation}
	\langle1;\bR |1;\bR\rangle =  \frac{1}{\cosh \frac{\pi t}{2}}
\end{equation}
%We would also like to compute
%\begin{equation}
%	\langle1;\bR |e^{2 \pi m H} |1;\bR\rangle 
%\end{equation}
%If we go back to the pre-averaged expression, that gives a Fourier transform of $\frac{1}{2\sqrt{\cosh 2 \pi \beta \cosh 2 \pi (\beta + m)}}$, which is a complicated expression. 

If we express $\langle1;\bR |a |1;\bR\rangle$ as 
\begin{equation}
	\langle1;\bR |a |1;\bR\rangle = \int_{-\infty}^\infty e^{2 \pi i \beta t}  \langle1;\bR |a e^{2 \pi \beta \mu} |1;\bR\rangle
\end{equation}
we can rewrite that as 
\begin{equation}
	\langle1;\bC |a |1;1\rangle = \int_{-\infty}^\infty e^{2 \pi i \beta t}  \langle1;\bC |a e^{2 \pi \beta \mu}  |1;1\rangle 
\end{equation}
reproducing the recipe to compute partition functions with Dirichlet boundary conditions for the gauge fields: average over the non-compact directions of the group. 
%Notice that the boundary state $|1;1\rangle$ is not $U(1)$ invariant, but is contracted with a $U(1)$-invariant state $\langle1;\bC |a e^{2 \pi \beta \mu}$. 

A simple consequence of this discussion is that $L^2(\bR^2)$ has been decomposed into spherical principal series representations for $SL(2,\bR)$, by Mellin transform.

\subsection{The $P=X^n$ boundary condition for $\bC^2$}
Next, we can look at an example of non-standard Hermiticity condition which is not associated to some $\tau$. 
We work backwards from an exponential boundary state, defining 
\begin{equation}
	\left(f,g\right) = \int |dz|^2 e^{- |z|^2+ \frac{\lambda}{n+1} z^{n+1} -  \frac{\bar \lambda}{n+1} \bar z^{n+1}} f(\bar z) g(z)
\end{equation}
More generally, we define 
\begin{equation}
	\left(f, P^b X^a  g\right) \equiv \left(p^b f(p), x^a  g(x)\right) 
\end{equation} 
We then derive $P g = \partial_x g(x) +\lambda  x^n g(x)$ and $f X = \partial_p f(p) - \bar \lambda p^n f(p)$. 

In other words, we take $M$ to be $\bC[x]$ with a deformed left action of the Weyl algebra and $\wt M$ to be $\bC[p]$ 
with a right action of the Weyl algebra deformed in the opposite way. 

The Gram-Schmidt procedure will give two collections of monic orthogonal polynomials $\wt m_{x^k} \equiv p_k(p)$ and $m_{x^k} \equiv q_k(x)$ such that 
\begin{equation}
	\int |dz|^2 e^{- |z|^2+ \frac{\lambda}{n+1} z^{n+1} -  \frac{\bar \lambda}{n+1} \bar z^{n+1}} p_k(\bar z) q_m(z) = h_k \delta_{k,m}
\end{equation}
Thus the expected positivity of inner products is expressed by the condition $h_k >0$. 

As an aside, notice that such orthogonal polynomials would appear in the solution of the complex matrix model 
\begin{equation}
	\int |dM|^{2 N^2} e^{- \tr_{N\times N} \bar M \, M+ \frac{\lambda}{n+1} \tr_{N\times N} M^{n+1} -  \frac{\bar \lambda}{n+1} \tr_{N\times N} \bar M^{n+1}} = \prod_{k=0}^{N-1} h_k 
\end{equation}
so $h_k>0$ is equivalent to positivity of the above matrix integral for all $N$.

A rotation $z \to e^{\frac{2\pi i}{n+1}}$ shows that only powers which differ by multiples of $(n+1)$ can mix. 
Hence $p_k$ contains powers $k - m(n+1)$. 
Furthermore $p_k$ and $q_k$ should be related by the $\bZ_2$ symmetry flipping the sign of the terms 
which differ by odd multiples of $(n+1)$ from the leading one. 

Because the action of $X$ on $g(x)$ maps in the pairing to the action of $(\partial_p f(p) - \bar \lambda p^n)$ on $f(p)$, we must have
\begin{equation}
	(X q_m)(z) = z q_m(z) = q_{m+1}(z) - \bar \lambda \frac{h_m}{h_{m-n}} q_{m-n}(z) 
\end{equation}
and analogously 
\begin{equation}
	(P q_m)(z) =(\partial_z + \lambda z^n) q_m(z) = \lambda q_{m+n}(z) + \frac{h_m}{h_{m-1}} q_{m-1}(z) 
\end{equation}
We thus have
\begin{equation}
	(\partial_z +z^n) (z q_m(z)) = \lambda q_{m+n+1}(z) + \frac{h_{m+1}}{h_{m}} q_{m}(z)  - \bar \lambda \lambda \frac{h_m}{h_{m-n}} q_{m}(z) + \bar \lambda \frac{h_m}{h_{m-n-1}} q_{m-n-1}(z) 
\end{equation}
to be compared to 
\begin{equation}
	z (\partial_z +z^n) q_m(z) = \lambda q_{m+n+1}(z) -\lambda \bar \lambda \frac{h_{m+n}}{h_{m}} q_{m}(z) + \frac{h_m}{h_{m-1}} q_{m}(z) + \bar \lambda \frac{h_m}{h_{m-n-1}} q_{m-n-1}(z) 
\end{equation}
We learn that
\begin{equation}
	\frac{h_{m+1}}{h_{m}} - \frac{h_m}{h_{m-1}} + \lambda \bar \lambda \left[\frac{h_{m+n}}{h_{m}} - \frac{h_m}{h_{m-n}} \right]  =1
\end{equation}

We also have dual relations 
\begin{equation}
	(p_m P)(p) = p p_m(z) = p_{m+1}(p) + \lambda \frac{h_m}{h_{m-n}} p_{m-n}(p) 
\end{equation}
and
\begin{equation}
	(p_m X)(p) =  (\partial_p - \bar \lambda p^n) p_m(z) =- \bar \lambda p_{m+n}(p) + \frac{h_m}{h_{m-1}} p_{m-1}(p)  
\end{equation}

If we define our hermitean conjugation as $q_k^\dagger = p_k$, then by definition
\begin{equation}
	X^\dagger q_m = (p_m X)^\dagger = - \lambda q_{m+n}+ \frac{h_m}{h_{m-1}} q_{m-1}
\end{equation}
and 
\begin{equation}
	P^\dagger q_m = (p_m P)^\dagger = q_{m+1} + \bar \lambda \frac{h_m}{h_{m-n}} q_{m-n}
\end{equation}

We are now ready to determine the relation between $X$, $P$ and $X^\dagger$ and $P^\dagger$. 
Inspired by the form of hyper-K\"ahler rotations, we can define 
\begin{equation}
	\hat P q_m \equiv (P-X^\dagger)/2 q_m =  \lambda q_{m+n}
\end{equation}
and 
\begin{equation}
	\hat X q_m \equiv (X+P^\dagger)/2 q_m =  q_{m+1} 
\end{equation}
We thus recover a quantization of the rotated Lagrangian $\hat P = \lambda \hat X^n$.

Analogously, we have conjugate relations
\begin{equation}
	\hat P' q_m \equiv (P+X^\dagger)/2 q_m =  \frac{h_m}{h_{m-1}} q_{m-1}
\end{equation}
and 
\begin{equation}
	\hat X' q_m \equiv (X-P^\dagger)/2 q_m = - \bar \lambda \frac{h_m}{h_{m-n}} q_{m-n} = -\bar \lambda (\hat P')^n q_m
\end{equation}

Hence the output of sphere quantization is a quantization of the sub-manifold 
\begin{equation}
	\frac{p - \bar x}{2} = \lambda \left(\frac{x - \bar p}{2}  \right)^n
\end{equation}
in $\bC^2$ using the pull-back of the symplectic form $dx d\bar x + dp d \bar p$. 

\subsection{Multiple hypermultiplets and positivity}
As discussed in detail in the Appendices, the positivity of the inner product defined by two-sphere partition functions for a 
2d Landau-Ginzburg with quasi-homogeneous superpotential results in rather non-trivial integral identities. 

The Gram-Schmidt procedure defines a finite collection of dual semi-orthogonal polynomials $\wt p_{\bar \alpha}$ and $p_\alpha$ labelled by an element $\alpha$ in the Jacobi ring of $W$, such that the inner product 
\begin{equation}
	\langle \alpha, \beta \rangle \equiv \int_{\bC^r} d^r\varphi \,d^r\bar \varphi \, \wt p_{\bar \alpha}(\bar \varphi) p_\beta(\varphi) e^{W(\varphi) - \wt W(\bar \varphi)} 
\end{equation}
vanishes unless $\alpha$ and $\beta$ have the same degree. The physics prediction is that this basis exists and the inner product for each degree is positive-definite. 
In the Appendix \ref{app:GS} we provide a proof based on ${\cal N}=4$ supersymmetric quantum mechanics.

We can formulate an analogous statement for LG boundary conditions of a 3d theory of free hypermultiplets, generalizing the previous example. The relevant pairing is now 
\begin{equation}
	\langle \alpha, \beta \rangle \equiv \int_{\bC^r \times \bC^k } d^r\varphi \,d^r\bar \varphi \,d^k X \,d^k \bar X \, \wt p_{\bar \alpha}(\bar \varphi,\bar X) p_\beta(\varphi,X) e^{-|X|^2 +W(\varphi,X) - \wt W(\bar \varphi, \bar X)} 
\end{equation}
Here $W$ is quasi-homogeneous of degree $2$ and $X_a$ are given degrees between $0$ and $2$. The $\alpha$ label denotes polynomials in $X$ and $\varphi$ modulo multiples of 
$\partial_\varphi W$. 

It would be interesting to give a proof of positive-definiteness analogous to the 2d case. 
\section{Open Questions and Future Directions}
Many results in super-symmetric quantum field theory can be efficiently recast in the language of twisted SQFTs. For example, the algebras $A[T]$ and $A[T^!]$ appear as algebras of local operators in the A- and B- topological twists of $T$. Twisted SQFTs are increasingly well-understood mathematically. It is not unreasonable to expect that such results may be proven in a rigorous mathematical fashion within that framework, without reference to the original physical theories.  

The constructions of this paper do not quite fit in such a vision, as they employ in a critical way the 
unitarity/reflection positivity properties of the underlying physical theory. A simple but important question arising from our work is how to identify a minimal framework which would allow a rigorous mathematical analysis of our setup through the methods of QFT. 

The relative simplicity of many of our results lead naturally to a somewhat opposite question:
do we really need QFT ideas to derive them or can they be proven systematically without any reference to 3d ${\cal N}=4$ SCFTs? Based on our analysis: 
\begin{enumerate}
	\item The existence and positivity of the traces on Higgs and Coulomb branches
	for Lagrangian gauge theories should admit simple mathematical proofs along the lines discussed in the main text. 
	\item It is difficult to make general predictions about non-Lagrangian theories without getting into the specifics of how they are defined. Manipulation of theories such as gauging or triggering RG flows through Higgs- or Coulomb vevs lead to simple rules to define new traces. In many cases we can sketch an argument for positivity, but not for the RG flow triggered by Coulomb vevs: the new algebra and trace is defined by quotienting the kernel of a non-positive trace. This claim may lead to non-trivial mathematical conjectures.
	\item Positivity of hemi-sphere correlation functions appears mathematically non-trivial, even for the case of a trivial 3d theory. We devoted our appendices to a discussion of simple non-trivial examples of that phenomenon. The associated quantization of real forms of the Higgs- and Coulomb branches appears mathematically rich and may greatly extend the theory of unitary representations of real semisimple groups $G_\bR$. 
\end{enumerate}

We can also list some more specific physical puzzles which appeared in the paper:
\begin{enumerate}
	\item The precise identification of the Higgs branch for $T_{\rho^\vee}[G]$ is still a somewhat open problem. The $\fg$ moment maps are expected to 
	map it to a specific (union of) nilpotent orbits \cite{Chacaltana:2012zy} in $\fg$, but it is not completely obvious that the map should be injective. The mathematical work 
	of \cite{2018arXiv181007625L} appears relevant to the question and to the sphere quantization of $T_{\rho^\vee}$. Furthermore, the analytic continuation of the trace for $T[G]$
	to special values of the FI parameters of the theory may also provide a route to identify the (quantization of) the Higgs branch algebra. Quantization of the Higgs branch of $T_{\rho^\vee}[G]$
	will produce unitary representations of $G_\bC$. 
	\item It would be interesting to formulate novel predictions for S-duality of half-BPS boundary conditions by using the theory of dual (hyper)spherical varieties \cite{spher}. 
	\item The representation-theoretic applications of hemisphere quantization are obstructed by our limited knowledge on the interplay of S-duality and 3d mirror symmetry and 
	$(2,2)$ boundary conditions. Brane configurations may potentially be used to predict such dualities. See \cite{Chung:2016pgt,Gaiotto:2019jvo,Okazaki:2020lfy}. 
	\item In particular, it would be interesting to employ S-duality considerations to predict the existence of interesting half-BPS boundary conditions for $T[G]$ or 
	 $T_{\rho^\vee}[G]$, which could be associated to various unitary representations of real forms $G_\bR$ of $G_\bC$. 
\end{enumerate}
In general, it appears that representation theory of complex and real Lie algebras and, more generally, Higgs and Coulomb branch algebras may be potentially employed to 
conjecture a variety of novel physical dualities. Such dualities, in turn, may be fed into the machinery of symplectic or Langland duality to generate new mathematical predictions,
e.g. following \cite{spher}.

Finally, we should mention that our construction can be extended to four-dimensional ${\cal N}=2$ theories
via correlation functions on $S^1 \times S^3$. This will be the subject of a companion paper.

\acknowledgments{We would like to thank David Kazhdan, Pavel Etingof, Alexander Braverman and Alexander Goncharov for useful conversations. This research was supported in part by a grant from the Krembil Foundation. DG is supported by the NSERC Discovery Grant program and by the Perimeter Institute for Theoretical Physics. Research at Perimeter Institute is supported in part by the Government of Canada through the Department of Innovation, Science and Economic Development Canada and by the Province of Ontario through the Ministry of Colleges and Universities.}
\appendix

\section{Two-sphere partition functions and positivity} \label{app:posi}
The supersymmetric two-sphere partition function and correlation functions are defined for any 2d $(2,2)$ SQFT which has at least one $U(1)_A$ R-symmetry acting on both chiral and anti-chiral supercharges \cite{Doroud:2012xw,Benini:2012ui}. We use conventions where this is the axial R-symmetry. The case of a vector symmetry is obtained from 2d mirror symmetry. 

The construction involves a careful deformation of the theory by certain relevant operators in the stress-tensor multiplet. The outcome has an $SU(2|1)_A$ ``super-isometry'' group, 
including $SU(2)$ rotations of the two-sphere. Localization calculations involve a SUSY generator which squares to a combination of $U(1)_A$ and of the rotation generator 
fixing the poles of the sphere. As a result, one can compute correlation functions of local operators at the North and South pole \cite{Gomis:2012wy,Ishtiaque:2017trm} which are annihilated by that SUSY generator, giving a pairing between two vector spaces $R$ and $\wt R$ defined as an equivariant cohomology of the space of local operators. 

The space $R$ is identified with the space of rotation-equivariant local operators in a topological B-twist \cite{Vafa:1990mu,Witten:1988xj} of the theory, i.e. the space of local operators at the origin in an $\Omega$ deformation of the physical theory \cite{Yagi:2014toa}. In particular, it is a deformation of the space $R^\cl$ of local operators in the B-twist. The space $\wt R$ is a deformation of the complex conjugate $\bar R^\cl$, but the sign of the $\Omega$ deformation is opposite to the one for $R$ and thus there is no natural identification between $\wt R$ and $\bar R$.

The sphere partition function gives a linear pairing 
\begin{equation}
(\cdot, \cdot): \wt R \otimes R \to \bC
\end{equation}
which is usually computable by finite-dimensional integrals. 

If the 2d theory is an SCFT, it also has an $U(1)_V$ vector R-symmetry. Furthermore, 
a conformal transformation can be employed to define two-point functions of local operators 
on the two-sphere, without deforming the theory. The group of super-conformal transformations contains the 
$SU(2|1)_A$ super-isometry group as a subgroup and one can consider correlation functions 
preserved by the same SUSY generator as before. 

The operators at the North pole will now belong to the chiral ring of the 2d SCFT, these at the South pole 
to the anti-chiral ring. The chiral ring operators are also naturally identified with the B-model ring $R^\cl$ and
the anti-chiral ones with $\bar R^\cl$. There is a positive-definite inner product
\begin{equation}
\langle \cdot, \cdot \rangle: \bar R^\cl \otimes R^\cl \to \bC
\end{equation}
making the space of chiral ring operators into an Hilbert space. 

The localization calculation is expected to reproduce this answer, but there is an obvious mismatch: it involves an $\Omega$-deformed space of local operators 
and a linear pairing between vector spaces which are not conjugate to each other. This is due to operator mixing: the localization calculation 
implicitly employs a computational scheme which treats local operators at the two poles in a different manner and does not preserve the full conformal symmetry. 

The operator mixing can be disentangled with the help of a Gram-Schmidt procedure, described in an analogous four-dimensional setup in 
\cite{Gerchkovitz:2016gxx} and in two dimensions by \cite{Ishtiaque:2017trm}. Recall that the space of chiral operators is graded by scaling dimension, which is identified with the weight 
for the $U(1)_V$ action. Consequently, $R^\cl$ is graded with components $R^\cl_d$. 
The same is true for $\bar R^\cl$, with the opposite sign of the $U(1)_V$ charge, so that 
$\bar R_d^\cl$ is conjugate to $R^\cl_d$ and the CFT inner product involves operators with the same weight. 

The crucial observation is that the equivariant rotation parameter, i.e. the inverse radius of the two-sphere,
is a dimensionful quantity.  In our conventions, it has weight $2$. The ambiguity in the definition of an operator of given scaling dimension thus only involves operators of smaller scaling dimension.
\footnote{Indeed, the it only involves operators whose scaling dimension/R-charge differ by even integral amounts.}
Correspondingly, the spaces $R$ and $\wt R$ are filtered: we can define subspaces 
$R_{\leq d}$ and $R_{<d}$ of operators of weight smaller or equal to $d$ of smaller than $d$ respectively. 
The associated graded 
\begin{equation}
\mathrm{gr} R_d \equiv R_{\leq d}/R_{<d}
\end{equation}
is canonically isomorphic to $R^\cl_d$. Analogous statements hold for $\wt R$ and $\bar R^\cl$. 

As a consequence, the SCFT inner product can be recovered from a Gram-Schmidt procedure, by defining operators $p_\alpha \in R_{\leq d}$ with highest weight part $\alpha \in R^\cl_d$ and
orthogonal to $\wt R_{<d}$ and $\wt p_\alpha \in \wt R_{\leq d}$ with highest weight part $\bar \alpha \in \bar R^\cl_d$ and orthogonal to $R_{<d}$. Then the inner product becomes
\begin{equation}
	\langle \alpha|\beta\rangle \equiv ( \wt p_\alpha, p_\beta )
\end{equation}
In concrete examples, it is far from obvious that the Gram-Schmidt procedure applied to the localization results will lead to a positive-definite inner product. In any specific localization setting, this 
appears to  be a rather non-trivial mathematical statement about a large class of finite-dimensional integrals. We will now review some examples and potential proof strategies. 

%We will sometimes be interested in situations where the Gram-Schmidt procedure is compatible with a pre-existing anti-linear map 
%$\tau: R \to \wt R$, with a classical , i.e. $\wt p_\alpha = \tau(p_\alpha)$. 

\section{LG partition functions}\label{app:LG}
The simplest localization formulae for sphere correlators occur in 2d LG theories, say 
with target $\bC^r$. 

The localization expression for sphere correlation functions of an LG theory defined by $r$ chiral multiplets $\varphi^i$ with holomorphic 
polynomials superpotential $W(\varphi)$ takes the form of a pairing
\begin{equation}
	(\wt f,g) \equiv \int_{\bC^r} d^r\varphi \,d^r\bar \varphi \, \wt f(\bar \varphi) g(\varphi) e^{W(\varphi) - \wt W(\bar \varphi)}
\end{equation}
Here we denoted the complex conjugate of the superpotential as 
\begin{equation}
	\wt W(\bar \varphi) \equiv \overline{W(\varphi)}
\end{equation}
The insertions $\wt f(\bar \varphi)$ and $g(\varphi)$ are (anti)holomorphic polynomials in the $\varphi^i$.

This integral is not absolutely convergent. We could try to make sense it in various ways. A natural way is to recast the integral
as a middle-dimensional contour integral in $\bC^r \times \bC^r$ 
\begin{equation}
	(\wt f,g) = \int_{\wt \varphi = \bar \varphi}  d^r \varphi \,d^r \wt \varphi \,\wt f(\wt \varphi) \,g(\varphi) e^{W(\varphi) - \wt W(\wt \varphi)}
\end{equation}
The integration contour can then be continuously deformed so that it goes to infinity in the region 
where $\mathrm{Re} \,W(\varphi)$ is positive and $\mathrm{Re} \,\wt W(\wt \varphi)$ is negative. 

The deformation can be implemented by a Morse flow for $\mathrm{Re} \,W(\varphi)- \mathrm{Re}\, \wt W(\wt \varphi)$. Recall that the Morse flow 
for the real part of an holomorphic function will keep the imaginary part unchanged and decrease the real part. 

Any amount of flow applied to the contour will be sufficient to {\it define} the original integral. More generally, Morse theory leads to a factorization of the integral
into a bilinear of exponential integrals 
\begin{equation}
	(\wt f,g) = \sum_{a,b} c_{ab} \left[\int_{\wt \varphi \in \wt \gamma_a} d^r \wt \varphi \,\wt f(\wt \varphi) e^{- \wt W(\wt \varphi)} \right]\left[\int_{\varphi \in \gamma_b}  d^r \varphi \,g(\varphi) e^{W(\varphi) } \right]
\end{equation}
for integers $c_{ab}$, expressing the decomposition of the $\wt \varphi = \bar \varphi$ contour into a basis of integration contours for the exponential integrals. We will come back to this momentarily. 

%Flowing by a large amount can give further insights. Critical points $(\varphi_*, \wt \varphi_*)$ of $W(\varphi) - \wt W(\wt \varphi)$ are fixed by the Morse flow, and the original contour passes through ``diagonal'' critical points $(\varphi_*, \wt \varphi_* = \bar \varphi_*)$. Regions of the original contour which lie away from critical points
%will be suppressed unless the flow hits other non-diagonal critical points, with lower values of $\mathrm{Re} W(\varphi)- \mathrm{Re} \wt W(\wt \varphi)$. 
%We will discuss this further after specializing to the class of superpotentials we are interested in. 

Another advantage of the Morse flow regularization is that it makes Ward identities manifest: the pairing annihilates polynomials of the form 
\begin{align}
	\wt f(\wt \varphi) &= \wt h^i(\wt \varphi) \partial_{\wt \varphi^i} \wt W(\wt \varphi) - \partial_{\wt \varphi^i} \wt h^i(\wt \varphi) \cr
	g(\varphi) &= h^i(\varphi) \partial_{\varphi^i} W(\varphi) + \partial_{\varphi^i} h^i(\varphi)
\end{align}
We denote these linear spaces of polynomials as $\wt I$ and $I$. These are deformations of the Jacobi ideals $\wt I_\cl$ and $I_\cl$ 
generated by $\partial_{\wt \varphi^i} \wt W(\wt \varphi)$ and $\partial_{\varphi^i} W(\varphi)$ respectively. 

The pairing $(\cdot, \cdot)$ is thus well-defined on the quotients $\wt R \equiv \bC[\wt \varphi]/\wt I$ and $R \equiv \bC[\varphi]/I$,
which are vector spaces deforming the Jacobi rings $\wt R_\cl$ of $\wt W(\wt \varphi)$ and $R_\cl$ of $W(\varphi)$. 

When $\wt W(\bar \varphi) = \overline{W(\varphi)}$, $\wt R_\cl$ and $R_\cl$ are canonically complex conjugate of each other,
i.e. complex conjugation gives an invertible anti-linear map $\rho_\cl: R_\cl \to \wt R_\cl$. On the other hand, there is no natural complex conjugation 
map relating $\wt R$ and $R$, due to the opposite sign in the deformation. 

If $W$ had isolated critical points, a natural basis would consist of thimbles labelled by critical points. We are interested in the opposite example where $W$ is a weight $2$ quasi-homogeneous polynomial, with the variables $\varphi^i$ having positive weights $\Delta_i$ and a single critical point at the origin. 

The Jacobi rings $R_\cl$ and $\wt R_\cl$ are graded by the weight. Denote as $R_\cl^{d}$ and $\wt R_\cl^{d}$ the subspaces of weight $d$. 
The spaces $R$ and $\wt R$ are only filtered. We can denote as $R^{\leq d}$ and $\wt R^{\leq d}$ the subspaces of weight smaller or equal to $d$
and as $R^{<d}$ and $\wt R^{<d}$ the subspaces of weight smaller than $d$. 

There are canonical isomorphisms $R_\cl^{d} \simeq R^{\leq d}/R^{<d}$ and $\wt R_\cl^{d} \simeq \wt R^{\leq d}/\wt R^{<d}$ between the associated graded of $R$ and $\wt R$ and the Jacobi rings. 

In this situation, it is possible to tentatively define an invertible anti-linear map $\rho: R \to \wt R$ via a Gram-Schmidt procedure.
The strategy is simple and recursive in the weight $d$: we build invertible maps $R^{d}_\cl \to R^{\leq d}$ which are the identity on $R^{\leq d}/R^{<d}$ and have an image orthogonal to 
all polynomials of lower degree, and analogously for $\wt R^{d}_\cl \to \wt R^{\leq d}$. We call these images ``orthogonal polynomials'' and denote them as $p_\alpha$ for $\alpha \in R_\cl$ and 
$\wt p_\alpha$ for $\alpha \in \wt R_\cl$. 

The pairing $(\wt p_\alpha, p_\beta)$ of orthogonal polynomials vanishes by definition if $\alpha$ and $\beta$ have different weight. The recursion works as long as the pairing 
$(\wt p_\alpha, p_\beta)$ between orthogonal polynomials of the same weight is non-degenerate.  The recursive step assume that we have built the orthogonal polynomials in weight smaller than $d$. For any $\alpha$ of weight $d$ we can pick some random representative $q_\alpha$ in $R^{\leq d}$ with leading term $\alpha$ and correct it by linear combinations of orthogonal polynomials in weight lower than $d$:
\begin{equation}
	p_\alpha \equiv q_\alpha - \sum_{\beta\,|\, \beta<\alpha} c^\beta_\alpha p_\beta\,.
\end{equation}
The correction is fixed uniquely by the linear equations obtained by setting to zero the pairing with orthogonal polynomials in $\wt R^{<d}$:
\begin{equation}
	 \sum_{\beta\,|\, \beta=\gamma} c^\beta_\alpha (\wt p_\gamma,p_\beta) =(\wt p_\gamma,q_\alpha) 
\end{equation} 
as long as the pairing of orthogonal polynomials is non-degenerate. 

The physical expectation is that the sesquilinear inner product $(\wt p_{\alpha^*}, p_\beta)$ should be positive definite and coincide with the 2d CFT positive Hermitian inner product $\langle \alpha, \beta \rangle$ on $R_\cl$. We will also define an anti-linear map $\rho$ from $R$ to $\wt R$ by $\rho(p_\beta) = \wt p_{\beta^*}$.

Notice that the form of $I$ and $\wt I$ allows one to define the weight modulo $2$ of elements in $R$ and $\wt R$, i.e. the operator 
$(-1)^d$. The filtrations can be defined within the subspaces of fixed weight modulo $2$, the 
inner product is only non-vanishing between elements of the same weight modulo $2$ and the orthogonal polynomials have definite weight modulo $2$. 

We can also consider the $i^d$ rotation, which flips the sign of the superpotential. We have 
\begin{equation}
	(\wt f,g) \equiv \int_{\bC^r} d^r\varphi d^r\bar \varphi \wt f(i^{-d} \bar \varphi) g(i^d \varphi) e^{-W(\varphi) + \wt W(\bar \varphi)}
\end{equation}
and thus
\begin{equation}
	(\wt f,g)^* \equiv \int_{\bC^r} d^r\varphi d^r\bar \varphi g^*(i^{-d} \bar \varphi) \wt f^*(i^{d}\varphi)  e^{W(\varphi) - \wt W(\bar \varphi)}
\end{equation}
equals the pairing of $g^*(i^{-d} \bar \varphi)$ and  $\wt f^*(i^{d}\varphi)$. In particular, 
$p_\alpha^*(i^{-d} \wt \varphi)$ must equal the dual orthogonal polynomial for the rotated, conjugated $\alpha$ and viceversa.  

\subsection{Relation to homology}
We can explore further the relation between the Gram-Schmidt procedure and the factorization of the localization expression into exponential contour integrals. 

Denote as $H$ the integral lattice of potential middle-dimensional integration cycles for the $g e^{W}$ contour integrals. The pairing 
between $H$ and $R$ is non-degenerate. Similarly, we can define a space $\wt H$ of potential middle-dimensional integration cycles for $\wt f e^{-\wt W}$.
There is a natural non-degenerate intersection product $\wt H \otimes_\bZ H \to \bZ$, where we take an integration contour $\wt \gamma \in \wt H$ and 
intersect with the complex conjugate $\bar \gamma$ of a cycle $\gamma \in H$ to produce $(\gamma, \wt \gamma)$. Indeed, $\wt \gamma$ goes to infinity 
in the region where $\wt W(\wt \varphi)$ has a large positive real part and $\bar \gamma$ in the region where $W(\bar \varphi)$ has a large negative real part. 
Hence we can intersect them with the $\bar \varphi = \wt \varphi$ identification. 

The contour deformation strategy we employed to define the pairing $(\wt f, g)$ can be implemented by deforming 
the $\bar \varphi = \wt \varphi$ contour to an element in $\wt H \otimes_\bZ H$. It is easy to see that this is just the inverse of the 
 $(\gamma, \wt \gamma)$ pairing. Indeed, if $\wt \gamma^a$ and $\gamma_a$ are dual bases in $\wt H$ and $H$,
 then we can read off the decomposition of $\bar \varphi = \wt \varphi$  by intersecting it with $\wt \gamma^a$ and $\gamma_b$.
 But this is just $(\gamma_b, \wt \gamma^a) = \delta_b^a$ and thus the integration cycle is just $\sum_a \wt \gamma^a \otimes_\bZ \gamma_a$:
 \begin{equation}
	(\wt f,g) = \sum_a \left[\int_{\wt \gamma^a} d^r \wt \varphi \wt f(\wt \varphi)  e^{- \wt W(\wt \varphi)} \right] \left[\int_{\gamma_a} d^r \varphi g(\varphi) e^{W(\varphi) } \right]
\end{equation}

The inner product of orthogonal polynomials becomes a bilinear of the period integrals 
\begin{align}
	\wt \cI_\alpha^a &\equiv \int_{\wt \gamma^a} d^r \wt \varphi \,\wt p_\alpha(\wt \varphi)  e^{- \wt W(\wt \varphi)} \cr
	\cI_{\alpha,a} &\equiv \int_{\gamma_a} d^r \varphi\, p_\alpha(\varphi) e^{W(\varphi) }
\end{align}

The filtrations on $R$ and $\wt R$ induce filtrations of the complexifications $H_\bC$ and $\wt H_\bC$,
by complex linear combinations of cycles which are orthogonal to all polynomials of degree smaller than $d$ or smaller or equal to $d$. 
These are sort of Hodge filtrations. 

By construction, $\sum_a \wt \cI_\alpha^a \gamma_a$ is precisely an element in $H_\bC$ orthogonal to polynomials of lower degree. Same for 
$\sum_a \cI_{\alpha,a} \gamma^a$. Indeed, they are the result of a Gram-Schmidt procedure on the filtrations for $H_\bC$ and $\wt H_\bC$.
 From that perspective, 
 \begin{equation}
 (\wt p_{\alpha}, p_\beta) = \sum_a \wt \cI_\alpha^a\cI_{\beta,a}
 \end{equation}
  is the inner product of these vectors induced from 
$\wt H \otimes_\bZ H \to \bZ$. 

%We can actually formulate the Gram-Schmidt procedure procedure directly in terms of $\rho$, without referring to $R_\cl$ and $\wt R_\cl$. We only need to 
%use the existence of an anti-linear identification $\rho_\cl$ between the associated graded of $R$ and $\wt R$. Then $\rho$ is defined as an invertible anti-linear map 
%from $R$ to $\wt R$ which agrees with $\rho_\cl$ and such that $(\rho(f),g)$ vanishes unless $f$ and $g$ have the same degree. 
%
%Indeed, we can determine $\rho(f)$ for $f$ of degree $d$ by picking any lift of $\rho_\cl([f])$ and adding some $\rho(f')$ for $f'$ of degree smaller than $d$:
%\begin{equation}
%	\rho(f) = \rho_\cl([f]) - \rho(f')
%\end{equation} 
%Imposing 
% \begin{equation}
%	 \left(\rho(f'),g \right) = \left( \rho_\cl([f]) ,g \right)
%\end{equation} 
%for all $g$ of degree smaller than $d$ fixes $f'$ uniquely as long as $\left(\rho(f'),g \right)$ is non-degenerate in degree smaller than $d$. 
%

\subsection{One-variable examples}
With a single field, we can only employ $W(\varphi) = \frac{1}{k+1} \varphi^{k+1}$. 
The Jacobi ring consists of $\varphi^a$ for $0\leq a<k$. In particular, there is no space for operator mixing:
weights $\frac{2a}{k+1}$ differ by less than $2$. Positivity, though, is still non-trivial. 

The integrals
\begin{equation}
	\int e^{\frac{1}{k+1} \varphi^{k+1} - \frac{1}{k+1} \bar \varphi^{k+1}} \bar\varphi^a \varphi^a |d\varphi|^2
\end{equation}
can be evaluated by doing first the angular integral to get
\begin{equation}
	2 \pi \int_0^\infty J_0 (2/(k+1) r^{k+1}) r^{2a+1} dr = \frac{\pi  \left(\frac{1}{k+1}\right)^{-\frac{2 (a+1)}{k+1}} \Gamma
   \left(\frac{a+1}{k+1}\right)}{k \Gamma \left(\frac{k-a}{k+1}\right)}
\end{equation}
which is indeed positive in the $0 \leq a <k$ range. 

%For future reference, we will also compute 
%\begin{equation}
%	\int e^{\frac{1}{k+1} \varphi^{k+1} - \frac{1}{k+1} \bar \varphi^{k+1}} \bar\varphi^{\bar a} \varphi^a |d\varphi|^2
%\end{equation}
%for $\bar a - a$ multiple of $k+1$. The angular integral gives {\bf fix}
%\begin{equation}
%	2 \pi \int_0^\infty J_{(\bar a - a)/(k+1)} (2/(k+1) r^{k+1}) r^{a+ \bar a+1} dr = \frac{\pi  \left(\frac{1}{k+1}\right)^{-\frac{2 (a+1)}{k+1}} \Gamma
%   \left(\frac{a+1}{k+1}\right)}{k \Gamma \left(\frac{k-\bar a}{k+1}\right)}
%\end{equation}

\section{The Gram-Schmidt procedure in ${\cal N}=4$ SUSY quantum mechanics} \label{app:GS}
In this section we will discuss a simple example where the Gram-Schmidt procedure allows one to recover a positive-definite 
inner product on cohomology representatives of ground states. The specialization to a LG SQM should give a proof of positive-definiteness in the 
sphere localization of 2d LG theories.

\subsection{Statement of the problem}
In order to see why this is a non-trivial result, recall the standard  ${\cal N}=2$ supersymmetric quantum mechanics setup: 
an Hilbert space $\cH$ equipped with two supercharges $Q$ and $Q^\dagger$ which satisfy
\begin{equation}
	Q^2=0 \qquad \qquad \{Q^\dagger, Q\}  = H \qquad \qquad (Q^\dagger)^2 = 0
\end{equation}
with $H$ being the Hamiltonian. As $H=(Q^\dagger + Q)^2$, it is non-negative. As long as the spectrum is discrete 
or at least has a gap, we can define a space $\cG$ of ground states annihilated by $H$ and also by both $Q$ and $Q^\dagger$. 
The space $\cG$ is still an Hilbert space, with an Hermitian inner product. 

States with energy above the gap are organized in pairs, as $Q$ and $Q^\dagger$ behave as a Clifford algebra. 
In each pair there is a state annihilated by $Q$ and a state annihilated by $Q^\dagger$. 
As a consequence, the pairs cancel out when we take the cohomology of $\cH$ with respect of $Q$, or with respect to $Q^\dagger$. 
In other word, the maps $\cG \to H^*(\cH,Q)$ and $\cG \to H^*(\cH,Q^\dagger)$ are isomorphisms of vector spaces.

The cohomology $H^*(\cH,Q)$ is typically easier to compute than $\cG$. Indeed, it is often the case that one can bring $Q$ to a simple form at the cost of 
making $Q^\dagger$ more complicated. This simplification, though, does not extend to the calculation of the inner product on $\cG$. 
The inner product in $\cH$ of a $Q$-closed vector with a $Q$-exact vector is not guaranteed to vanish and thus it does not descend naturally to an inner product on 
$H^*(\cH,Q)$. 

Instead, in order to compute the inner product inherited from $\cG$ one has to explicitly invert the isomorphism: 
lift a class $[\alpha] \in H^*(\cH,Q)$ to some $Q$-closed $\alpha \in \cH$ and solve the equation 
\begin{equation}
	Q^\dagger (\alpha - Q \alpha^{(1)}) = 0
\end{equation}
i.e.  $Q^\dagger Q \alpha^{(1)} = Q^\dagger \alpha$. Then the inner product of cohomology classes 
$[\alpha]$ and $[\beta]$ is defined as the inner product between $\alpha - Q \alpha^{(1)}$ and any lift $\beta$ of $[\beta]$. 

The inner product descends naturally to a well-defined sesqui-linear pairing $\overline{H^*(\cH,Q^\dagger)} \otimes H^*(\cH,Q) \to \bC$ instead.
Indeed, $Q^\dagger$-closed vectors are orthogonal to $Q$-exact vectors and vice-versa. In order to reconstruct the inner product from this pairing, though,
we would need again to find the ground state $\alpha - Q \alpha^{(1)}$ representing a given $Q$-cohomology class $[\alpha]$ and project that to $H^*(\cH,Q^\dagger)$.

\subsection{${\cal N}=4$ Supersymmetric Quantum Mechanics}
The situation is much better for quantum mechanical systems endowed with ${\cal N}=4$ supersymmetry and a certain amount of R-symmetry.
 
Specifically, consider a quantum mechanical system with four super-charges, which we denote as $Q^\alpha$ and $Q_\alpha^\dagger$ with $\alpha = \pm 1$. We 
require commutation relations  
\begin{equation}
	\{Q^\alpha, Q^\beta \} =0 \qquad \qquad \{Q_\alpha^\dagger, Q^\beta\}  = H \delta^\beta_\alpha \qquad \qquad \{Q^\dagger_\alpha, Q^\dagger_\beta \} =0 
\end{equation}
where $H$ is the Hamiltonian. All operators act on the Hilbert space $\cH$. 

We will assume that we have an $U(1)$ symmetry under which $Q^\alpha$ has weight $1$ and $Q_\alpha^\dagger$ has weight $-1$. 
We can decompose $\cH$ into subspaces $\cH_d$ of weight $d$. 

In many important examples we also have an $SU(2)$ symmetry rotating the Greek indices. In the following we will only really 
need a symmetry acting as $Q^1 \to Q^2$ and $Q_2 \to -Q_1$. An action of the Cartan of $SU(2)$ will be useful as a cohomological grading,
but not necessary. 

In such a situation, ground states $\cG$ are annihilated by all super-charges. The ground states can also be decomposed into 
subspaces $\cG_d$ of weight $d$. States above the gap transform in quadruplets:
one state annihilated by both $Q^\alpha$ and the images under the action of $Q^\dagger_\alpha$'s. 
A consequence is that we always have ``descent relations''. E.g. if a state $|\rangle$ is $Q^1$-closed, then $Q^\dagger_2|\rangle$
is $Q^1$-exact.

One can identify $\cG$ with the cohomology of any nilpotent linear combination of supercharges. The cohomology $H^*(\cH_d,Q^1)$ of individual supercharges such as 
$Q^1$ can be computed within each $\cH_d$. As a vector space, this is isomorphic to $\cG_d$.  Our objective is to recover the induced Hilbert space structure on $H^*(\cH_d,Q^1)$
using only cohomological operations, without solving for the actual ground states. 

The trick will be to deform the cohomology of $Q^1$ to the cohomology of linear combinations $Q^1 \pm Q^\dagger_2$ of supercharges with opposite weight. 
This is useful because the natural sesqui-linear pairing $\overline{H^*(\cH,Q^\dagger_1+  Q^2)} \otimes H^*(\cH,Q^1+  Q^\dagger_2) \to \bC$ can be 
combined it with an $SU(2)$ rotation identifying $H^*(\cH,Q^\dagger_1+  Q^2) \simeq H^*(\cH,-Q^\dagger_2+ Q^1)$
to get a natural pairing 
\begin{equation}
( \cdot, \cdot ): \,\overline{H^*(\cH,Q^1- Q^\dagger_2)} \otimes H^*(\cH,Q^1+ Q^\dagger_2) \to \bC. 
\end{equation}

The cohomology of $Q^1 \pm Q^\dagger_2$ cannot be graded by weight. We can still filter it by highest weight, though,
and define cohomologies $H^*(\cH_{\leq d},Q^1\pm Q^\dagger_2)$ of weight less or equal to $d$  and $H^*(\cH_{<d},Q^1\pm Q^\dagger_2)$
of weight less than $d$. These will be isomorphic as vector spaces to the corresponding $\cG_{\leq d} \equiv \oplus_{d' \leq d} \cG_d$ and $\cG_{< d} \equiv\oplus_{d' < d} \cG_d$
subspaces of $\cG$. 

Crucially, the associated graded of $H^*(\cH_{\leq d},Q^1 \pm Q^\dagger_2)$ is canonically identified with  $H^*(\cH_{d},Q^1)$. Indeed, the highest weight 
piece of a vector in $H^*(\cH_{\leq d},Q^1\pm Q^\dagger_2)$ must be $Q^1$-closed, and the highest weight element of any exact vector is $Q_1$-exact. 
Conversely, we can use descent to complete a class in $H^*(\cH_{d},Q^1)$ to a class in $H^*(\cH_{\leq d},Q^1 \pm Q^\dagger_2)$. 
This gives us some partial relation between $H^*(\cH_{\leq d},Q^1+ Q^\dagger_2)$ and  $H^*(\cH_{\leq d},Q^1- Q^\dagger_2)$ which does not require 
lifting classes to $\cG$. 

The Gram-Schmidt procedure allows one to reconstruct the positive-definite inner product on $H^*(\cH_{d},Q^1)$ precisely from the 
data of the filtrations $H^*(\cH_{\leq d},Q^1\pm Q^\dagger_2)$ and their pairing. Indeed, the concrete meaning of the filtration of 
$H^*(\cH,Q^1- Q^\dagger_2)$ is that a cohomology representative $\alpha_d + \alpha_{d-1} + \cdots$ of highest weight $d$ will contain 
a vector in $\cG_d$ represented by an element $[\alpha_d]$ of $H^*(\cH_{d},Q^1)$, but may also contain elements of $\cG_{<d}$ of lower weight. 
Such elements, though, would be detected by the inner product with elements in $H^*(\cH_{<d},Q^\dagger_1+  Q^2)$. 

We can fix this mixing problem by looking at elements $p_{[\alpha]} \in H^*(\cH_{\leq d},Q^1+ Q^\dagger_2)$
which have highest weight $[\alpha] \in H^*(\cH_{d},Q^1)$ and are orthogonal to $H^*(\cH_{<d},Q^1- Q^\dagger_2)$ under the natural pairing $( \cdot, \cdot )$. 
These are guaranteed to contain only the ground state labelled by $[\alpha]$. We can analogously define $\tilde p_{[\alpha]}\in H^*(\cH_{\leq d},Q^1- Q^\dagger_2)$, orthogonal to $H^*(\cH_{<d},Q^1+ Q^\dagger_2)$ under $( \cdot, \cdot )$.

Then $(\tilde p_{[\alpha]}, p_{[\beta]})$ recovers the positive definite inner product on $H^*(\cH_{d},Q^1)$ induced from $\cG_d$ and is computed 
through cohomological means only, without the need of computing actual ground states. \footnote{
The cohomology spaces $H^*(\cH,Q^1+ \zeta Q^\dagger_2)$ form an holomorphic bundle on the twistor $\bC P^1$ parameterized by $\zeta$. 
The inner product on $\cH$ descends to  pairings $\overline(H^*(\cH,Q^1- \bar \zeta^{-1} Q^\dagger_2)) \otimes H^*(\cH,Q^1+ \zeta Q^\dagger_2) \to \bC$
between the cohomologies at antipodal points on $\bC P^1$. Alternatively, it gives an identification between a fiber and its anti-Hermitean dual 
at antipodal points. Ground states give global sections of $H^*(\cH,Q^1+ \zeta Q^\dagger_2)$ over $\bC P^1$.
In principle, it should be possible to reconstruct the ground states and their inner product by trivializing the $H^*(\cH,Q^1+ \zeta Q^\dagger_2)$ bundle 
globally, even in the absence of an $U(1)$ symmetry. }

\subsection{Example: de Rham and Dolbeault cohomology. }
The classical example of $N=2$ SQM is the de Rham model, with $Q=d$ and $Q^\dagger = d^\dagger$ acting on forms on some Riemannian manifold
with the $L^2$ norm 
\begin{equation}
	\langle \alpha, \beta \rangle = \int \beta \wedge *\bar \alpha 
\end{equation}
The ground states are Harmonic forms, annihilated by $d$ and $d^\dagger$. 

On a Kahler manifold we can decompose $d = \partial + \bar \partial$ and define $Q^1 = \bar \partial$, $Q^\dagger_2 = \partial$. 
The weight of a $(p,q)$ form is $q-p$. The $SU(2)$ symmetry relating $Q^1 = \bar \partial$ and $Q^2 = \partial^\dagger$ 
is the Lefschetz symmetry, with raising and lowering operators defined by wedging and contracting with the K\"ahler form $\omega$. 

If the manifold is compact, the Hamiltonian will have a discrete spectrum. 

The decomposition of the space of ground states $\cG$ into weight spaces (and cohomological degree) gives harmonic $(p,q)$ forms. 
The space $H^*(\cH,Q^1+ Q^\dagger_2 = d)$ consists of de Rham cohomology. The Lefschetz symmetry 
allows us to identify $H^*(\cH,d^\dagger)$ with $H^*(\cH,\bar \partial - \partial)$, with the same weight but complementary form degree. 
The pairing between $H^*(\cH,\bar \partial + \partial)$ and $H^*(\cH,\bar \partial - \partial)$ becomes something like
\begin{equation}
	( \alpha, \beta )= i^{p-q} (-1)^{(p+q)(p+q-1)/2} \int \beta \wedge \bar \alpha \wedge \omega^{D-p-q}
\end{equation}
and indeed $\bar \partial + \partial$ is adjoint to $\bar \partial - \partial$ under this product up to an overall phase. 

Applying the Gram-Schmidt procedure gives de Rahm representatives for harmonic $(p,q)$ forms. The procedure 
is independent of the detailed choice of metric on the manifold. 

\subsection{Example: de Rham and Dolbeault cohomology with superpotential. }
The $N=2$ SQM de Rham model can be deformed by a function $h$, with $Q=d + dh \wedge$ and $Q^\dagger = (d+ dh \wedge)^\dagger$.

On a K\"ahler manifold, with $h$ being the real part of an holomorphic function $W$, we can split $d + dh \wedge$ into 
$Q^1 = \bar \partial + \partial W \wedge$ and $Q^\dagger_2 = \partial+ \bar \partial W \wedge$. 
We still have the Lefschetz symmetry. The $U(1)$ R-symmetry is preserved iff $W$ is quasi-homogeneous
(possibly up to constant shifts)

The presence of $W$ allows the Hamiltonian to have a discrete spectrum even if the manifold is non-compact, as long as the critical locus of $W$ is compact. 
We can apply the Gram-Schmidt procedure in this context. 

For simplicity, we now restrict to $\bC^r$ and a polynomial quasi-homogeneous $W$, as in the setup for the sphere correlation functions in a 2d LG theory.
We also assume that the representatives for the $Q^1$ cohomology take the form $g(\varphi) \,d^r \varphi$ modulo exact elements of the form $\partial W \wedge h$ for an holomorphic form $h$,
i.e. are identified with the Jacobi ring. We can also build representatives $g(\varphi) \,d^r \varphi \,e^{- \wt W(\bar \varphi)}$ for the $Q^1+ Q^\dagger_2$ cohomology,
modulo exact elements of the form  $(\partial h + \partial W \wedge h)\,e^{- \wt W(\bar \varphi)}$. We recover $R_\cl$, $R$ and the filtration structure from the previous Appendix \ref{app:LG}. 

Analogously, we can build representatives $\wt f(\bar \varphi) \,d^r \bar \varphi \,e^{W(\varphi)}$ for the $Q^1- Q^\dagger_2$ cohomology and pair them up 
to identify the 2d sphere partition function with the pairing between $H^*(\cH,\bar \partial + \partial)$ and $H^*(\cH,\bar \partial - \partial)$. 
The Gram-Schmidt procedure in 2d thus reproduces the inner product on $\cH$ and is in particular positive-definite.

\bibliographystyle{JHEP}

\bibliography{mono}

\end{document}